%% file: cpcosmics.tex
\documentclass[epj,nopacs]{svjour}
\usepackage{cite,mcite}
\usepackage[pdftex]{graphicx}
\usepackage{subfigure}
\usepackage{atlasphysics}
\usepackage{amsmath}
\usepackage{amssymb}
\usepackage{mathptmx}
\usepackage{helvet}
\usepackage[switch]{lineno}
\usepackage{hyperref}
\newenvironment{spaced_items}{

\begin{itemize}
  \setlength{\itemsep}{6pt}
  \setlength{\parskip}{0pt}
  \setlength{\parsep}{0pt}
}{\end{itemize}}

\begin{document}

\date{\today}
\title{Studies of the performance of the ATLAS detector using cosmic-ray muons}
 \author{The ATLAS Collaboration}
 \institute{}
%
%
\abstract{
Muons from cosmic-ray interactions in the atmosphere provide a high-statistics source of particles
that can be used to study the performance and calibration of the ATLAS detector.  
Cosmic-ray muons can penetrate to the cavern and deposit energy in all detector subsystems. Such events 
have played an important role in the commissioning of the detector since the start of the installation 
phase in 2005 and were particularly important for understanding the detector performance in the
time prior to the arrival of the first LHC beams.
Global cosmic-ray runs were undertaken in both 2008 and 2009 and these data have been used through to the 
early phases of collision data-taking as a tool for calibration, alignment and detector monitoring. 
These large datasets have also been used for detector performance 
studies, including investigations that rely on the combined performance of different 
subsystems. This paper presents the results of performance studies related to combined tracking,
lepton identification and the reconstruction of jets and missing transverse energy. Results are 
compared to expectations based on a cosmic-ray event generator and a full simulation 
of the detector response.
}
\maketitle
%
%

\section{Introduction}

The ATLAS detector~\cite{Atlas_detector} was constructed to provide excellent physics 
performance in the difficult environment of the Large Hadron Collider (LHC) at CERN~\cite{Evans:2008zzb}, 
which will collide protons at center-of-mass energies up to 14 TeV, with unprecedented 
luminosity. It is designed to be sensitive to any experimental signature that might be associated 
with physics at this new high-energy frontier. This includes precision measurements of high $p_{\rm T}$ leptons 
and jets, as well as large transverse-energy imbalances attributable to 
the production of massive weakly interacting particles. Such particles are predicted in numerous theories 
of physics beyond the Standard Model, for example those invoking weak-scale supersymmetry or the
existence of large extra dimensions.

Prior to the start of data-taking, understanding of the expected performance of individual subsystems 
relied on beam test results and on detailed GEANT4~\cite{Agostinelli:2002hh,Allison:2006ve} 
simulations\cite{Collaboration:2010wq}, 
including the modeling of inactive material both in the detector 
components and in the detector services and support structure. While extensive beam testing provided 
a great deal of information about the performance of the individual detector subsystems, a detailed 
understanding of the full detector could only be achieved after the system was in place and physics 
signals could be used for performance studies and for validation or tuning of the simulation.

\begin{figure*}[tbhp]
  \begin{center}
    \resizebox{0.98\textwidth}{!}{%
    \includegraphics{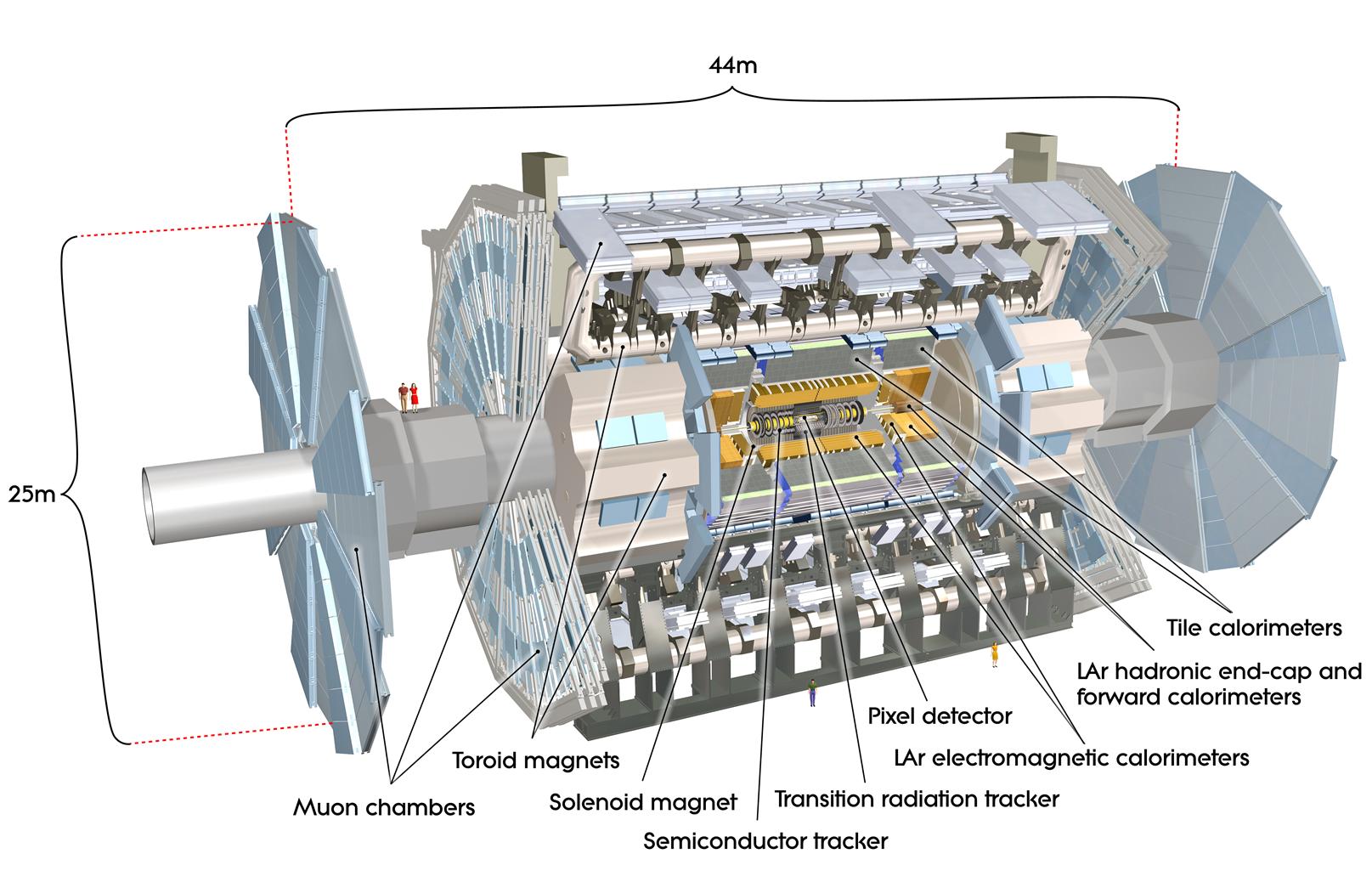}
}
  \end{center}
  \caption{The ATLAS detector and subsystems.
    \label{fig:detector}}
\end{figure*}

In both 2008 and 2009 the ATLAS 
detector collected large samples of cosmic-ray events. These extended periods 
of operation allowed for the training of shift crews, the exercising of the trigger and  data 
acquisition systems as well as of other infrastructure such as the data-handling system,
reconstruction software, and tools for hardware and data-quality monitoring. The large data samples 
accumulated have also been used for  a number of commissioning studies. 
Because cosmic-ray muons interact with the detector mainly as minimum-ionizing particles (MIPs), most 
traverse all of the subdetectors along their flight path. So, in addition to subdetector-specific cosmic-ray 
studies, these cosmic-ray data samples provide the first opportunity to study the combined performance of 
different detector components. Subsystem-specific cosmic-ray commissioning results have been documented 
in a series of separate publications\cite{IDCosmic,LArCosmic,TileCosmic,MSCosmic}. 
This paper presents the results of studies relevant to combined tracking performance, 
lepton identification and calorimeter performance for the reconstruction of jets and missing transverse energy. 
Where simulation results are available, results are compared to expectations based on a dedicated cosmic-ray 
event generator, implemented in the detector simulation.

\section{The ATLAS detector}

The ATLAS detector is described in detail elsewhere~\cite{Atlas_detector} and illustrated in Figure~\ref{fig:detector}. 
ATLAS uses a right-handed coordinate system with its origin at the nominal interaction point (IP).
The beam direction defines the $z$-axis, the positive $x$-axis points from the IP towards 
the center of the LHC ring and the positive $y$-axis points upwards. Cylindrical coordinates ($r$,$\phi$) are 
used in the transverse plane and the pseudorapidity $\eta$ is defined in terms of the polar angle $\theta$ as 
$\eta = -{\rm ln~tan}(\theta/2)$.

The ATLAS detector is made up of a barrel region and two endcaps, with each region consisting of several 
detector subsystems. Closest to the interaction point is the Inner Detector (ID), which performs charged 
particle tracking out to $|\eta| $ of 2.5. It consists of two silicon detectors -- the Pixel Detector and 
the SemiConductor Tracker (SCT) -- and the Transition Radiation Tracker 
(TRT), all immersed in a 2T axial magnetic field provided by a superconducting solenoid magnet. The
TRT is based on individual drift tubes with radiators, which provide for electron identification. The ID is 
surrounded by barrel and endcap liquid argon electromagnetic (EM) calorimeters which provide coverage
out to $|\eta|$ of 3.1. These, in turn, are surrounded by hadronic calorimeters. In the barrel region, 
the Tile Calorimeter is composed of steel and scintillating tiles, with a central barrel and two 
extended-barrel regions providing coverage out to $|\eta|$ of 1.7. In the endcap region the Hadronic 
Endcap Calorimeter (HEC) is based on liquid argon and covers the region $1.5 < |\eta| < 3.2$. The calorimetric 
coverage is extended into the region $3.2 < |\eta| <4.9$ by a liquid argon Forward Calorimeter (FCal) which 
occupies the same cryostat as the endcap EM calorimeter and the HEC. Beyond the calorimeter system is the 
Muon Spectrometer (MS), which relies on a set of massive superconducting air-core toroid magnets to produce 
a toroidal magnetic field in the barrel and endcap regions. In both regions, planes of interleaved muon 
detectors provide tracking coverage out to $|\eta|$ of 2.7 and triggering to $|\eta|$ of 2.4. The tracking 
studies presented in this paper are restricted to the barrel region of the detector, where precision 
measurements of the $(r,z)$ hit coordinates are provided by the Monitored Drift Tube (MDT) system. The remaining 
$\phi$ coordinate is measured by the Resistive Plate Chambers, or RPCs, which are primarily used for triggering. 

ATLAS employs a three-level trigger system, with the Level-1 (L1) trigger relying primarily on information from the 
Muon and Calorimeter systems. For cosmic-ray running there was additionally a TRT-based trigger at L1\cite{TRT-fast-or}. 
There is also a trigger based on signals from scintillators mounted in the endcap region, which are intended for 
triggering of collision events during the initial low-luminosity data-taking. This, however, plays no significant role
in the triggering of cosmic-ray events. 
For the MS, the triggering in the barrel region of 
the detector is based on hits in the RPCs; 
in the endcap region, the Thin Gap Chambers (TGCs) are used. The L1 Calorimeter trigger (L1Calo) is based on analog 
sums provided directly from the calorimeter front-end readout, from collections of calorimeter cells forming roughly 
projective trigger towers. In each case, the L1 trigger identifies a region of interest (ROI) and information from this ROI is 
transmitted to L2. In normal operation, events accepted by the L2 trigger are sent to the Event Filter which performs 
the L3 triggering, based on full event reconstruction with algorithms similar to those used offline. The L2 and L3 
trigger systems are jointly referred to as the High Level Trigger, or HLT. For the cosmic-ray data taking, events were 
triggered only at L1. Information from the HLT was used only to split the data into different samples.

\subsection{Tracking in ATLAS}
The two tracking systems, the ID and the MS, provide precision measurements of charged 
particle tracks. Reconstructed tracks are characterized by a set of parameters ($d_0$, $z_0$, $\phi_0$, $\theta_0$, $q/p$)
defined at the perigee, the point of closest approach of the track to the $z$-axis. The parameters $d_0$ and $z_0$
are the transverse and longitudinal coordinates of the perigee, $\phi_0$ and $\theta_0$ are the azimuthal and polar
angles of the track at this point, and $q/p$ is the inverse momentum signed by the track charge.
%
Analyses typically employ track quality cuts on the number of hits in a given tracking subsystem. The track 
reconstruction algorithms account for the possibility of energy loss and multiple scattering both in the material 
of the tracking detector itself, and in the material located between the tracking system and the particle production 
point. For the combined tracking of muons, which reconstructs the particle trajectories through both the ID and the 
MS, this requires an accurate modeling of the energy losses in the calorimeter. This will be discussed in 
section~\ref{s:muloss}.

\section{Cosmic-ray events in ATLAS}

\begin{figure}[t]
  \begin{center}
    \resizebox{0.49\textwidth}{!}{\includegraphics{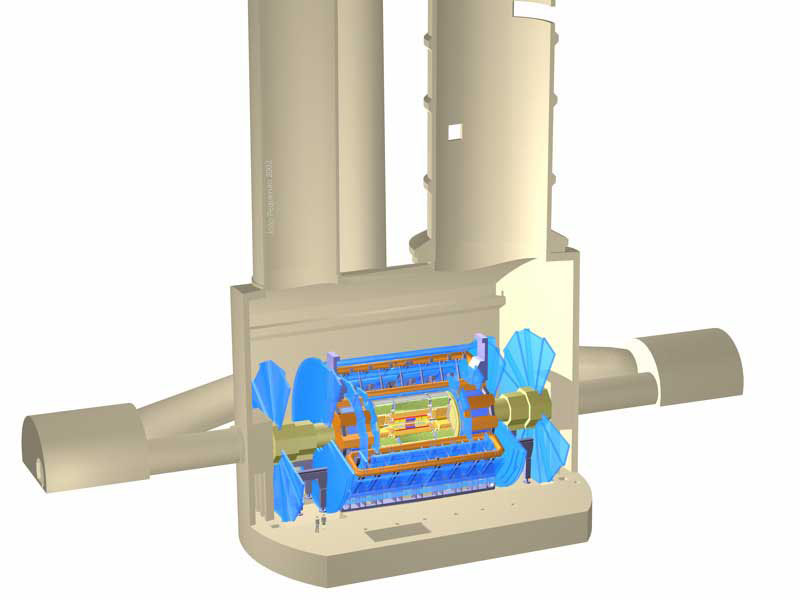}}
  \end{center}
  \caption{The ATLAS detector in the experimental cavern. Above the cavern are
the two access shafts used for the detector installation.
    \label{fig:cavern}}
\end{figure}

Cosmic rays in ATLAS come mostly from above, and arrive mainly 
via two large access shafts used for the detector installation, as illustrated in Figure~\ref{fig:cavern}.

In proton-proton collisions, the actual beam-spot position varies from the nominal IP by distances that are 
of order mm in the transverse plane and cm along the beam direction. Tracks produced in proton-proton collisions 
at the IP are said to be projective, that is, emanating from (or near, in the case of particles arising from 
secondary vertices) the IP. Cosmic-ray muons passing through the volume of the detector do not normally mimic 
such a trajectory. However, in a large sample of events, some do pass close to the center of the detector. 
By placing requirements on track impact parameters with respect to the nominal IP, it is possible to select a sample 
of approximately projective muons from those passing through the barrel region of the detector. Such cosmic-ray 
muons are referred to below as pseudo-projective. Due to the typical downward trajectory of the incoming cosmic-ray 
muons this cannot be done for those passing through the endcap region. For that reason, for those analyses presented 
here that 
rely on tracking, there is a requirement that the muons pass through the Inner Detector, which occupies a volume 
extending to about $1.15\,{\rm m}$ in radius and $\pm 2.7\,{\rm m}$ in $z$. The rate of such cosmic-ray muons is of 
order several~Hz. Most analyses further restrict the acceptance to the barrel region of the ID, which has a smaller 
extent, in $z$, 
of $\pm 71.2\,{\rm cm}$. Some analyses additionally place requirements on the presence of hits in the SCT or Pixel 
detectors, further restricting the volume around the nominal IP through which the cosmic-ray muons are required to pass.
Track-based event selection criteria are not applied in the case of the jet and missing transverse energy studies presented in 
section~\ref{sec:jets}, which focus on the identification of fake missing transverse energy due to cosmic-ray events 
or to cosmic-ray interactions that overlap with triggered events. 
While calorimeter cells are approximately projective towards the 
IP~\footnote{This is not the case for the FCal, which covers $3.2<|\eta|<4.9$, but that is not relevant to the analyses
presented here.}, energy deposits in the calorimeter can come from muons that pass through the calorimeter at any angle, 
including, for example, the highly non-projective up-down trajectory typical of cosmic muons passing through the endcap. 
While muons usually traverse the detector as MIPs, leaving only small energy deposits 
along their paths, in rare events they leave a larger fraction of energy in the detector, particularly in the case of 
energy losses via bremsstrahlung. These can be particularly important in the case of 
high-energy muons, which can lose a significant amount of energy between the two tracking detectors. Such events 
have been previously exploited for pulse shape studies of the LAr calorimeter
and as a source of photons used to validate the photon-identification capabilities of the ATLAS EM calorimeter~\cite{LArCosmic,LArEMDrift}. 

The reconstruction of cosmic-ray events is also complicated by the fact that they occur at random times with respect to the 
40~MHz readout clock, which is synchronized to the LHC clock during normal operation. For each subsystem, reconstruction 
of these events therefore first requires some measure of the event time with respect to the readout clock. An added 
complication, particularly for tracking, is that in the upper half of the detector, cosmic-ray muons travel from 
the outside in, rather than from the inside out, as would be the case for collisions. These differences can be addressed in 
the event reconstruction and data analysis. The modifications required for reconstruction of these events in the different 
detector components are discussed in the subsystem-specific cosmic-ray commissioning 
papers~\cite{IDCosmic,LArCosmic,TileCosmic,MSCosmic}. 

\subsection{Data samples}

ATLAS recorded data from global cosmic-ray runs during two extended periods, one in the fall of 2008 and another in 
the summer and fall of 2009. The analyses presented in this paper are each based on particular subsets of the available 
data. 


For studies involving only the calorimeter, events triggered by L1Calo are used. Studies relying on  
tracking require that both the MS and ID were operational, and that the associated toroidal and solenoid 
fields both were at nominal strength. All L1-triggered events taken under those conditions were checked 
for the presence of a track in the ID. Events with at least one such track were streamed by 
the HLT to what is referred to here as the Pseudo-projective Cosmic-ray Muon (PCM) dataset, which forms the 
basic event sample for all of the studies presented in section~\ref{sec:leptonID}. These events are mainly 
triggered at L1 by the RPCs. 
Hundreds of millions of cosmic-ray events were recorded during the 2008 and 2009 cosmic-ray runs. However, 
the requirement of a track in the Inner Detector reduces the available statistics 
dramatically, as does the requirement of nominal magnetic field strengths for the MS and ID, which is necessary 
for studies of the nominal tracking performance.

\subsection{Cosmic-ray event simulation}

Cosmic-ray events in ATLAS are simulated using a dedicated event generator and the standard GEANT4 detector 
simulation, with the modeling of the readout electronics adapted to 
account for the difference in timing. The simulation includes the cavern overburden, the layout of the access 
shafts and an approximation of the material of the surface buildings. The event generator is based on flux 
calculations in reference~\cite{Dar:1983pt} and uses a standard cosmic-muon momentum spectrum~\cite{PDG2008}. 
Single muons are generated near ground level, above the cavern in a $600\,{\rm m}\times600\,{\rm m}$ region centered above 
the detector, with angles up to 70$^\circ$ from vertical. Muons pointing to the cavern volume are  
propagated through up to $100\,{\rm m}$ of rock overburden, using GEANT4. Measurements of the cosmic-ray flux at 
different positions in the cavern were used to validate the predictions of this simulation. Once a muon has
been propagated to the cavern, additional filters are applied; only events with at least 
one hit in a given volume of the detector are retained, depending on the desired event sample.  Note that only single-muon 
cosmic-ray events are simulated. No attempt is made to model events in which cosmic-ray interactions produce an air shower 
that can deposit large amounts of energy in the detector. However, the rate of such events (in data) has been shown to be 
sufficiently low that they do not produce significant discrepancies in, for example, the agreement between data and Monte 
Carlo (MC) for the distribution of the summed transverse energy in cosmic-ray events\cite{HidekiOkawa:2010}.

\section{Lepton identification and reconstruction studies using cosmic-ray events}
\label{sec:leptonID}

Cosmic-ray muons are an important tool for the commissioning of the muon spectrometer, which is the largest 
ATLAS subsystem, occupying over 95\% of the total detector volume. 
As the rate of production of high-$p_{\rm T}$ muons in collision
events is rather low, the cosmic-ray data will continue to be relevant to the MS commissioning for some time to come. 
ATLAS continues to record data from cosmic-ray interactions when LHC beams are not present.

While the cosmic rays are primarily a source of muons, analysis of these data also allows for checks of the algorithms 
used to identify other leptons. The cosmic-ray muons serve as a source of electrons, mainly 
$\delta$-electrons but with smaller contributions from the conversion of muon 
bremsstrahlung photons and muon decays in flight. The identification of a sample of electrons allowed for an
examination of the performance of the electron identification algorithms, prior to first collisions. 
Similarly, although no 
$\tau$-leptons are expected in the cosmic-ray data sample, the tools designed for $\tau$-identification have been 
exercised using these data and checked against the simulation. 

The analyses discussed in this section rely on the PCM dataset described earlier, which contains cosmic-ray muon events
with tracks reconstructed in the ID. Most analyses also 
require the presence of hits in the Pixel Detector. These differ slightly for different analyses, 
as will be described below. 

\subsection{Combined muon tracking performance}
\label{sec:muons}

\begin{figure}[b!]
  \begin{center}
    \resizebox{0.49\textwidth}{!}{\includegraphics{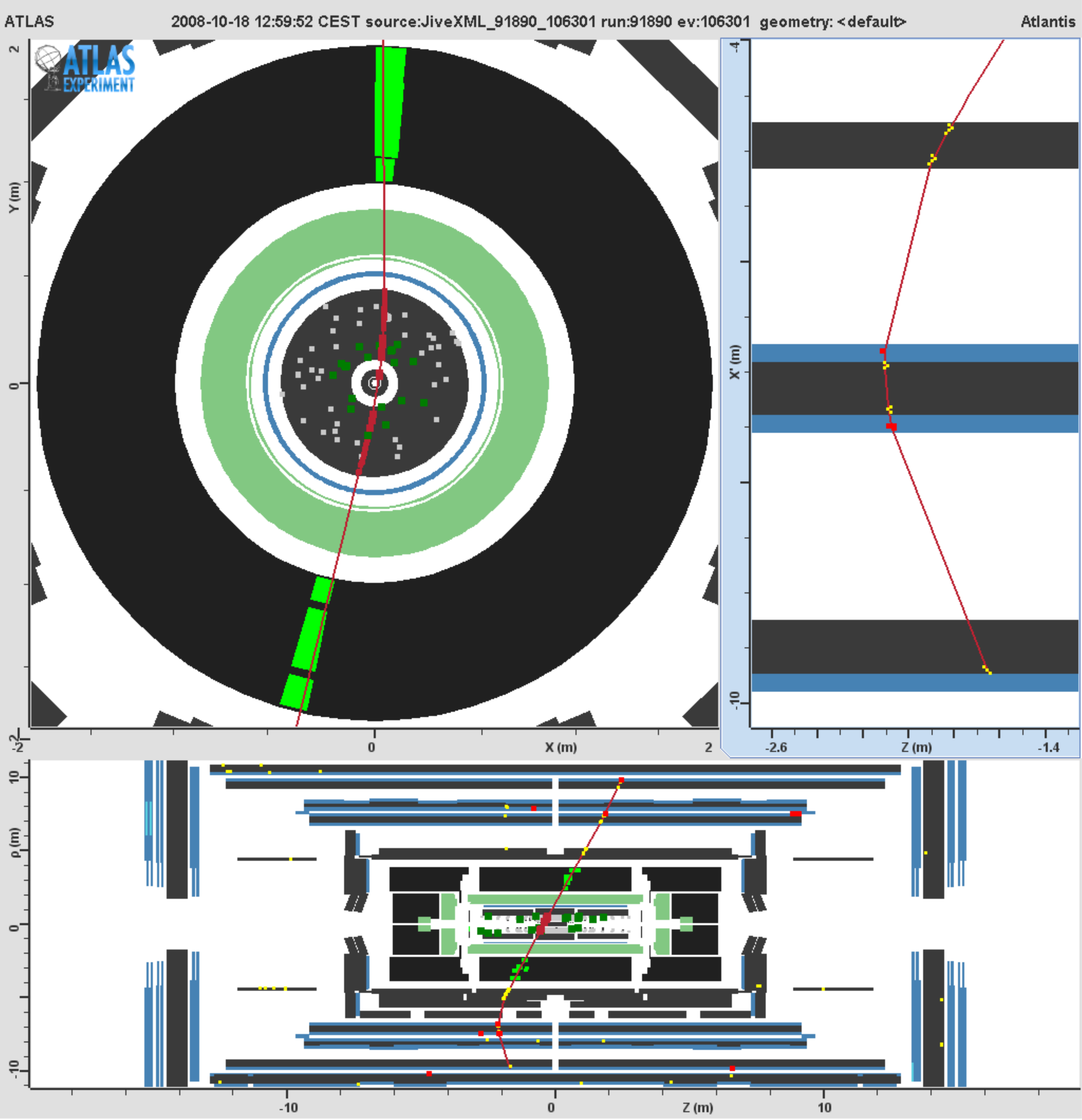}}
  \end{center}
  \caption{Event display of a cosmic-ray muon crossing the entire ATLAS detector, close to the nominal IP, 
leaving hits in all tracking subsystems and significant energy deposits in the calorimeter. The upper left
view shows the projection into the $r\phi$ plane. The lower plot shows the projection in the $rz$ plane.
The upper right projection is a longitudinal slice through the central part of the Muon Spectrometer at 
the $\phi$ value of the MDT planes in which the muon hits were recorded. 
    \label{fig:muon1}}
\end{figure}

This section describes studies of the performance of the combined tracking for muons, using cosmic-ray data recorded in 2009. 
The investigation uses the PCM dataset in order to have tracks that resemble, as much as possible, tracks from collision 
data. Selected events are required to have a topology consistent with that 
expected for the passage of cosmic-ray muon through the detector, which is illustrated by a typical event in 
Figure~\ref{fig:muon1}. The requirements are: 
\begin{itemize}
\item exactly 1 track reconstructed in the ID
\item 1 or 2 tracks reconstructed in the MS
\item exactly 1 combined track crossing both subdetectors
\end{itemize}
A special ID pattern recognition algorithm was used to reconstruct cosmic-ray muons as a single tracks. 
Because of the topology of these events, the analysis is restricted to the barrel region of the detector. 
Good quality ID and MS tracks are ensured using requirements on the number of hits in the different subsystems.
Events are required to have been triggered by the RPC chambers, since these also provide measurements along the 
$\phi$ coordinate ($\phi$ hits), which is not measured by the MDTs. Following the procedure used in the ID 
commissioning with cosmic-ray muons~\cite{IDCosmic}, a requirement is also placed on the timing from the TRT, 
to ensure that the event was triggered in a good ID time window.


The track parameter resolutions for Combined Muon (CM) tracks have been investigated in the same manner as used for similar 
studies of the ID\cite{IDCosmic} and MS\cite{MSCosmic} performance, by comparing the two reconstructed tracks left by a single 
cosmic-ray muon passing through the upper and then the lower half of the detector. In the case of the ID 
and combined tracks, this involves separately fitting the hits in these two regions, to form what are referred to
below as ``split tracks'' from the track created by the passage of a single muon.

Prior to a study of combined tracking, it is necessary to establish that the relative alignment of the two tracking 
systems is adequate. Checks were performed by comparing the track parameters for standalone tracks reconstructed 
by the two separate tracking systems, in the upper and lower halves of ATLAS. Tracks in the MS were reconstructed using 
a least-squares method that directly incorporates the effects of the 
material that sits between the MS detector planes and the point at which the track parameters are defined\cite{MS_MOORE}. 
ID tracking was also performed by standard tracking algorithms\cite{NewTrackingPubNote,GlobalChi2TrackFitter}.  

The alignment check relies on the study of three different classes of tracks: split ID tracks, MS standalone 
tracks, and split CM tracks. In what follows these will be referred to simply as ID, MS and CM tracks, 
respectively.
\begin{figure*}[phtb]
  \begin{center}
    \resizebox{0.46\textwidth}{!}{\includegraphics{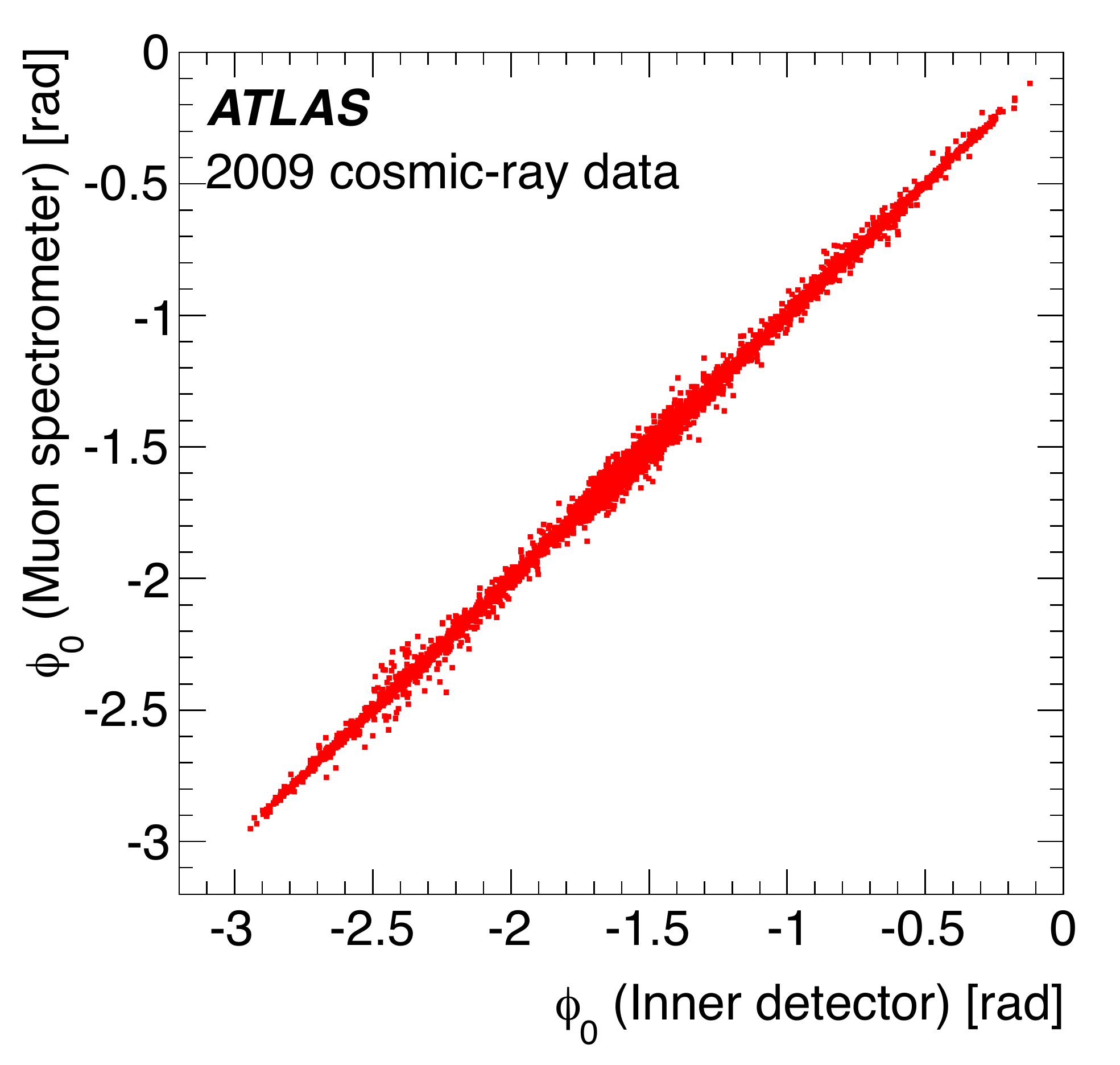}}
    \resizebox{0.46\textwidth}{!}{\includegraphics{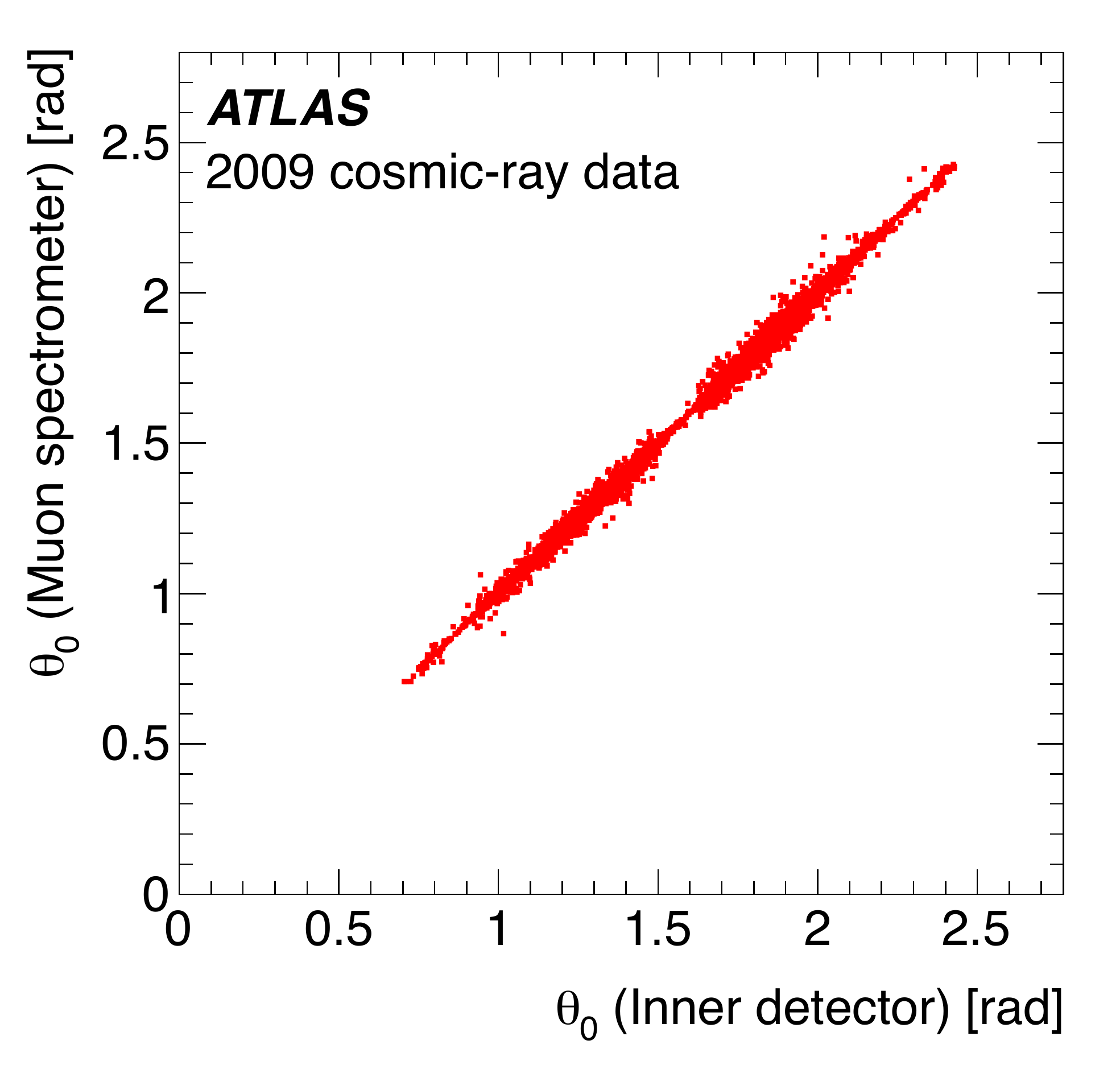}}
  \end{center}
  \caption{Correlations between the track parameters $\phi_0$ and $\theta_0$ obtained from 
standalone ID and MS tracks, in the bottom half of ATLAS. 
    \label{fig:muon2}}
\end{figure*}
\begin{figure*}[phtb]
  \begin{center}
    \resizebox{0.46\textwidth}{!}{\includegraphics{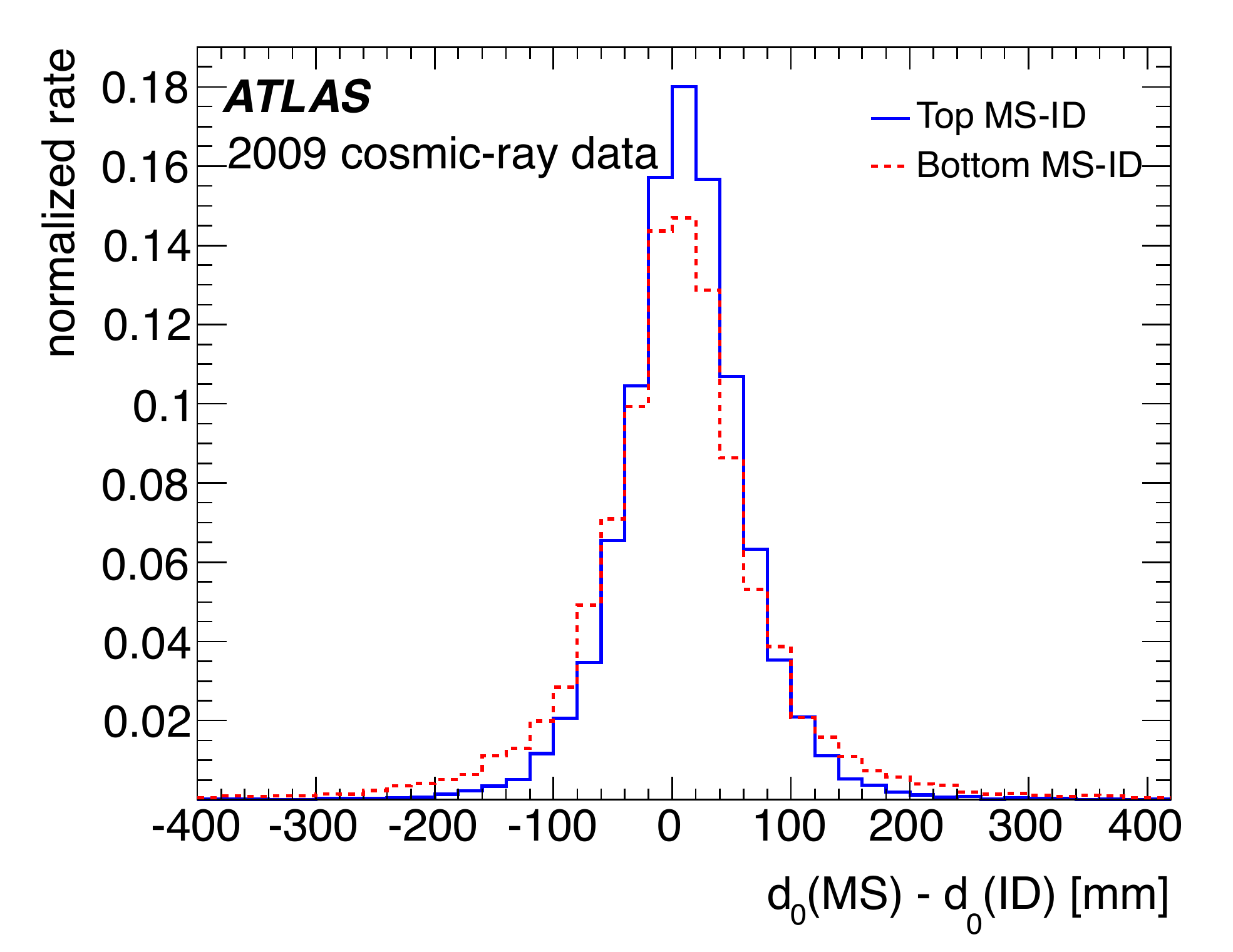}}
    \resizebox{0.46\textwidth}{!}{\includegraphics{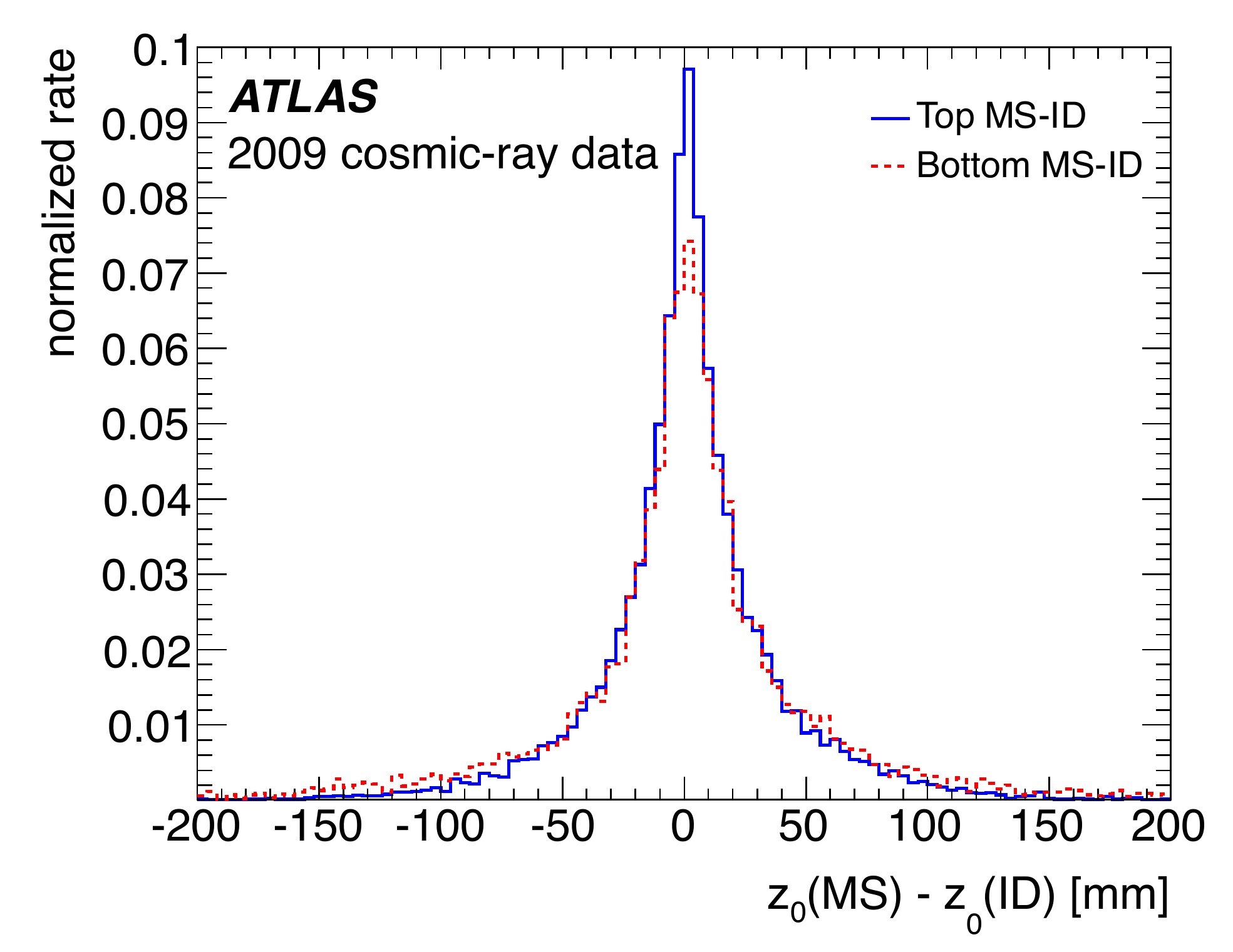}}
    \resizebox{0.46\textwidth}{!}{\includegraphics{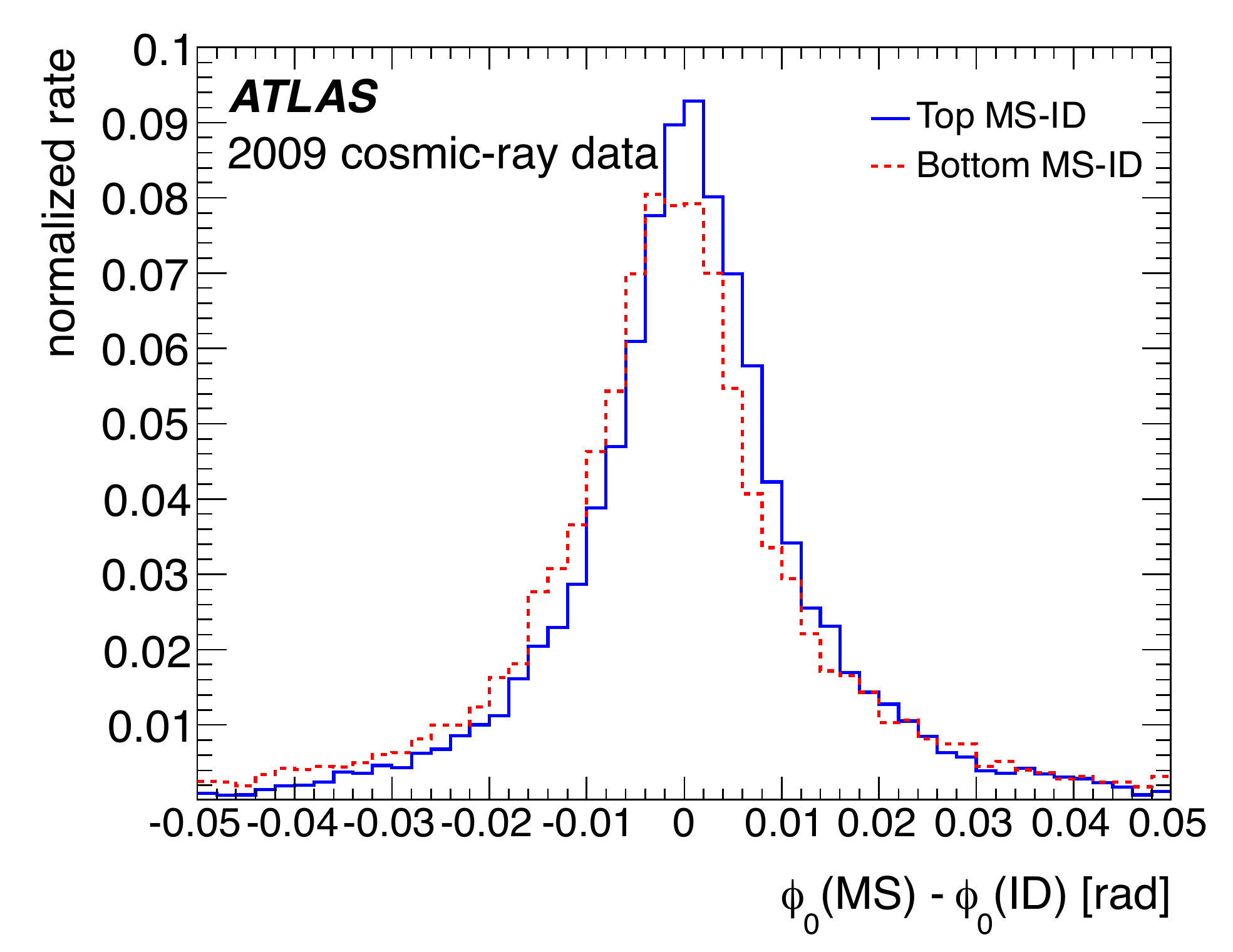}}
    \resizebox{0.46\textwidth}{!}{\includegraphics{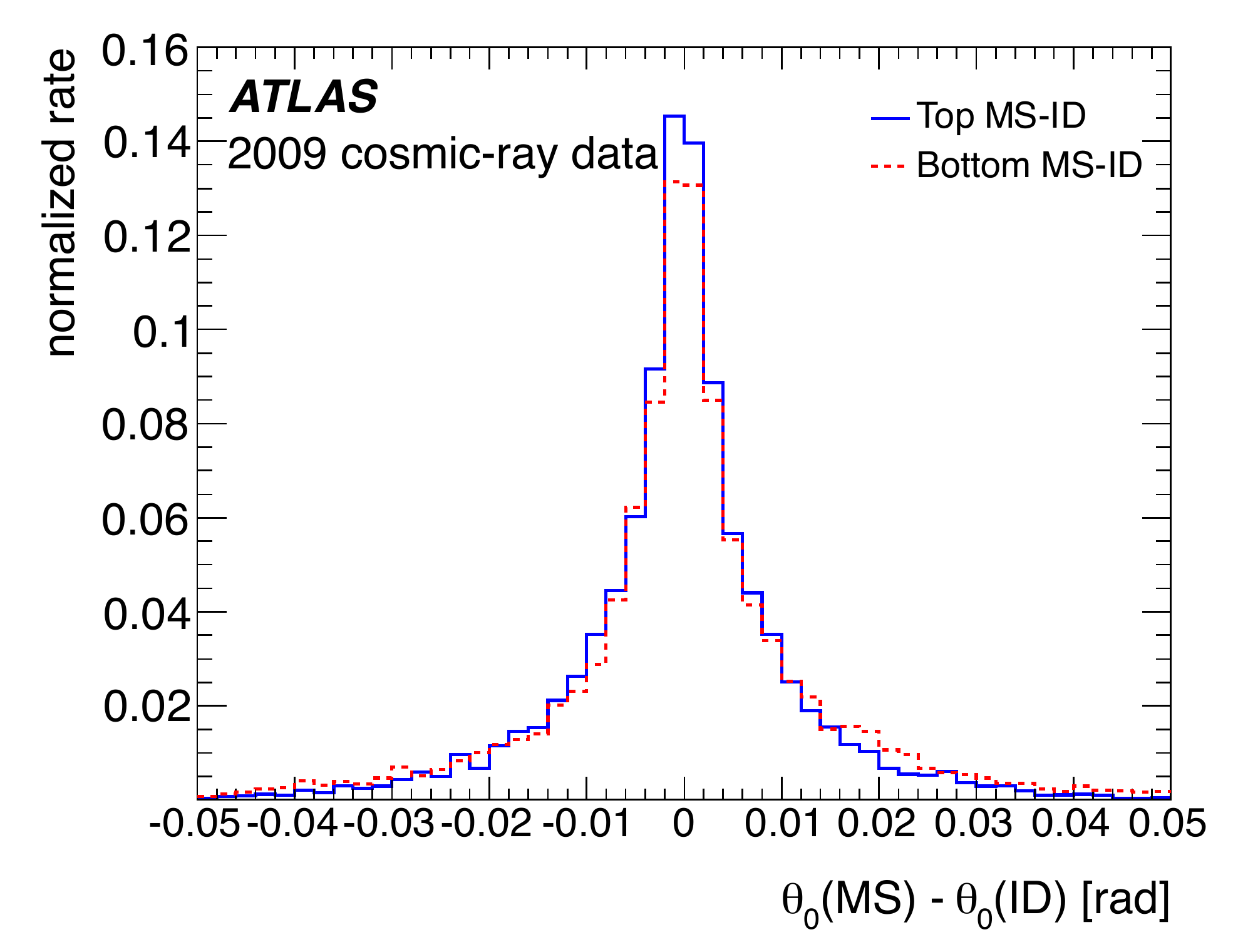}}
  \end{center}
  \caption{
Difference distributions of the track parameters, $d_0$, $z_0$, $\phi_0$ and $\theta_0$ obtained 
from standalone ID and MS tracks, for the top and bottom halves of the detector. 
    \label{fig:muon3}}
\end{figure*}
\begin{figure*}[p!]
  \begin{center}
    \resizebox{0.44\textwidth}{!}{\includegraphics{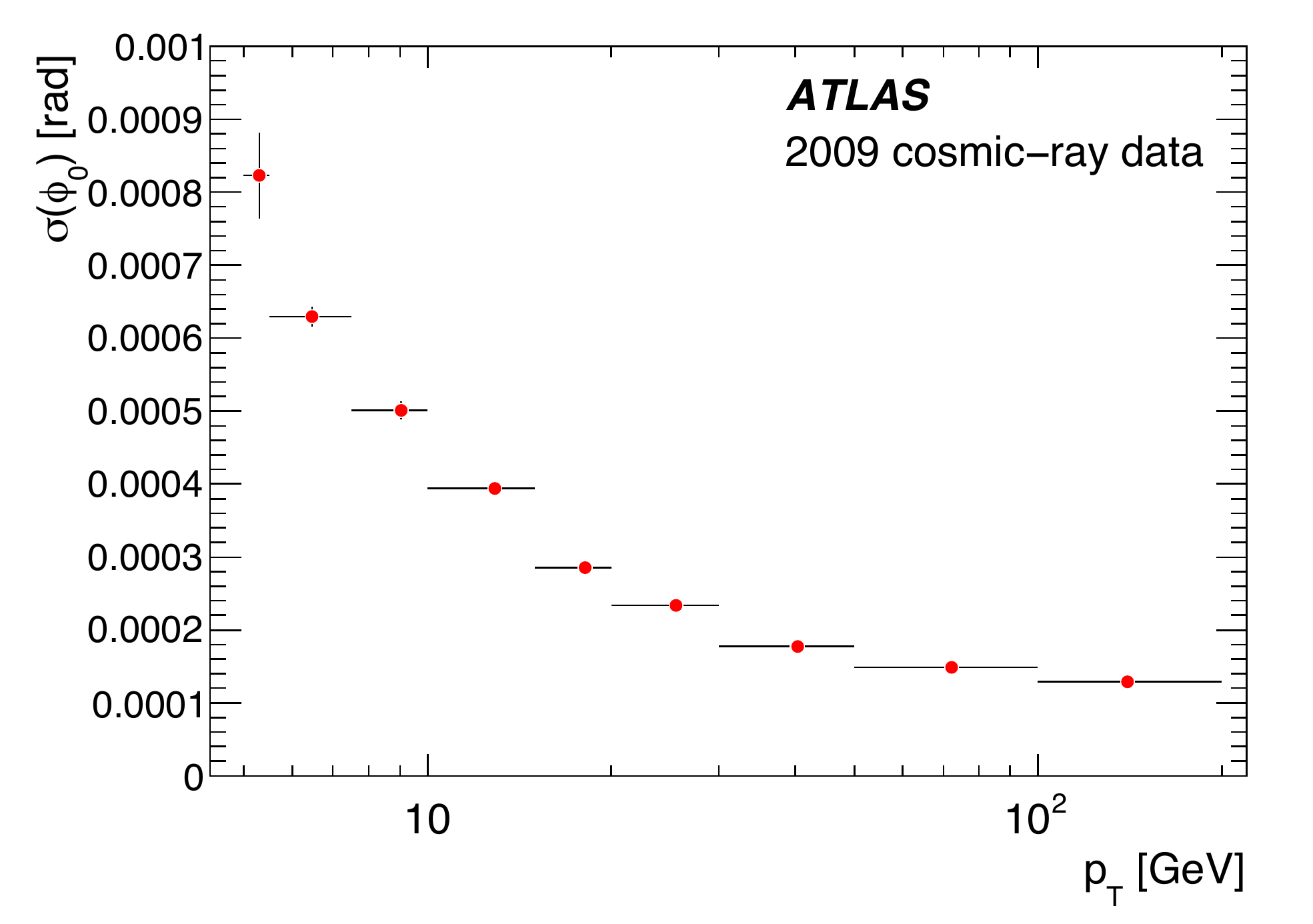}}
    \resizebox{0.44\textwidth}{!}{\includegraphics{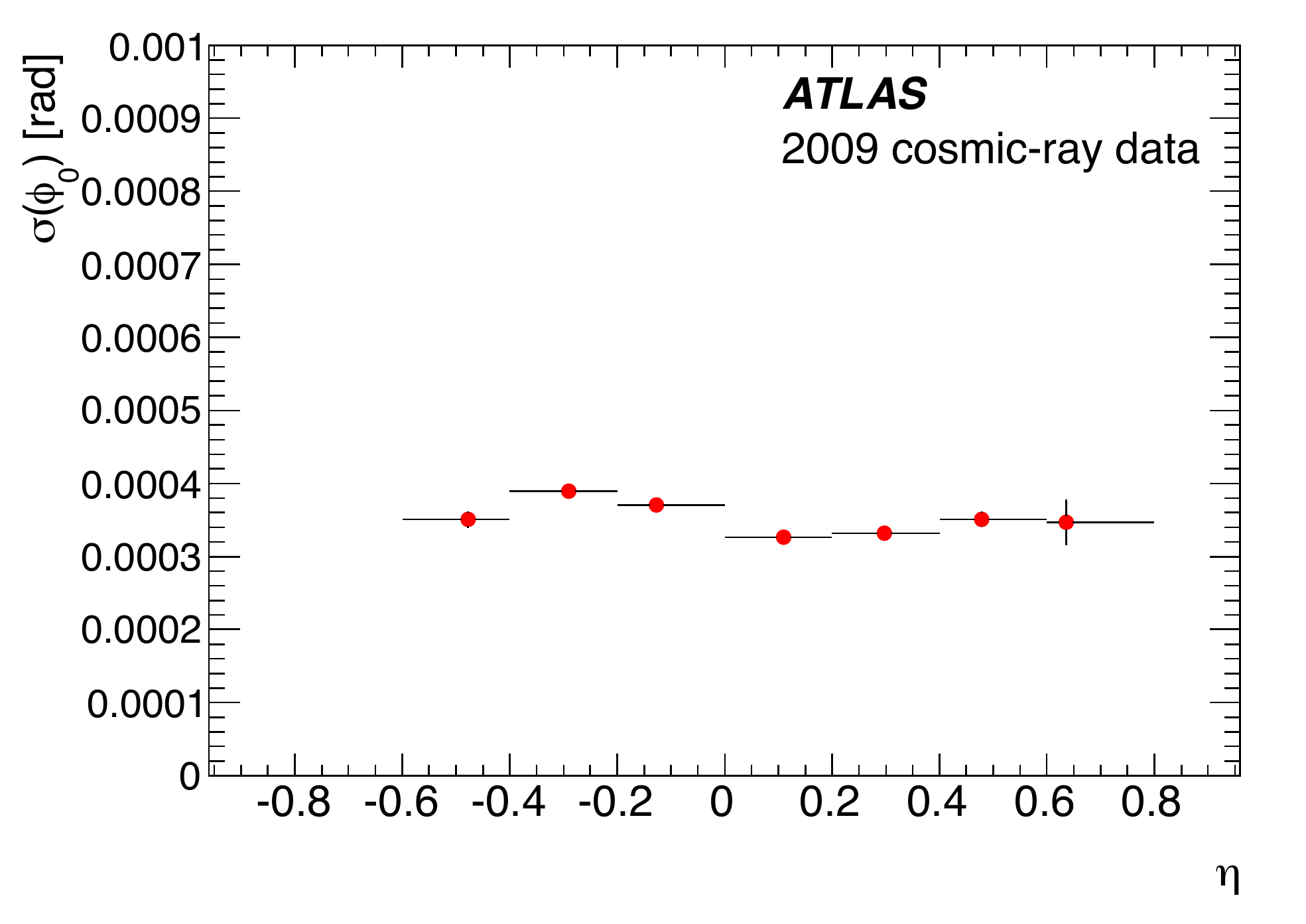}}
    \resizebox{0.44\textwidth}{!}{\includegraphics{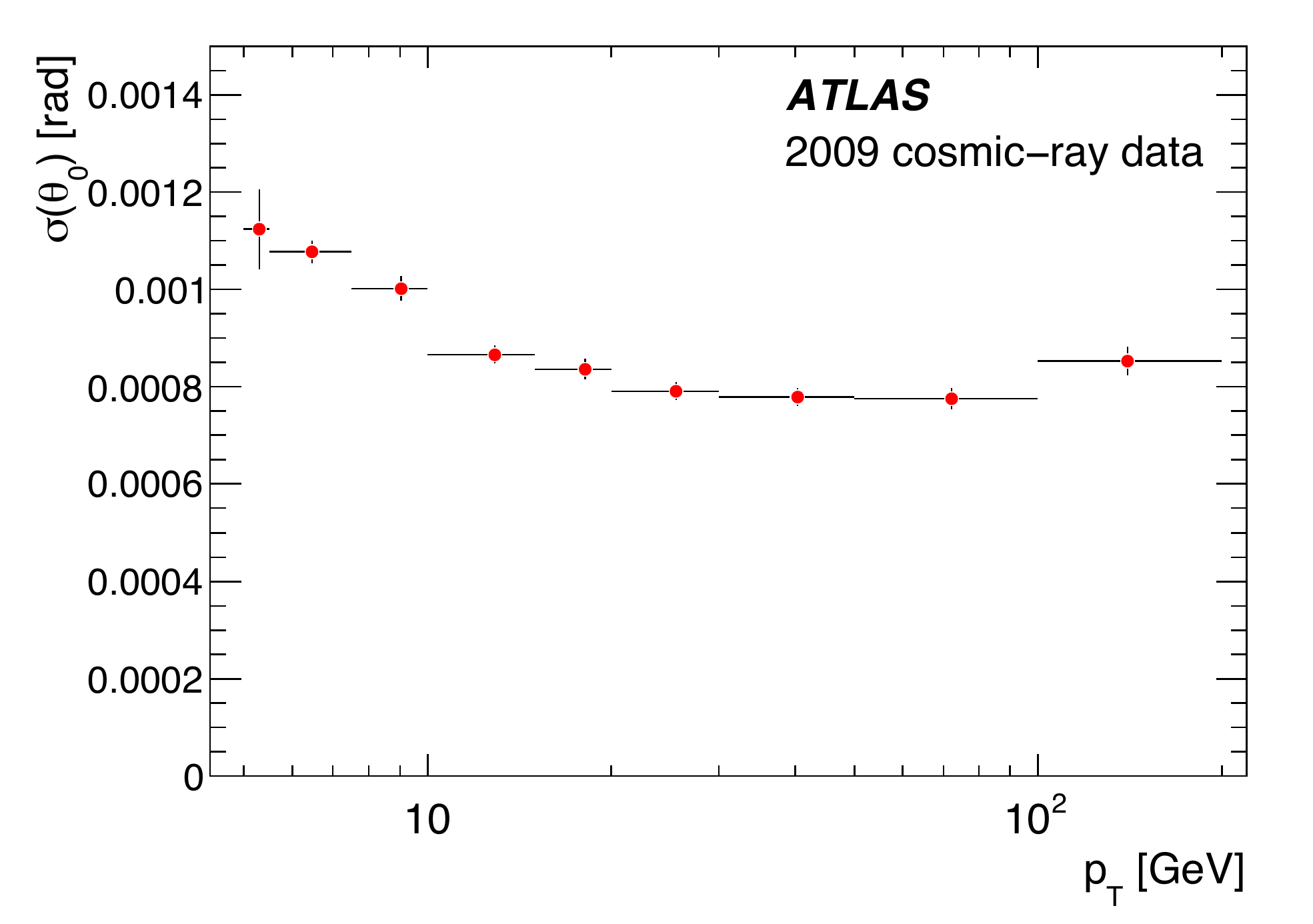}}
    \resizebox{0.44\textwidth}{!}{\includegraphics{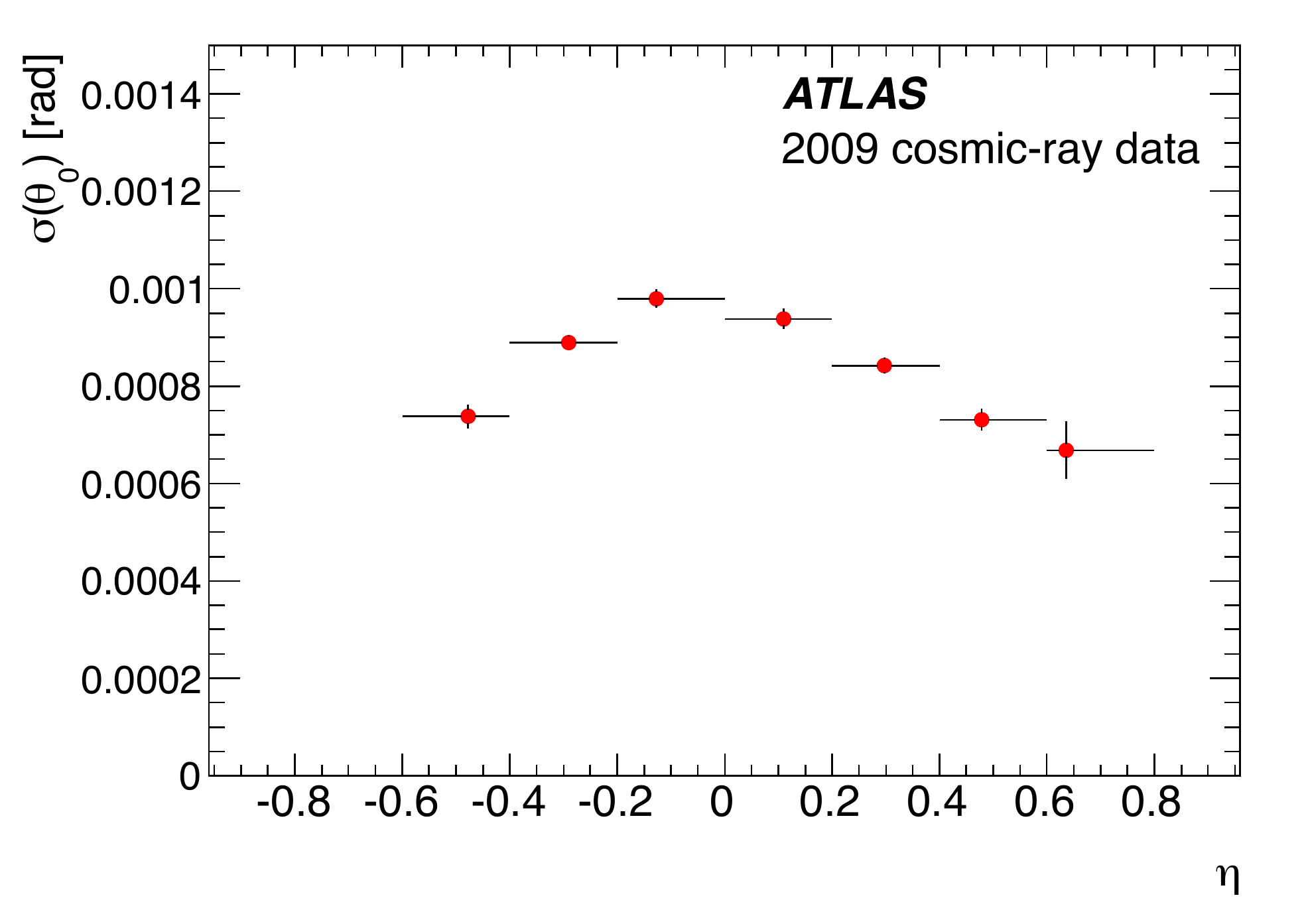}}
    \resizebox{0.44\textwidth}{!}{\includegraphics{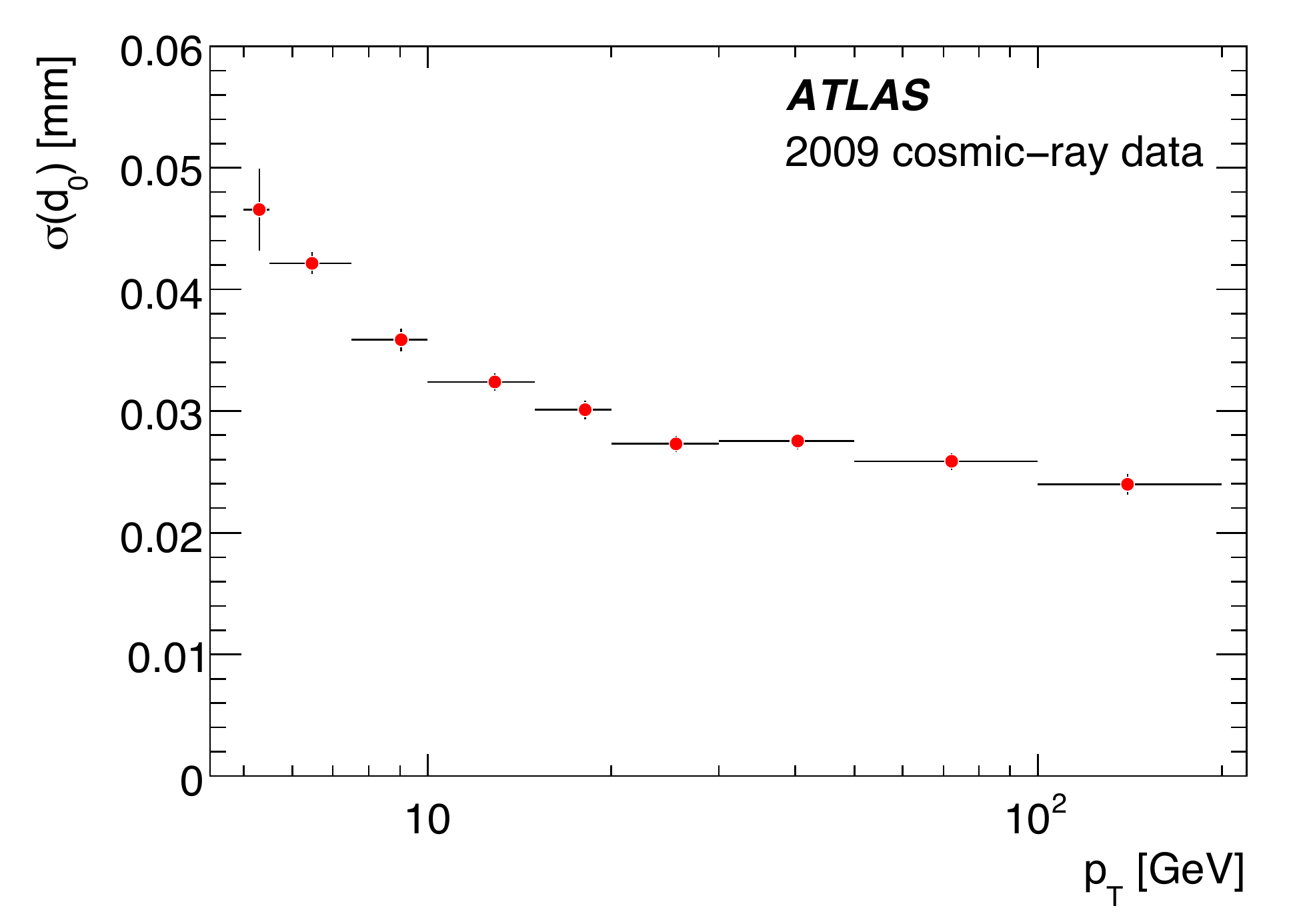}}
    \resizebox{0.44\textwidth}{!}{\includegraphics{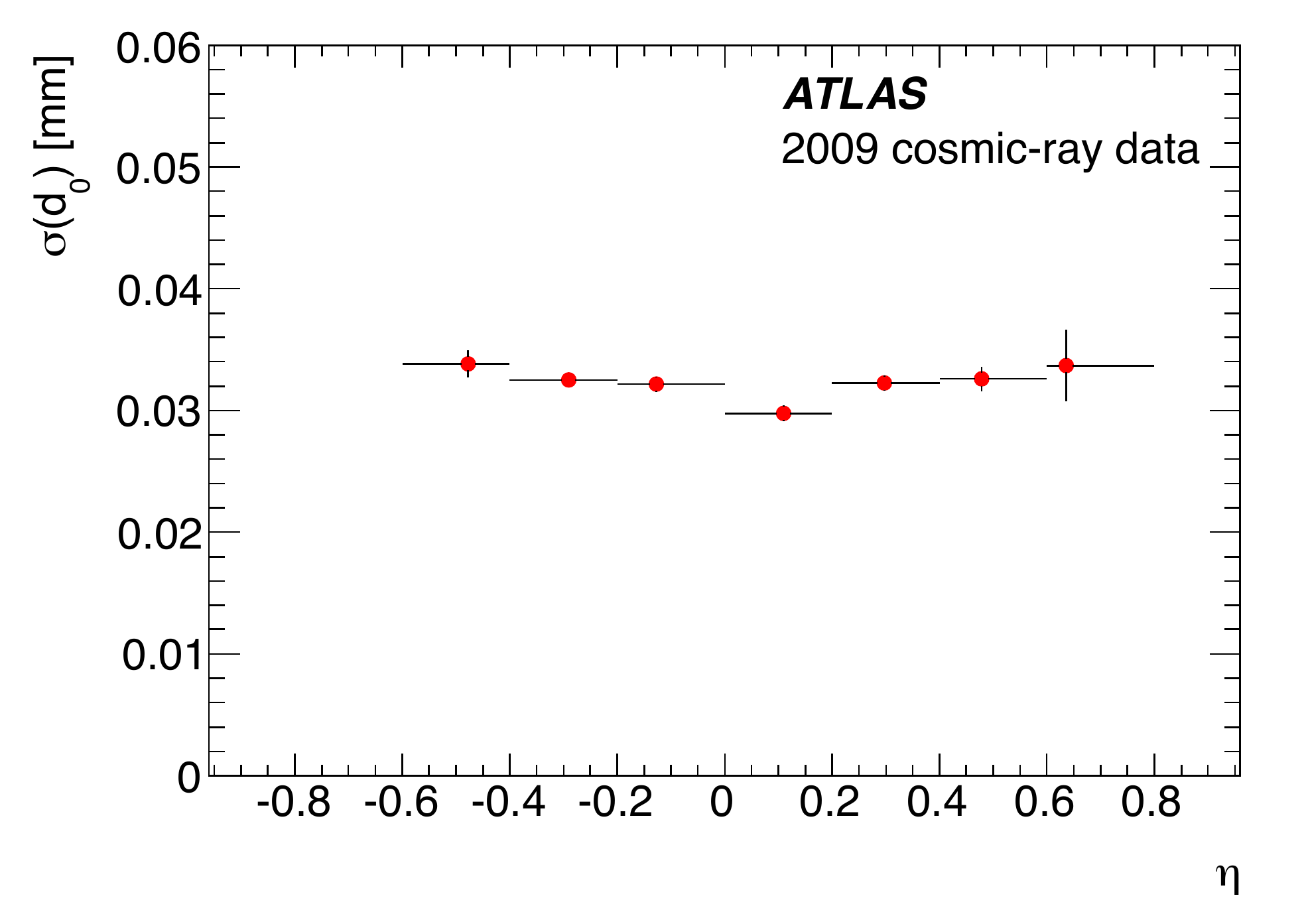}}
    \resizebox{0.44\textwidth}{!}{\includegraphics{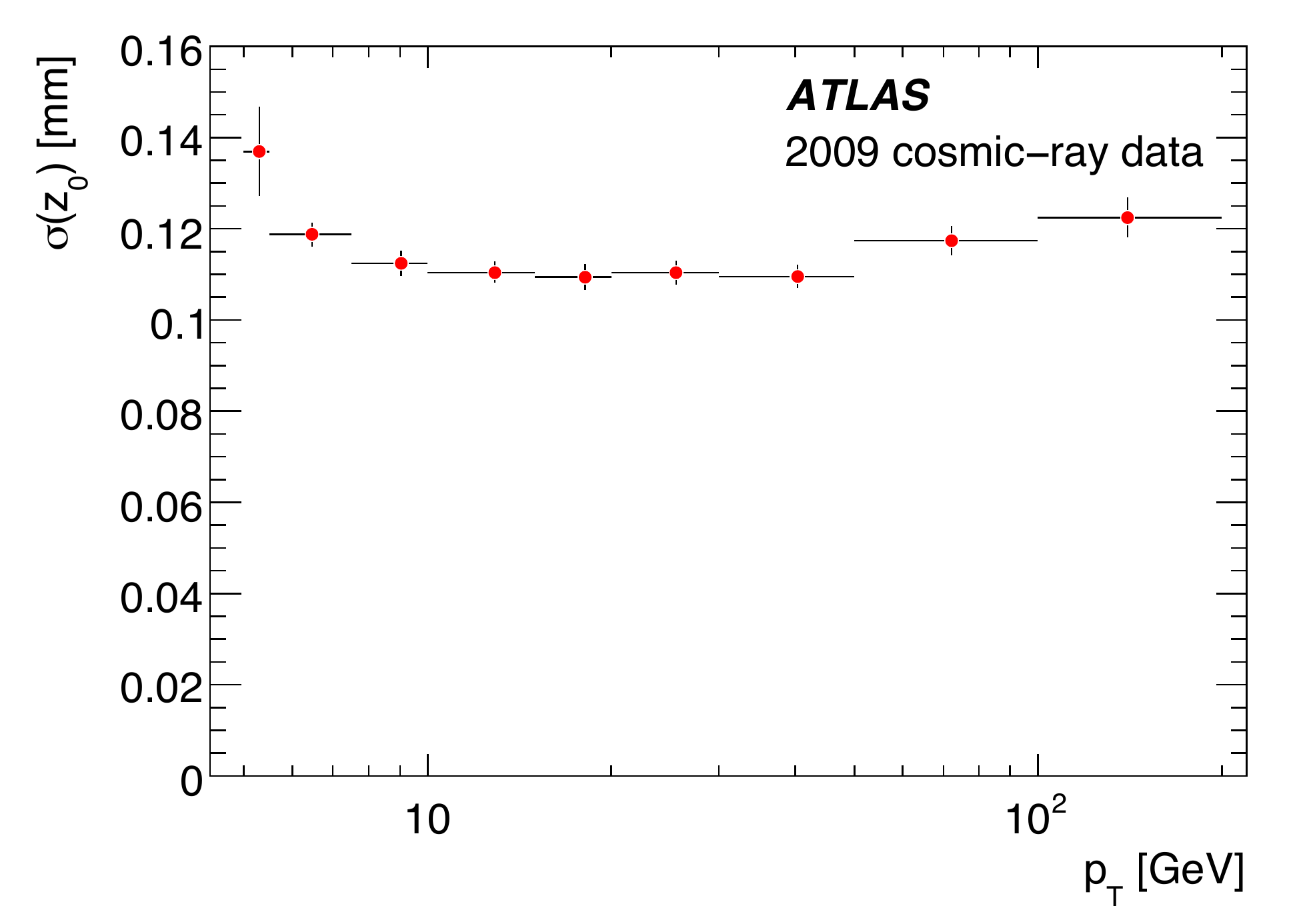}}
    \resizebox{0.44\textwidth}{!}{\includegraphics{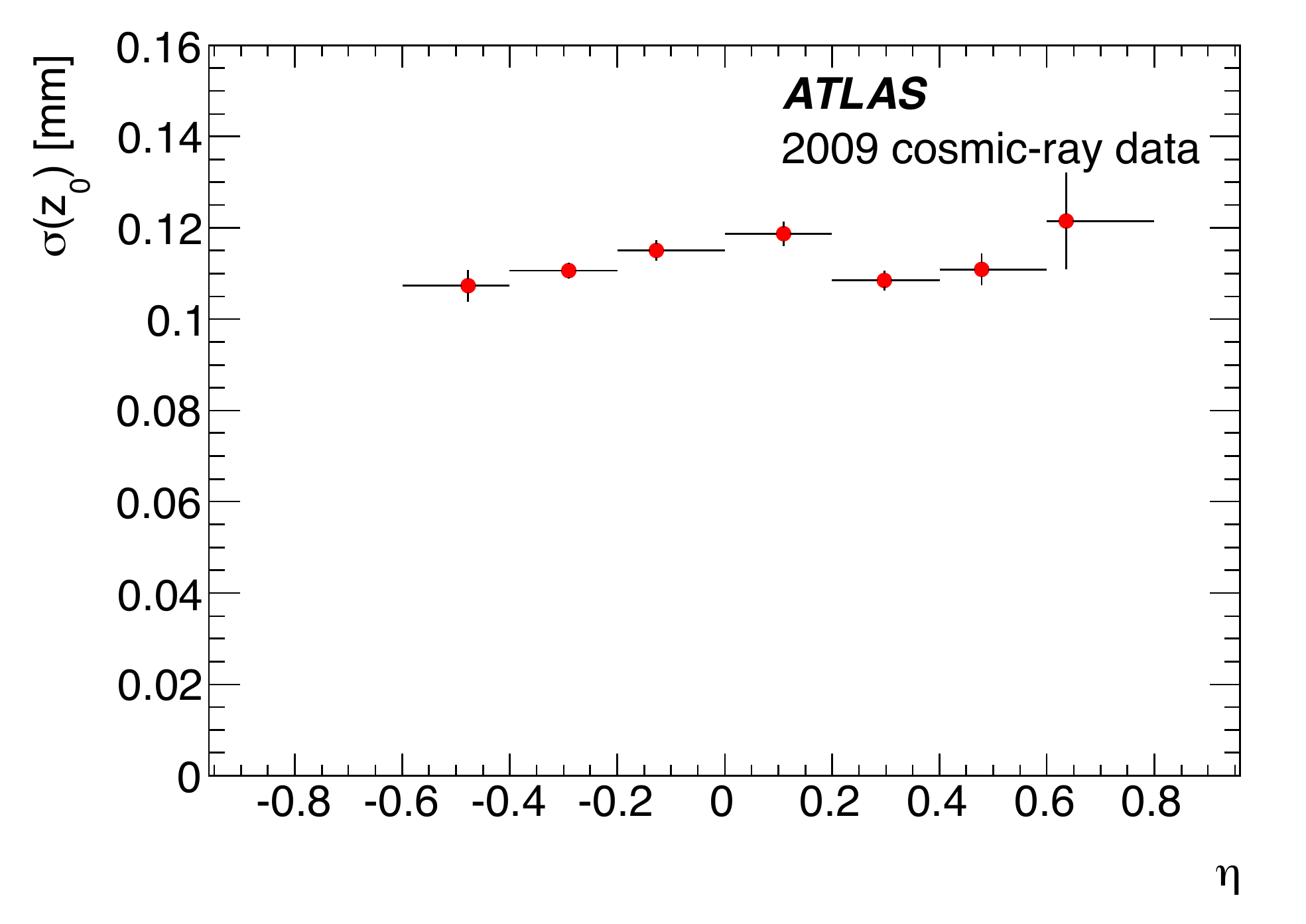}}
  \end{center}
  \caption{Resolution on track parameters $\phi_0$, $\theta_0$, $d_0$ and $z_0$, obtained from split tracks, 
as a function of $p_{\rm T}$ (left column) and $\eta$ (right column).
    \label{fig:muon5}}
\end{figure*}
Different quality cuts are placed on the three track types. For ID and CM tracks $|d_0|$ and $|z_0|$ are required to 
be less than $400\,{\rm mm}$ and $500\,{\rm mm}$ respectively. For MS tracks, for which these parameters must be extrapolated
from the MS back to the perigee, the requirements are $|d_0| < 1000\,{\rm mm}$ and  $|z_0| < 2000\,{\rm mm}$. ID and CM 
tracks are required to have at least 1, 6 and 20 hits in the Pixel, SCT and TRT detectors, respectively. MS and CM tracks 
are required to have hits in all three MS layers, with more than four RPC hits, at least two of which are $\phi$ hits, 
and a $\chi^2$ per degree of freedom less than 3. All tracks are required to have momentum larger than 5~GeV.

Figure~\ref{fig:muon2} shows the correlation between the $\phi_0$ and $\theta_0$ parameters determined from MS and ID 
tracks in the bottom half of ATLAS. Very good consistency is evident and similar results are obtained in the other hemisphere. 
The level of agreement between the two systems is better quantified by distributions of the difference between the track parameters 
obtained from the two systems. These are shown in Figure~\ref{fig:muon3} for $d_0$, $z_0$, $\phi_0$ and $\theta_0$, separately 
for tracks in the upper and lower halves of the detector. The somewhat narrower distributions obtained from 
the upper half of the detector are attributed to the higher average momentum of the cosmic-ray muons in this part of the detector, 
since those in the bottom have lost energy passing through the lower half of the calorimeter before reaching the MS. Small biases 
are observed for the $d_0$ and $\phi_0$ parameters. These are consistent with a slight translational 
misalignment between the MS and ID that is of order 1 mm. However, the combined tracking study presented 
below was performed without any relative ID-MS alignment corrections.


The track parameter resolutions for 
combined tracking have been investigated in the manner discussed above, using CM tracks passing through the barrel 
part of the detector, which are split into separate tracks in the upper and lower halves. The two resulting tracks 
are then fitted using the same combined track fit procedure. For studies of the angular and impact parameter resolution, 
the track quality cuts are tightened somewhat, with the requirements of at least two pixel hits, 
$|d_0|< 100\,{\rm mm}$ and $|z_0|<400\,{\rm mm}$. 
An estimate of the resolution on each track parameter, $\lambda$, is obtained from the corresponding distribution of the 
difference in the track parameters obtained from the two split tracks, $\Delta\lambda = \lambda_{\rm up}-\lambda_{\rm low}$. 
Each such distribution has an expectation value of 0 and a variance equal to two times the square 
of the parameter resolution: $var(\Delta\lambda)=2\sigma^2(\lambda)$. For each parameter, the mean and resolution of this 
difference distribution have been studied in bins of $p_{\rm T}$ and $\eta$. Since the cosmic-ray muon momentum distribution is a 
steeply falling function, the $p_{\rm T}$ value for each bin is taken as the mean of the $p_{\rm T}$ distribution in that bin.
For the resolutions, the results are shown in Figure~\ref{fig:muon5}. The means are roughly independent of 
$p_{\rm T}$ and $\eta$ and show no significant bias, with the exception of the $z_0$ distribution. That shows a small bias that 
varies with  $\eta$, but with a magnitude that is less than about 60 $\mu$m over the $\eta$-range investigated. This is 
negligible relative to the MS-ID bias already discussed. The means and resolutions obtained from tracks with 
$p_{\rm T}>30$~GeV are shown in Table~\ref{tab:muon2}. 

\begin{table}[t!]
\begin{center}
\begin{tabular}{|l|c|c|}\hline 
 Parameter & Mean & Resolution \\ \hline
$\phi_{0}$(mrad) & -0.053 $\pm$ 0.005 & 0.164$\pm$ 0.004 \\ \hline
$\theta_{0}$(mrad) & 0.27 $\pm$ 0.03 & 0.80 $\pm$ 0.02 \\ \hline
$d_{0}$($\mu$m) &  -0.9 $\pm$0.7  & 26.8 $\pm$ 0.8 \\ \hline
$z_{0}$($\mu$m) & 2.0 $\pm$ 3.7  & 116.6 $\pm$ 2.9 \\ \hline
\end{tabular}
\caption{Overview of the track parameter bias and resolution for CM tracks obtained with the track-splitting method for 2009 
cosmic-ray data, for tracks with $p_{\rm T}>30$~GeV.}
\label{tab:muon2}
\end{center}
\end{table}

\begin{figure}[phbt]
  \begin{center}
    \resizebox{0.49\textwidth}{!}{\includegraphics{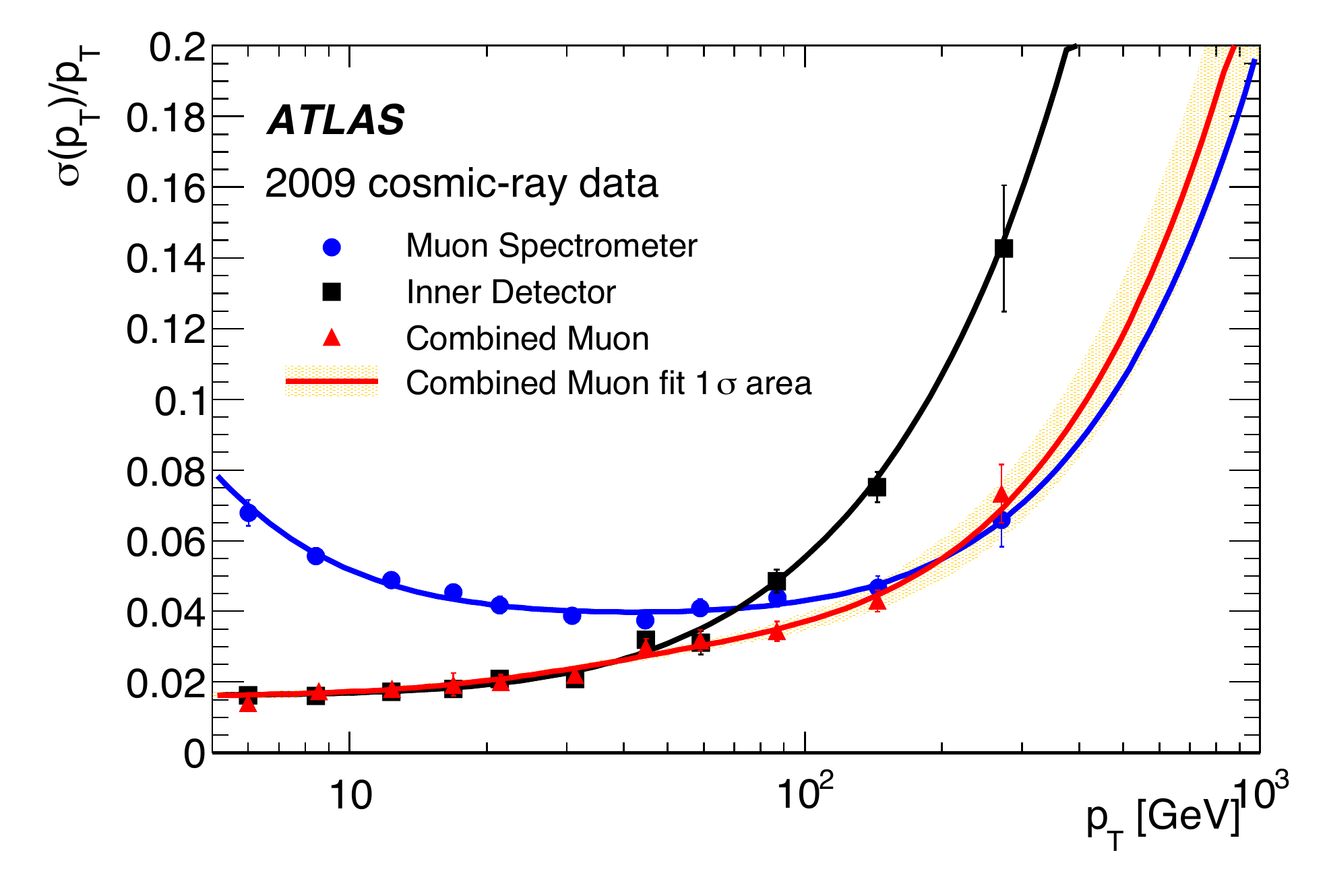}}
  \end{center}
  \caption{Resolution on relative $p_{\rm T}$ as a function of $p_{\rm T}$ for ID and MS standalone tracks, and
for CM tracks. The shaded region shows the $\pm 1\sigma$ region of the fit to the resolution
curve for the CM tracks.
    \label{fig:muon7}}
\end{figure}

A similar study of the track momentum reconstructed in the upper and lower halves of the detector shows that the 
mean of the momentum-difference distribution $(p_{\rm up}-p_{\rm low})$ is consistent with zero and flat as a function 
of $p_{\rm T}$ and $\eta$.  For studies of the $p_{\rm T}$ 
resolution, slightly looser cuts are employed in order to increase the statistics, particularly in the 
high-momentum region. For tracks having momenta above 50~GeV the requirement of a pixel hit is removed 
and the cuts on $|d_0|$ and $|z_0|$ are loosened to $1000\,{\rm mm}$. Figure~\ref{fig:muon7} shows the 
relative $p_{\rm T}$  resolution for ID, MS and CM tracks as a function of $p_{\rm T}$.
For each pair of upper / lower tracks, the value of the transverse momentum was evaluated 
at the perigee. The difference between the values obtained from the upper and lower parts of the detector, divided 
by their average
\[ \frac{\Delta p_{\rm T}}{p_{\rm T}} = 2\ \frac{p_{Tup} - p_{Tdown}}{p_{Tup} + p_{Tdown}} \]
was measured and plotted in eleven bins of $p_{\rm T}$. 
As above, the plotted $p_{\rm T}$ value is the mean of the $p_{\rm T}$ distribution in that bin.
The results of this procedure have been fitted to parametrizations appropriate to each particular track class. For 
the ID the fit function was:
\begin{equation}
  \frac{\sigma_{p_{\rm T}}} {p_{\rm T}} = \ P_{1} \ \oplus \ P_{2}\times p_{\rm T}
  \nonumber
\end{equation}
where $P_{1}$ is related to the multiple scattering term and $P_{2}$ to the ID intrinsic resolution.
For the MS tracks, the same function is used but with an additional term (coefficient $P_{0}$) related 
to uncertainties on the energy loss corrections associated with the extrapolation 
of the MS track parameters to the perigee:
\begin{equation}
  \frac{\sigma_{p_{\rm T}}} {p_{\rm T}} = \frac{P_{0}}{p_{\rm T}} \ \oplus \ P_{1} \ \oplus \ P_{2}\times p_{\rm T}.
  \nonumber
\end{equation}
For the combined resolution a more complex function is needed:
\begin{equation}
  \frac{\sigma_{p_{\rm T}}} {p_{\rm T}} = \ P_{1}  \ \oplus    \frac{\ P_{0} \times p_{\rm T} }  { \sqrt{1 + (\ P_{3} \times p_{\rm T} )^2 }  }  \ \oplus \ P_{2}\times p_{\rm T}
  \nonumber
\end{equation}
where $P_{1}$ is related to the multiple scattering term, $P_{2}$ to the intrinsic resolution at very high momentum and the  
$P_{3}$ term describes the intermediate region where ID and MS resolutions are comparable.


Table~\ref{table:CombRes} compares the fitted sizes of the multiple scattering and intrinsic resolution terms for the ID, MS and CM  
tracks.  For the CM tracks the multiple scattering term is determined mainly by the ID contribution while the intrinsic high-energy 
resolution comes mainly from the MS measurement.

Extrapolation of the fit result yields an ID momentum resolution of about 1.6\% at low momenta and of about 50\% at 1 TeV. 
The MS standalone results are improved over those previously obtained~\cite{MSCosmic}: the resolution extrapolated to 1 TeV is about 20\%. 
As expected the ID and MS systems dominate the resolution at low and high $p_{\rm T}$, respectively. However, at intermediate momenta from 
about 50 to 150~GeV both systems are required for the best resolution. The $\pm 1\sigma$ region returned by the fit to the 
resolution for the CM tracks is shown as the shaded region in Figure~\ref{fig:muon7}. 

\begin{table}[t!]
\centering
\begin{tabular}{|c|c|c|}
\hline
Fitted Resolution  &  $P_1$  & $P_2$ \\
\hline \hline
Inner Detector & 1.6 $\pm$  0.1 \% & $(53 \pm 2) \times 10^{-5}$ GeV$^{-1} $ \\
\hline
Muon Spectrometer & $3.8 \pm 0.1$ \% & $(20 \pm 3) \times 10^{-5}$ GeV$^{-1} $ \\
\hline
Combined Muon & $1.6 \pm 0.1$ \% & $(23 \pm 3)\times 10^{-5}$ GeV$^{-1} $ \\
\hline
\end{tabular}
\caption{Fitted values of the multiple scattering and intrinsic momentum resolution terms (as described in the text) for ID, MS and CM 
tracks.}
\label{table:CombRes}
\end{table}

\subsection{Muon energy loss in the ATLAS calorimeters}
\label{s:muloss}

Muons traverse more than 100 radiation lengths between the two tracking systems. Interactions
with the calorimeter material result in energy losses. These losses are typically around 3~GeV,  mainly due to ionization,
but are subject to fluctuations, especially for high momentum muons which can deposit a large fraction of their energy 
via bremsstrahlung. Muon reconstruction in collision events depends on a correct accounting for these losses, as does 
determination of the missing transverse energy in the event. A parametrization of these losses is normally used for extrapolating 
the track parameters measured by the MS to the perigee where they are defined. However, since 80\% of the material 
between the trackers is instrumented by the calorimeters, studies of the associated energy deposits in the calorimeter  
should allow improvements to the resolution in the case of large losses.

\begin{figure}[tbph]
\centering
\resizebox{0.49\textwidth}{!}{\includegraphics{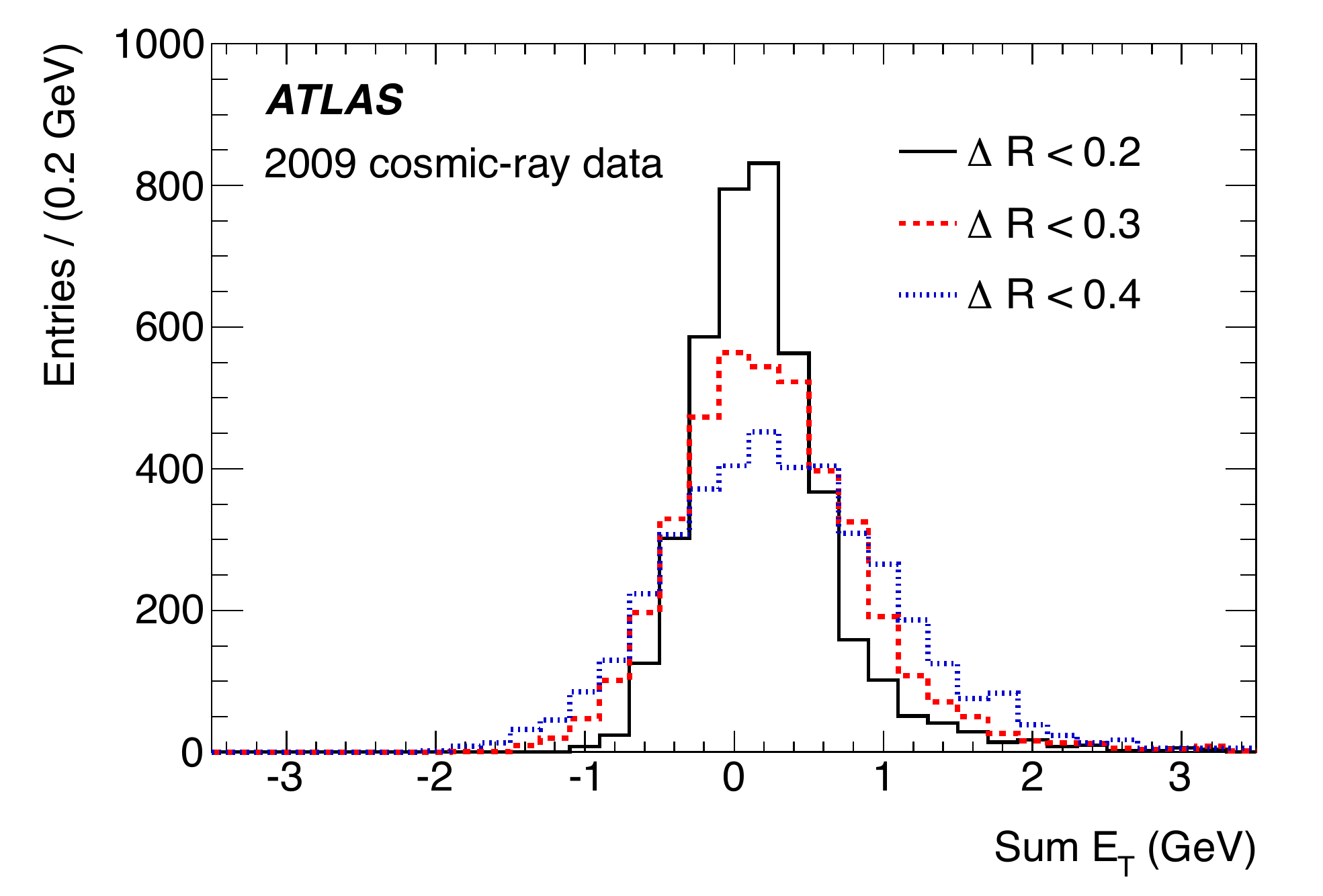}}
\resizebox{0.49\textwidth}{!}{\includegraphics{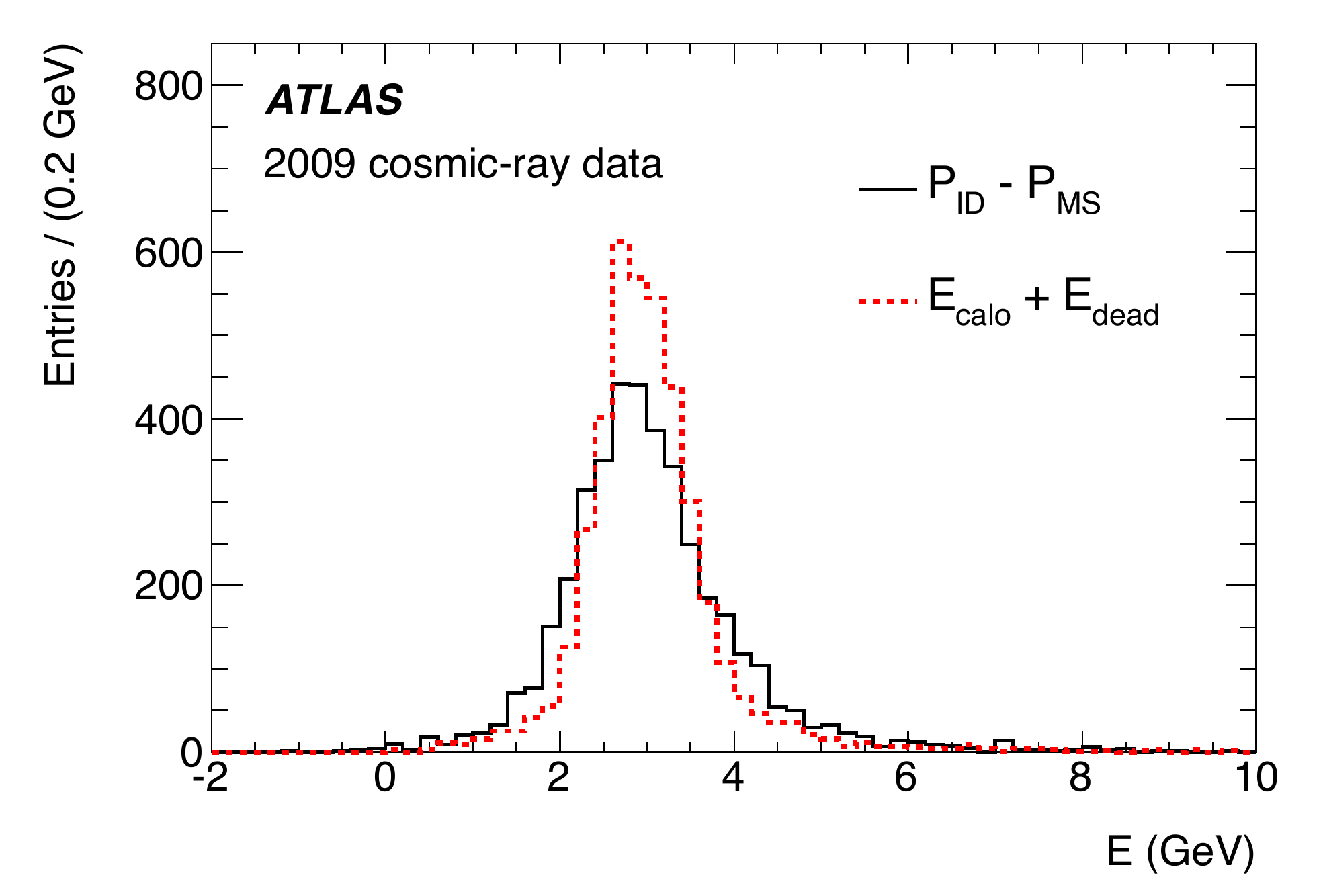}}
\resizebox{0.49\textwidth}{!}{\includegraphics{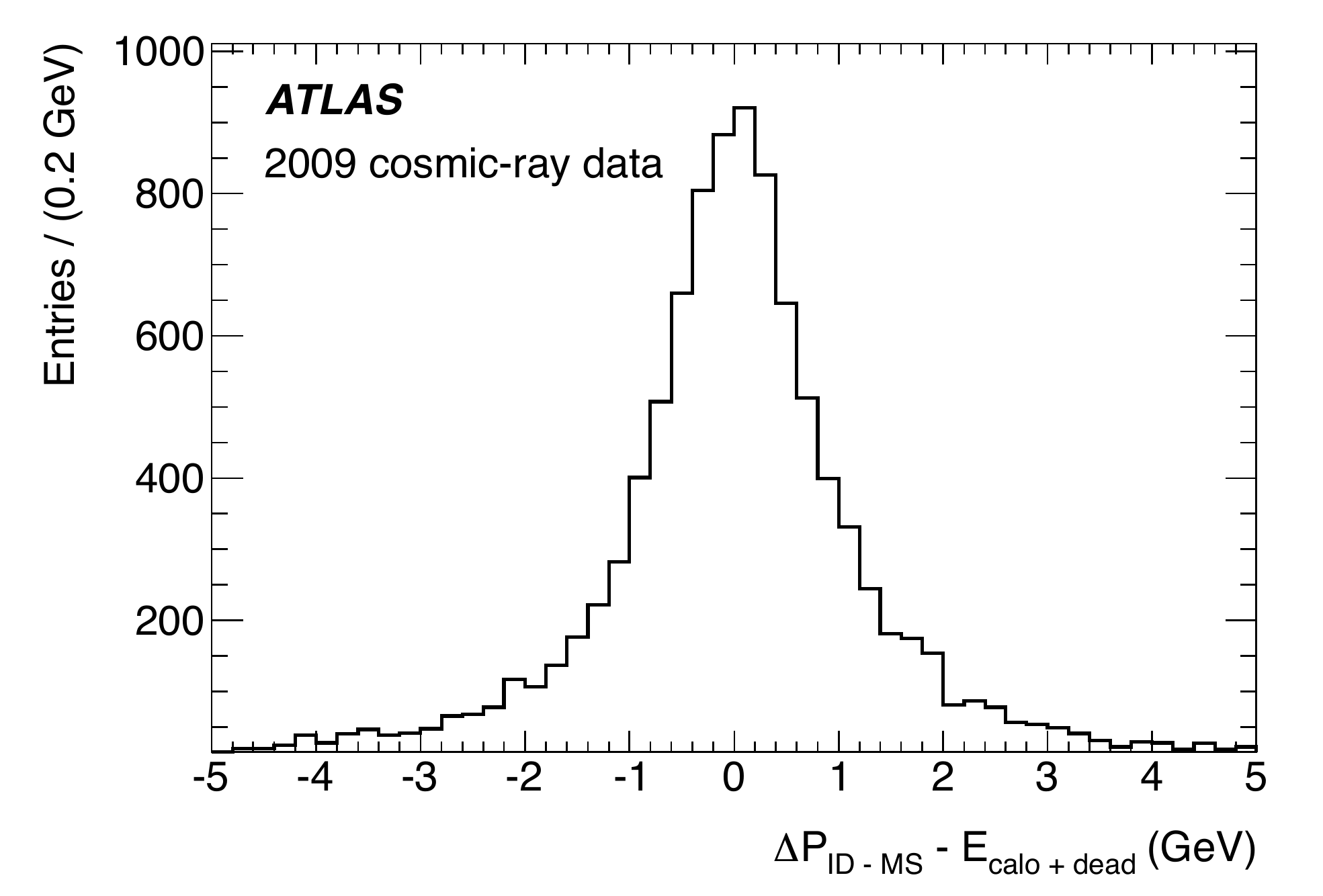}}
\caption{\label{fig:bl23a} The upper plot shows the sum of the transverse energy around muon tracks, outside the core region, 
for cones of $\Delta R$ = 0.2, 0.3 and 0.4. The middle plot compares the momentum difference between Inner Detector and Muon 
Spectrometer tracks $(P_{\rm ID} - P_{\rm MS})$ with the sum of the energy loss measured in the calorimeters, $E_{\rm calo}$, and the 
parametrized energy loss in the inert material, $E_{\rm dead}$. The lower plot shows the distribution of 
$(P_{\rm ID} - P_{\rm MS})-(E_{\rm calo}+E_{\rm dead})$.}
\end{figure}

This possibility has been investigated using cosmic-ray muons traversing the barrel part of ATLAS.
The analysis is based on the PCM sample from a single 2009 cosmic-ray run, consisting of about one million events. 
Strict criteria were applied to ensure pseudo-projective trajectories that are well measured in the relevant tracking subsystems: 
the SCT and the TRT in the Inner Detector, and the MDT and RPC systems in the Muon Spectrometer. The analysis was restricted to 
tracks crossing the bottom part 
of the Tile Calorimeter, in the region $|\eta| < 0.65$. A track-based algorithm\cite{Dinos_TICT,TICT_proceeding,BLenzi:2010} was 
used to collect the muon energy deposits in the calorimeters. The trajectory of the particle was followed using the ATLAS 
extrapolator~\cite{ATLAS_extrapolator}, which, using the ATLAS tracking geometry \cite{ATLAS_tracking_geometry},
takes into account the magnetic field, as well as material effects, to 
define the position at which the muon crossed each calorimeter layer. The cells within a predefined 
`core' region around these points were used for the measurement of the energy loss. This region was optimized according to the 
granularity and the geometry of each calorimeter layer. Only cells with $|E|~>~2\sigma_{\rm noise}$ were considered. 
Here $\sigma_{\rm noise}$ is the electronics noise for the channel and $|E|$ is used instead of $E$ to avoid biases.
As a check 
that this procedure properly reconstructs the muon energy deposits, the total transverse energies reconstructed in 
calorimeter cells within cones of $\Delta R = \sqrt{(\Delta \eta)^2 + (\Delta \phi)^2}$ of 0.2, 0.3 and 0.4, around the 
particle trajectory, were determined. From these, the sum of the transverse energy inside the core region 
($E^{\rm core}_{\rm T}$) was subtracted. In collision events these quantities can be used to define the muon isolation, while in this analysis 
they indicate how much energy is deposited outside the core. The distributions of these quantities, shown in the top plot of Figure~\ref{fig:bl23a}, 
are reasonably centered around zero with widths that increase with the cone size, as expected due to the inclusion of a larger number 
of cells. Small energy losses outside the $E^{\rm core}_{\rm T}$ region shift the distributions to slightly positive values, due to 
either uncertainties in the extrapolation process or to radiative losses.

As a measure of the energy deposited by the muon, $E^{\rm core}_{\rm T}$ is used with no additional correction. Monte Carlo 
simulations of single muons in the barrel region show that this method provides a nearly unbiased energy determination, 
with 2\% scale uncertainty and 11\% resolution for the energy deposited by 100~GeV muons.
To allow comparison of these losses with the difference between the momenta reconstructed in the two tracking systems, 
a parametrization of the losses in the dead (uninstrumented) material, $E_{\rm dead}$, is added to the calorimeter measurement. 
The tracking geometry provides this information in combination with the extrapolator. 
The energy measured in the calorimeter, corrected for the dead material, is compared with the momentum difference between 
Inner Detector and Muon Spectrometer tracks in the middle plot of Figure~\ref{fig:bl23a}. The mean values of the 
momentum-difference and energy-sum distributions are 3.043~GeV and 3.044~GeV respectively.
The typical momentum of the selected tracks is 16 (13)~GeV in the Inner Detector (Muon Spectrometer), 
measured (see Figure~\ref{fig:muon7}) with a resolution of about 2\% (4\%), while the energy collected in the calorimeters, 
$E_{\rm calo}$, is on average 2.4~GeV, with a precision of about 10-20\%. The RMS values of the 
two distributions are 1.081~GeV and 0.850~GeV respectively. In simulation the two distributions have means of 3.10~GeV and 3.12~GeV
compared to a true energy loss distribution with a mean of 3.11~GeV and an RMS of 0.750~GeV. The resolutions were 0.950~GeV and 
0.820~GeV, respectively, roughly consistent with the measured values. The bottom plot in figure~\ref{fig:bl23a} shows the 
distribution of $(P_{\rm ID} - P_{\rm MS})-(E_{\rm calo}+E_{\rm dead})$, which has a mean of -0.012~GeV and an RMS of 1.4~GeV. This
distribution is dominated by contributions from rather low-momentum tracks. Restricting to the momentum region of 10-25~GeV 
retains about 40\% of the statistics and yields a distribution with mean and RMS of -0.004~GeV and 1.0~GeV respectively.
 
Although the tracking systems are relatively more precise than the 
calorimeters, in both data and Monte Carlo simulation, the RMS of the energy-sum distribution from the calorimeter is smaller 
than that of the momentum-difference distribution from the tracking systems. Use of the calorimeter information may therefore 
allow future improvements to the combined tracking for muons.

\subsection{Identification of electrons}
\label{sec:electrons1}

\begin{figure}[phtb]
  \begin{center}
    \resizebox{0.32\textwidth}{!}{\includegraphics{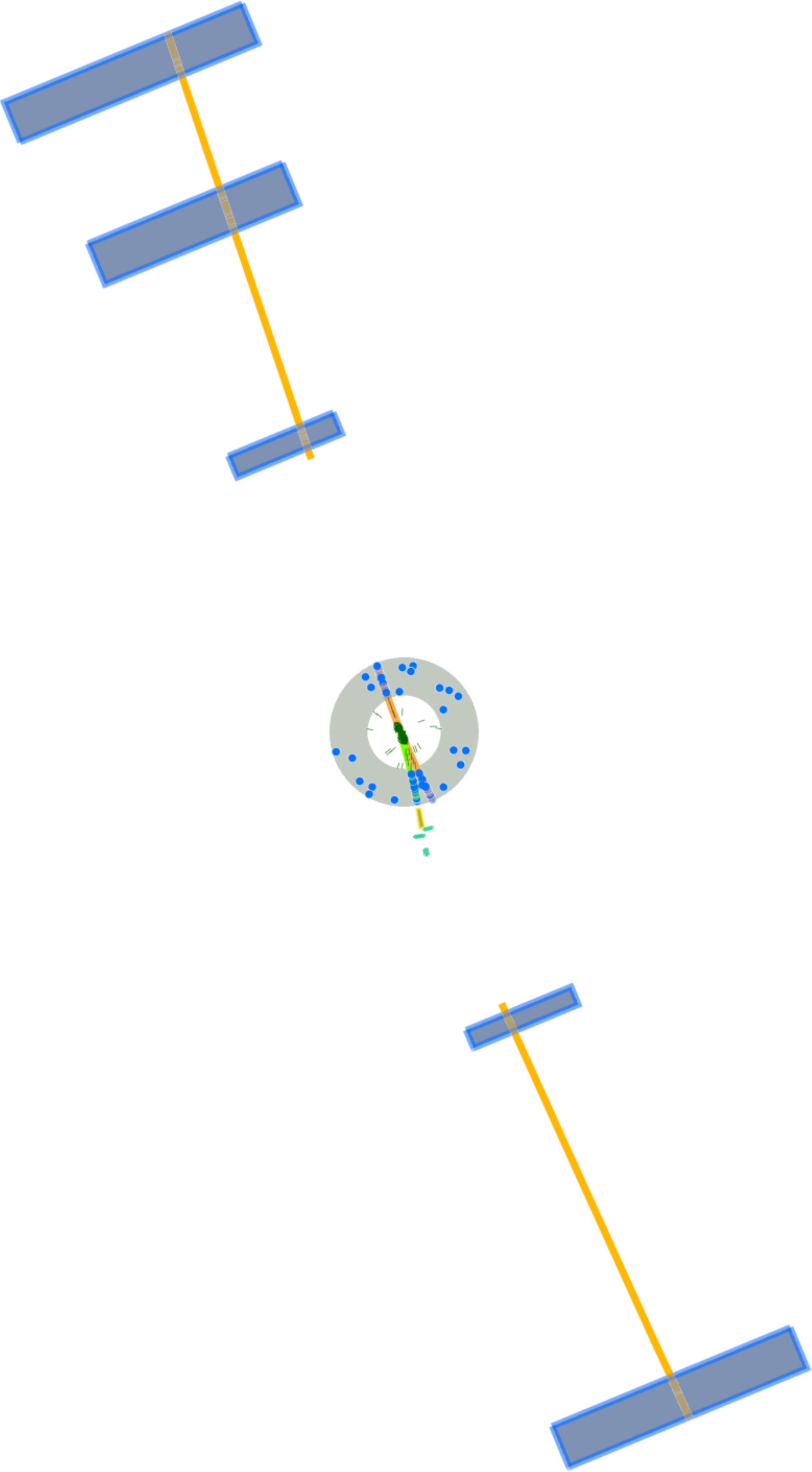}}
    \resizebox{0.44\textwidth}{!}{\includegraphics{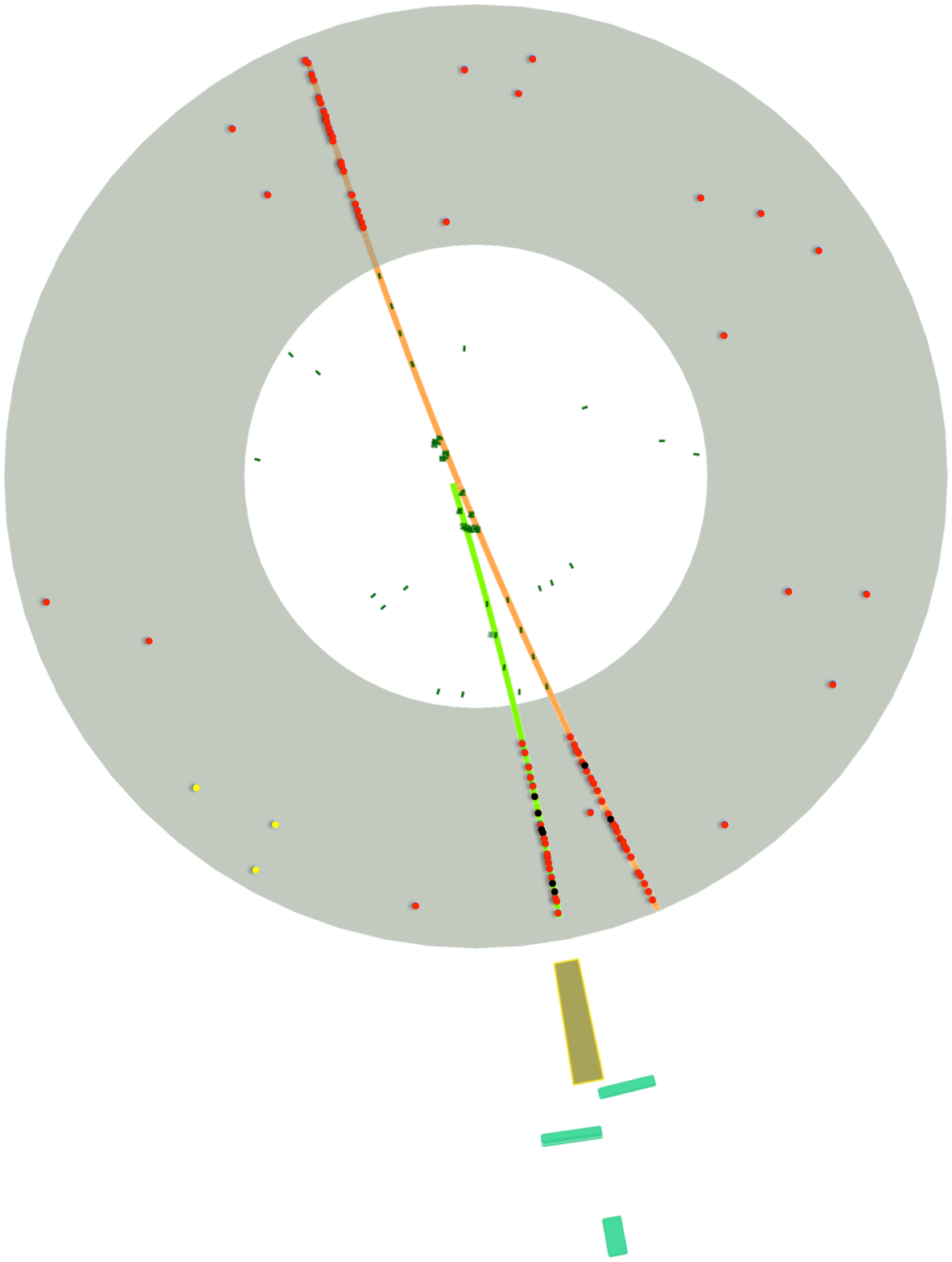}}
  \end{center}
  \caption{Event display of a typical $\delta$-electron candidate event. The upper figure shows
a view that includes the three layers of muon detectors on either side, while the lower plot
shows a close-up view of the Inner Detector. The shaded region represents the volume of the TRT, while
the inner region is occupied by the SCT and Pixel detectors. The two ID tracks, and associated hits, are 
clearly visible. High- and low-threshold TRT hits are displayed with the dark and light markers, respectively. 
The calorimeter cluster associated with the electron candidate is also shown.
    \label{fig:electron1}}
\end{figure}

The identification of electrons is performed by algorithms relying on information from both the EM calorimeter 
and the ID. Two methods are used, one seeded by tracks and the other by EM calorimeter clusters. The 
cluster-based algorithm is the standard identification tool, with clusters seeded using a sliding-window 
algorithm~\cite{Aad:2009wy}. This algorithm, used only for the identification of electromagnetic $(e/\gamma\,)$ objects in the EM 
calorimeter, uses a fixed grid of calorimeter cells in $\eta\times\phi$, centered on a seed cell having a signal-to-noise ratio 
exceeding a set threshold. For a cluster to form an electron candidate, there must normally be an ID track nearby in 
$\eta$ and $\phi$. However, in cosmic-ray events many tracks have only barrel TRT ($r-\phi$) hit information, in which case the 
association is done only in $\phi$.  The threshold for the reconstruction of an $e/\gamma$ object with the standard selection 
is about 3~GeV. To improve the identification of electrons with lower $p_{\rm T}$, a track-seeded algorithm is employed. This first 
searches for tracks in the ID with $p_{\rm T} > 2$~GeV and hits in both the SCT and Pixel Detectors. These tracks are 
extrapolated to the second layer of the EM calorimeter and a $3\times 7$ 
($\eta\times\phi$) cell cluster is formed about this point; the cell size in this layer varies with $\eta$ but is  
$0.025\times 0.025$ in $\eta\times\phi$ over the acceptance for this analysis.   
In both algorithms, the track momentum and the energy of the 
associated calorimeter cluster are required to satisfy $E/p < 10$. This section describes the use of these standard techniques 
for the selection of a sample of $\delta$-electrons, which are used to investigate the calorimeter response to 
electrons with energies in the 5~GeV range. Section~\ref{sec:electrons2} will describe an alternative low-$p_{\rm T}$ selection 
which can identify electrons down to $p_{\rm T}$ of about 500 MeV, using a more sophisticated clustering algorithm for determination of 
the energy of the associated electromagnetic calorimeter cluster. 

Electron identification relies in part on the particle identification abilities of the TRT.
Transition radiation (TR) is produced by a charged particle crossing the boundary between two materials having different 
dielectric constants. The probability of producing TR photons depends on the Lorentz factor ($\gamma= E/m$) of the particle. 
The effect commences at $\gamma$ factors around 1000 which makes it particularly useful for electron identification, since 
this value is reached for electrons with energies above about 500 $\MeV$. For muons, these large $\gamma$ factors occur only for  
energies above about $100\:\GeV$.  The TR photons are detected by absorption in the chamber gas which is a xenon mixture 
characterized by a short absorption length for photons in the relevant energy range. The absorption leads to high electronic 
pulses; pulses due to energy deposits from particles which do not produce transition radiation are normally much lower. 
A  distinction between the two classes of particles can therefore be made by comparing the pulse heights against high and 
low thresholds, and looking at the fraction of high-threshold hits for a given track. This fraction is referred to below as 
the TR ratio.

The production of electrons in cosmic-ray events is expected to be dominated by knock-on or $\delta$-electrons 
produced by ionization caused by cosmic-ray muons. The energy distribution of such electrons is typically rather soft, but has
a tail extending out into the GeV region, where the standard electron identification tools can be employed. 
The experimental signature of such an event consists of a muon track traversing the muon chambers at the top and 
bottom of the detector, having corresponding MIP-like energy deposits in the calorimeters, accompanied by a second 
lower-momentum track in the ID associated with a cluster in the EM calorimeter, as illustrated by the event 
displayed in Figure~\ref{fig:electron1}. In the upper view, the incoming and outgoing muon tracks, are seen to leave hits 
in three muon layers on the top of ATLAS and in two layers below, as well as in the Inner Detector.
In the lower, expanded view of the ID region the muon track and the electron candidate track are shown with the associated hits 
in the silicon detectors as well and those in the TRT, which are illustrated by either light or dark markers, depending on whether 
they are low- or high-threshold. The candidate electron track clearly displays a larger number of high-threshold TRT hits, 
as expected for an electron, as well as an association to a cluster of energy in calorimeter (at the bottom).

The search  was performed using data from the PCM sample obtained from cosmic-ray running in the fall of 2008.
Based on the expected topology, events were selected if they satisfied the following requirements:
\begin{itemize}
\item 2 or more ID tracks
\item 1 electron in the bottom of the detector (since the muons come from above).
\item 1 or more muon tracks: if there is more than one there must be at least one track 
in the top and bottom halves of the detector, consistent with coming from a muon of 
the same charge.
\end{itemize}
The events so selected are referred to below as the signal sample, or the ionization sample.

There is one important background for which this selection can lead to the identification of fake electron candidates. 
A highly energetic muon can emit a bremsstrahlung photon that does not convert within the ID. This photon will produce a 
cluster in the EM Calorimeter that can be incorrectly associated with the muon track if the track and cluster are 
nearby, creating a fake electron candidate. The signature for this process is one incoming and one outgoing muon track in 
the MS, one track in the ID and a cluster in the lower part of the EM Calorimeter. This signature can be clearly 
distinguished from the true electron production processes by the number of tracks in the ID (except for muon decays in flight 
which are expected to contribute only a very small fraction of the electrons of interest in this analysis). 
Nevertheless, for muon bremsstrahlung events, an additional (fake) track may be reconstructed leading to an event with 
the same signature as the signal process. For these fakes, equal numbers of electron and positron candidates are expected, 
in contrast to true $\delta$-electrons, where only negatively charged electrons are produced. To study this, a background 
sample depleted in $\delta$-electrons and enriched in background events due to muon bremsstrahlung, was selected using the 
requirements:
\begin{itemize}
\item exactly 1 ID track
\item 1 electron in the bottom of the detector 
\item 1 or more muon tracks
\end{itemize}
In the analysis of the signal and background samples, slightly modified versions of standard algorithms were used
to identify electrons. The standard selection~\cite{Atlas_detector} defines three classes of candidates: loose, medium and 
tight, according to increasingly stringent cuts on the typical properties of electron tracks and their associated EM 
showers, particularly quantities related to the longitudinal and transverse shower development. For the analysis discussed
here, a ``modified medium'' selection is adopted, which is a combination of selection criteria applied in the standard 
medium and tight selections, with slight modifications to allow for the different topology of the cosmic-ray muon events. 
In particular, since most of the muons do not pass through the SCT or Pixel Detector, requirements on the number of 
hits in the silicon detectors are replaced with quality cuts based on the number of TRT barrel hits and the $\phi$ matching
of the electron track to the EM cluster. A cut on $|z_0|$ is made to ensure that
tracks are in the barrel part of the TRT. 

\begin{figure*}[thbp]
  \begin{center}
    \resizebox{0.49\textwidth}{!}{\includegraphics{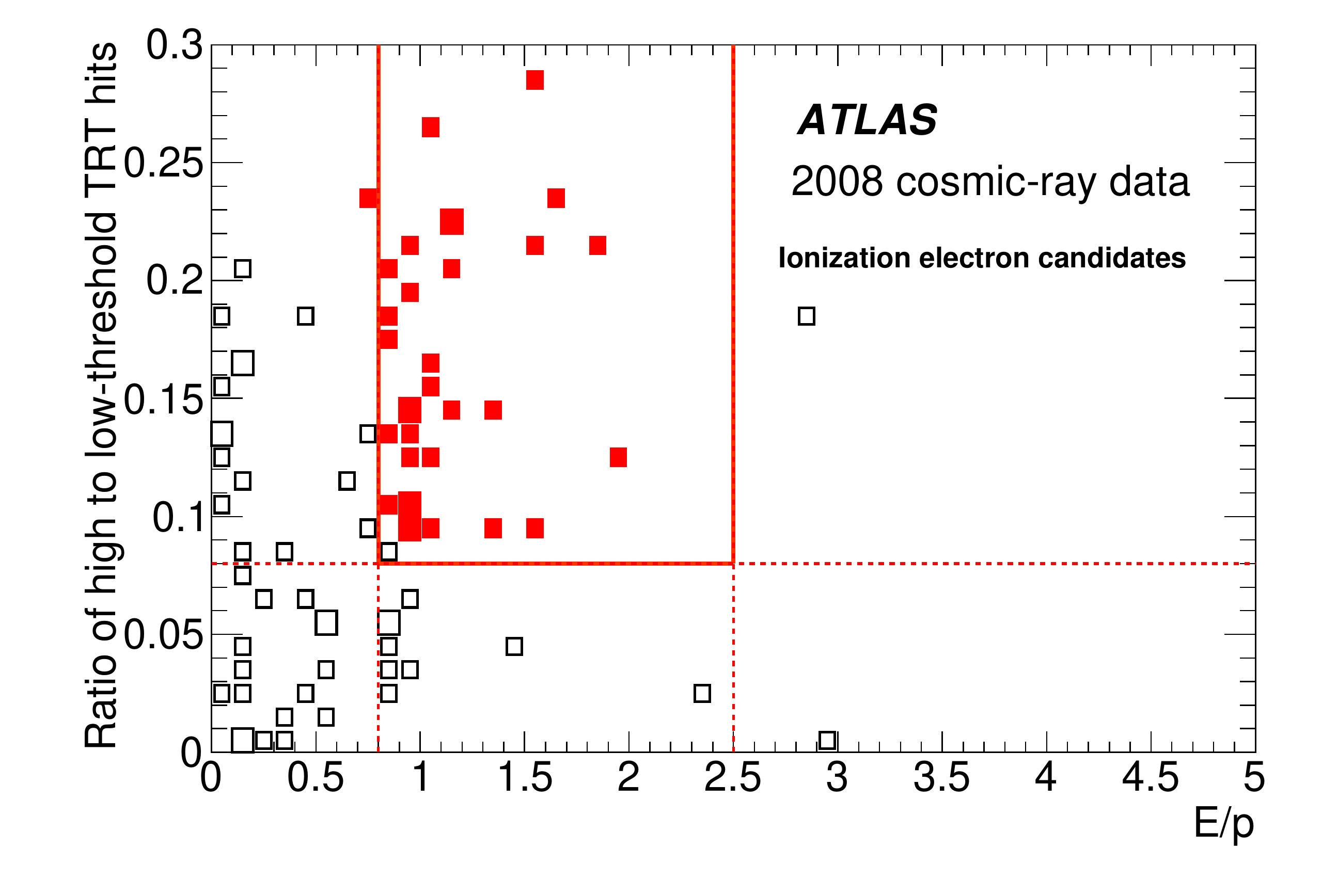}}
    \resizebox{0.49\textwidth}{!}{\includegraphics{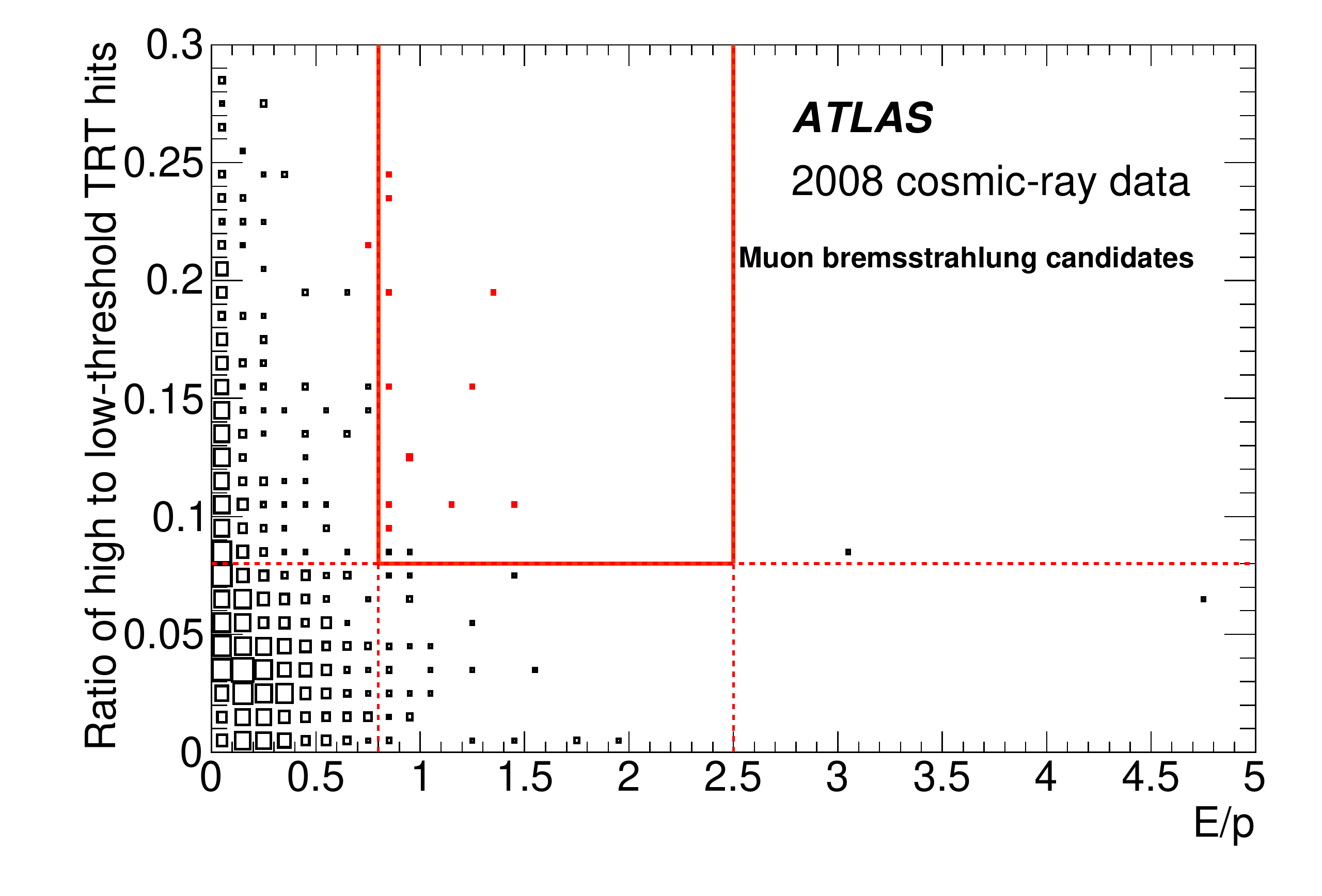}}
    \resizebox{0.49\textwidth}{!}{\includegraphics{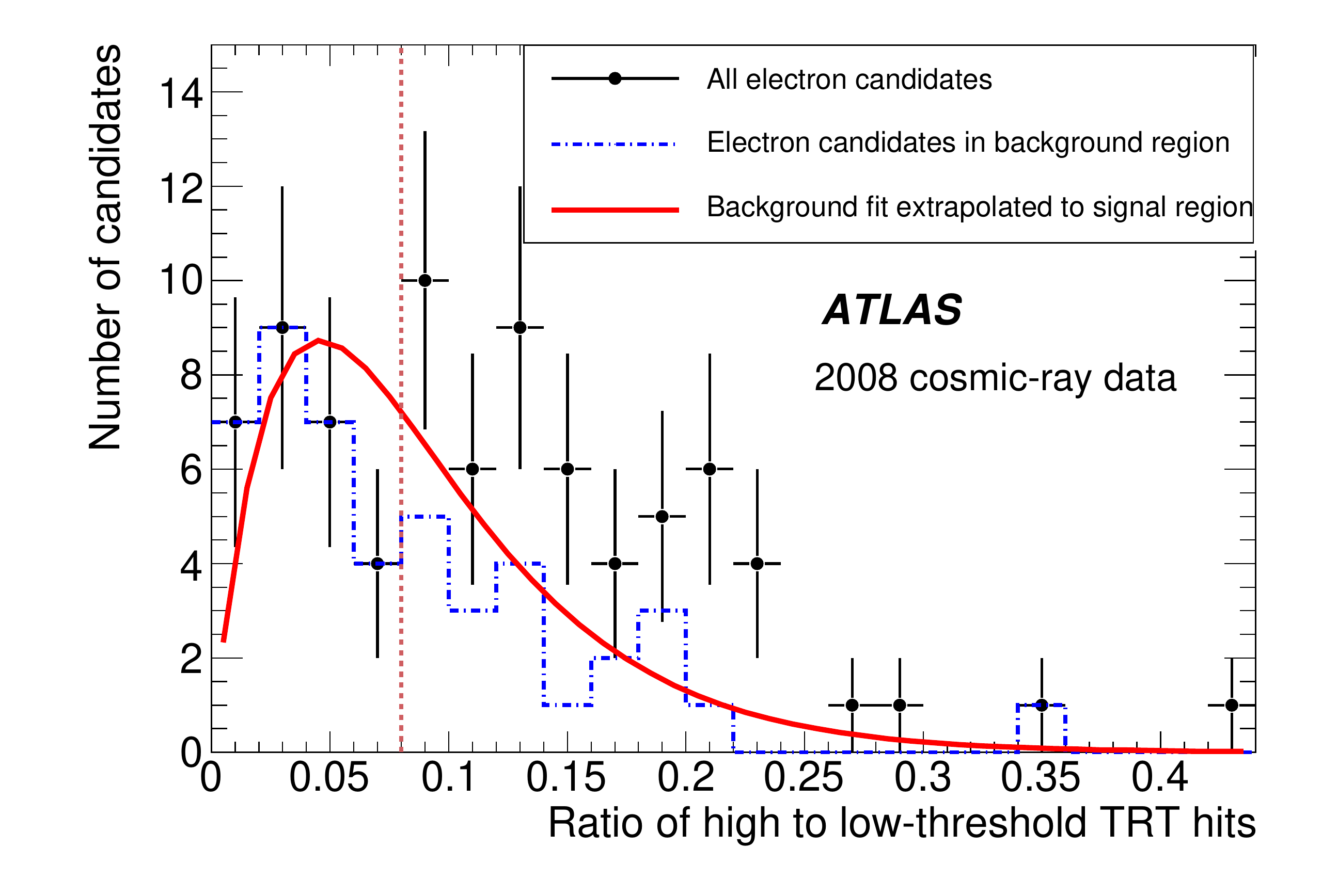}}
    \resizebox{0.49\textwidth}{!}{\includegraphics{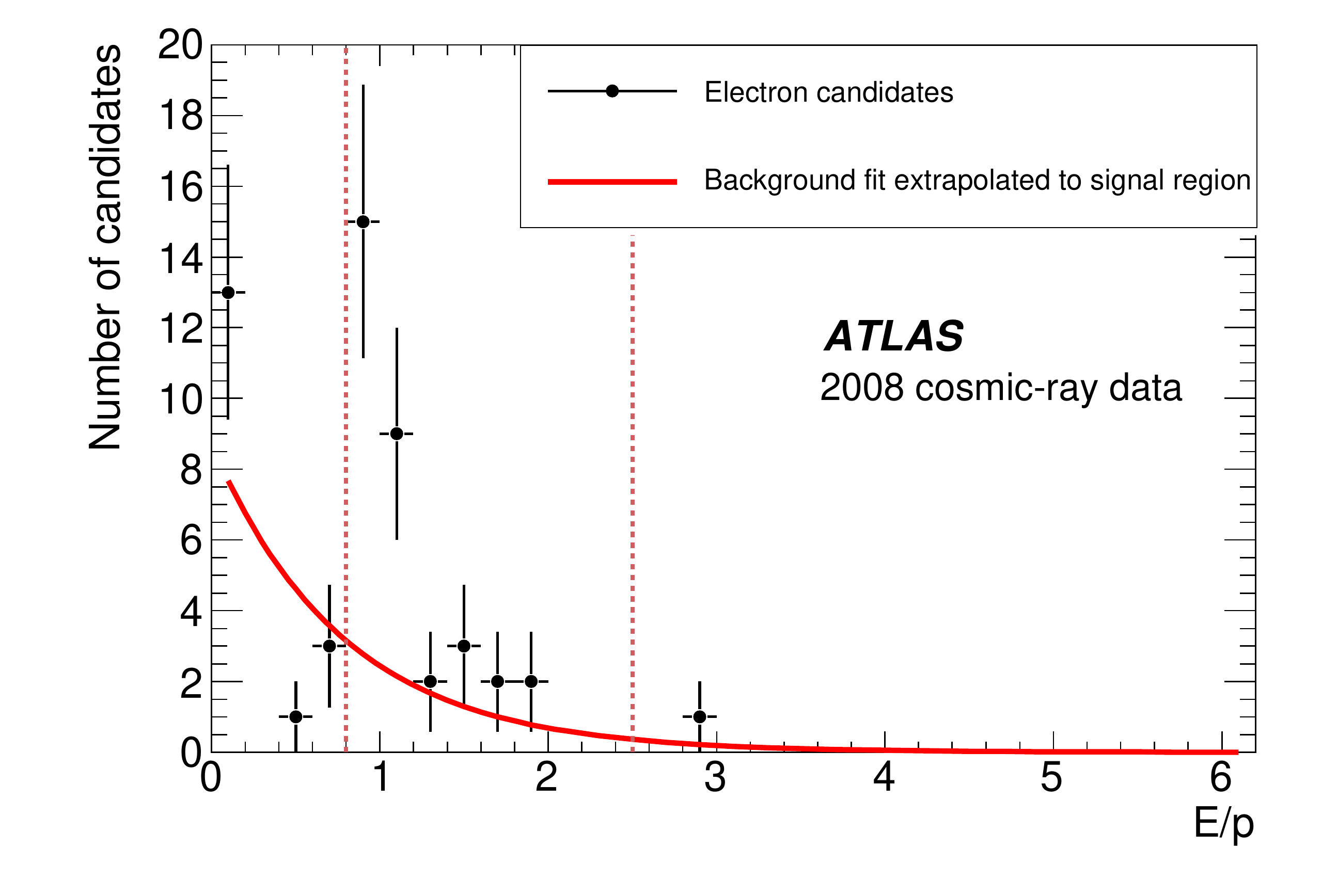}}
  \end{center}
  \caption{The upper plots show the two-dimensional distributions of the TR ratio vs. $E/p$ 
for the ionization sample (left) and the background 
sample (right). The open (black) boxes show the distribution of electron candidates passing the modified medium cuts. The 
solid (red) boxes indicate the electron candidates which also survive the tight selection. The dotted lines show the cuts 
applied to most of the events having $\eta \approx 0$ and low transverse energy: $0.8 < E/p < 2.5$ and TR ratio $>0.08$. The 
solid lines indicate the signal region. Two outliers at high TR ratio (1 in signal, 1 in background region), and two outliers 
at high $E/p$ are not shown. The lower plots show projections of the fit result for the ionization sample. The left plot shows the 
distribution of the TR ratio for all 81 electron and positron candidates after the modified medium cuts (points with error bars). The 
dashed histogram shows the 47 events in the background region and the curve shows the projection of the two-dimensional binned maximum 
likelihood fit. The dotted vertical line indicates the lower selection cut applied to the bulk of events.  The right plot shows the 
distribution of $E/p$ for all modified medium electron candidates after the additional application of the tight-selection cut on the 
TR ratio. The curve shows the projection of the two-dimensional background fit from which the number of background events under the 
signal region is estimated. The dotted vertical lines represents the upper and lower selection cuts on $E/p$, applied to the bulk of 
the data. 
    \label{fig:electron2}}
\end{figure*}

In addition to  this modified medium selection, a tight selection is defined by 
two additional requirements:
\begin{itemize}
\item $0.8 < E/p < 2.5$
\item TR ratio $> 0.08$
\end{itemize}
Note that both of these cuts are actually slightly $\eta$-dependent, following the standard tight selection.
The values quoted above are those applied over most of the acceptance.
After application of the modified medium selection, there are 81 events in the signal sample and 1147 in the background 
sample. 
Since the background candidates arise dominantly from the case where the EM cluster is associated to the cosmic-ray muon, 
this sample can be used to model the properties of the corresponding background events in the signal sample, in which the 
requirement of an additional ID track greatly reduces the number of events.
Because $E/p$ and the TR ratio are correlated, these quantities are shown plotted against
one another in the upper plots of Figure~\ref{fig:electron2}, separately for the signal and background samples. 
The open (black) boxes
show the distribution of candidates passing the modified medium selection. For candidates that also
survive the tight selection the distribution is shown using solid (red) boxes. In each plot, the dotted lines show the cuts 
applied (as quoted above) on each quantity, for the majority of the candidates. These define the signal region which is enclosed
by the overlaid solid lines. The open markers in the signal region and solid markers in the background region arise due 
to the slight $\eta$-dependence of the cuts. There are 34 events from the signal sample passing all cuts, compared to 13 
from the background sample. Of the 34 events in the signal region, 4 are positively charged. 

\begin{figure*}[tbph]
  \begin{center}
    \resizebox{0.49\textwidth}{!}{\includegraphics{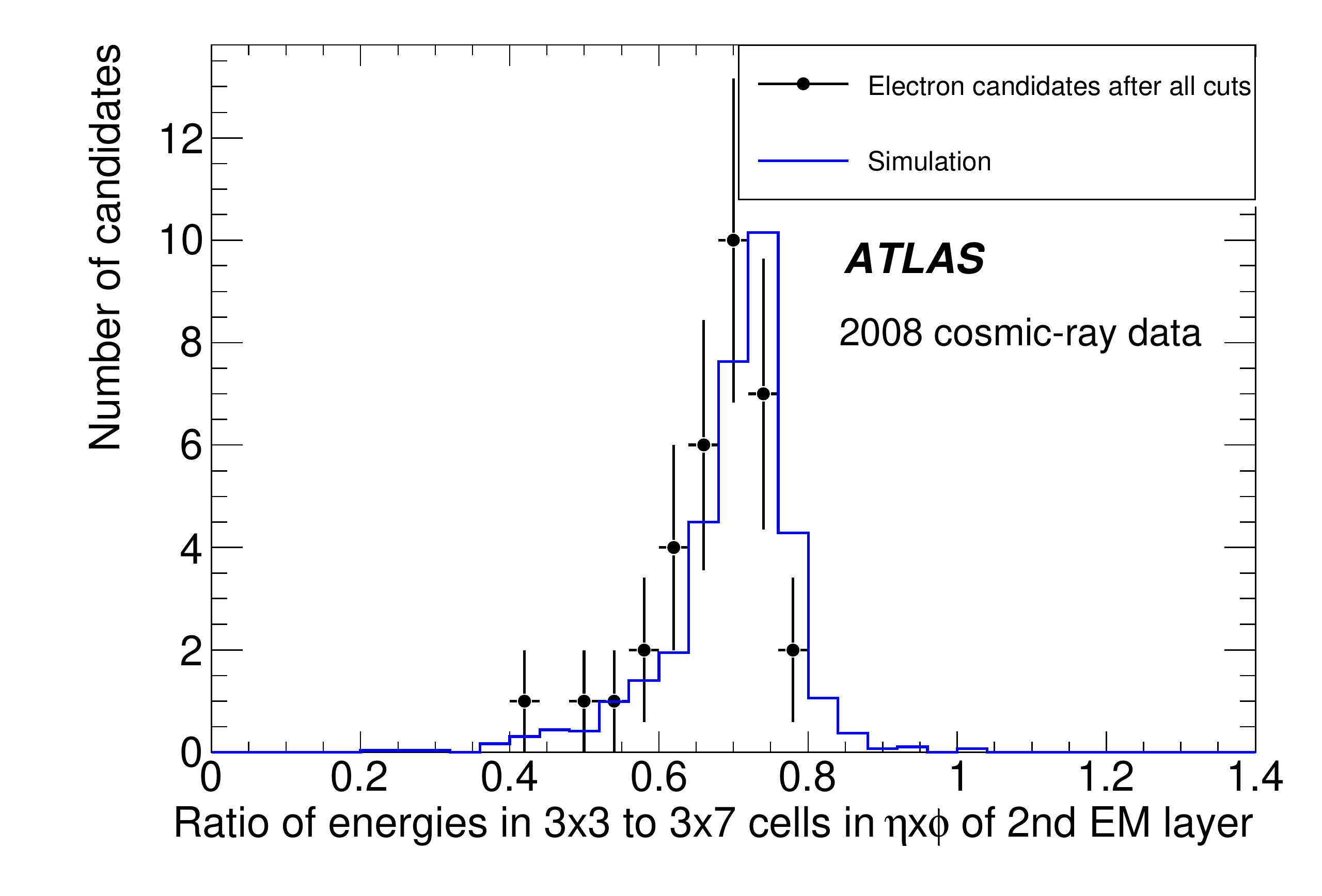}}
    \resizebox{0.49\textwidth}{!}{\includegraphics{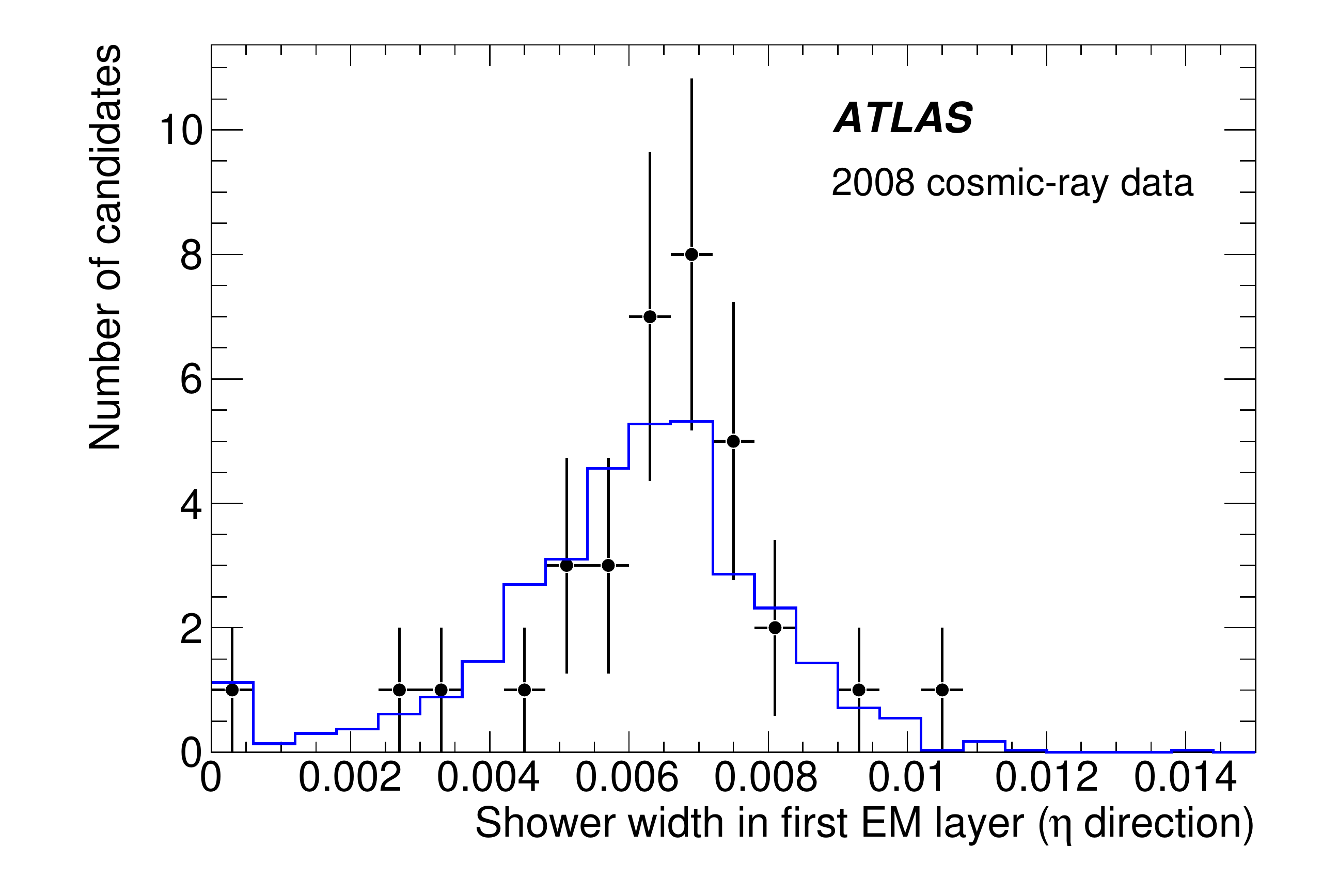}}
    \resizebox{0.49\textwidth}{!}{\includegraphics{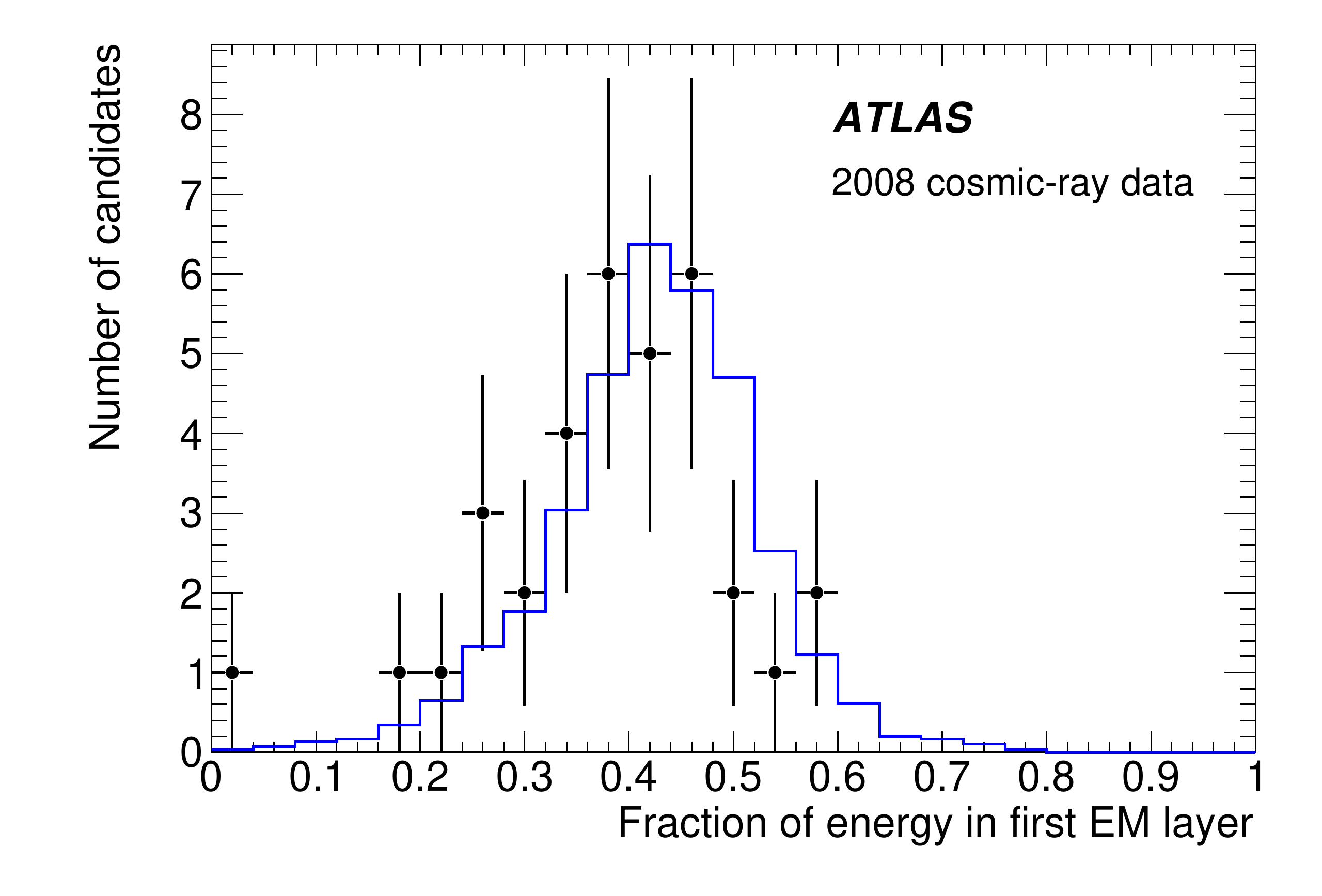}}
    \resizebox{0.49\textwidth}{!}{\includegraphics{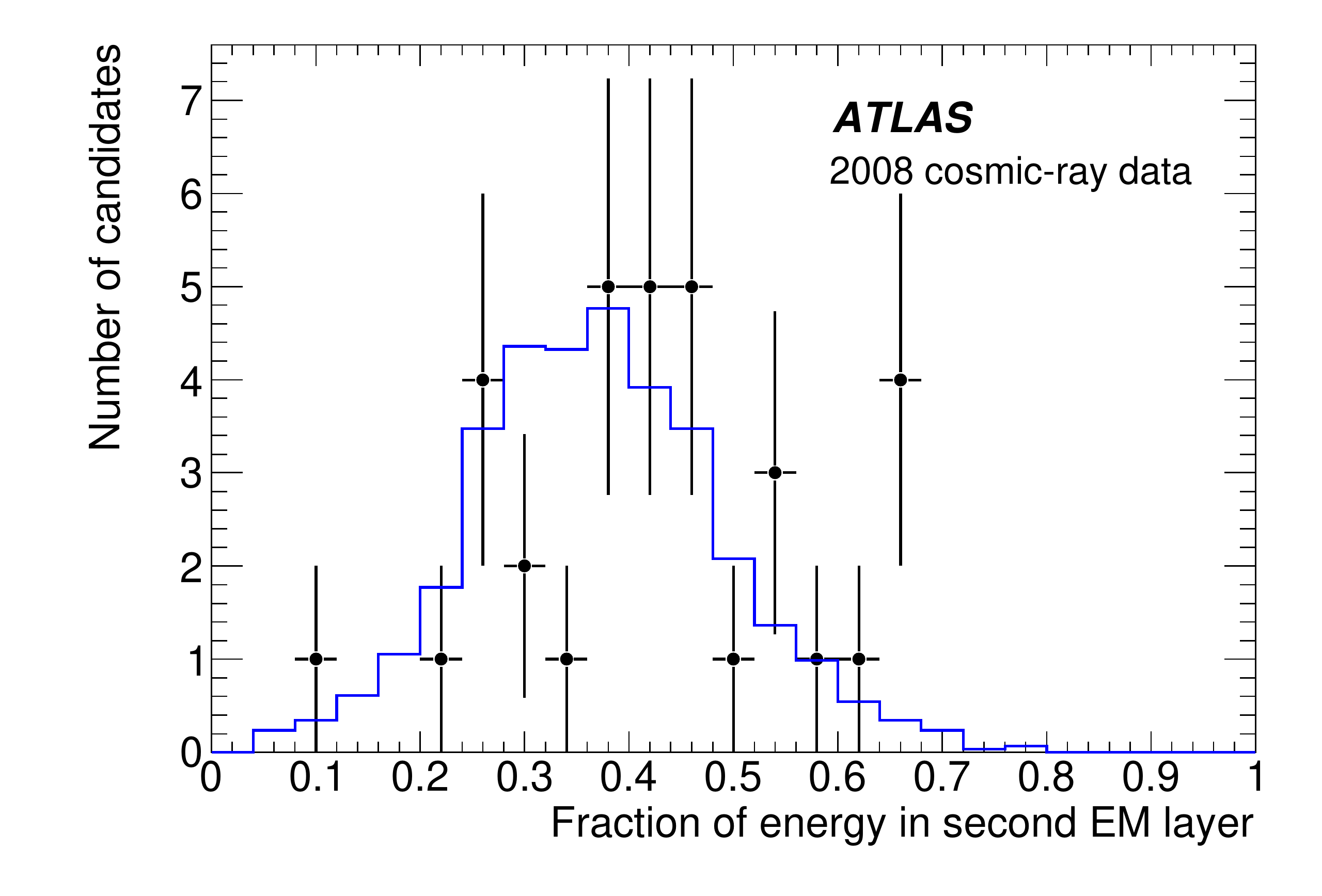}}
  \end{center}
  \caption{Comparison of shower profiles for all 34 ${\rm e}^{\pm}$ candidates to those from simulated projective electrons with
a transverse energy of 5 $\GeV$ and $\left| \eta \right | <0.8$. The data points indicate the electrons from the cosmic-ray 
data, while the histograms indicate distributions obtained from the simulated electrons. The upper left plot shows the ratio of 
energies in $3\times 3$ over $3\times 7$ cells in $\eta\times\phi$ in the second layer of the EM 
calorimeter. The upper right plot shows the energy-weighted shower width in $\eta$, in the 
first layer of the EM calorimeter. The lower left (right) plot shows the distribution of the fraction of energy in the first
(second) layer of the EM calorimeter. The Monte Carlo distributions are normalized to the number of data events. 
    \label{fig:electron4}}
\end{figure*}

The sample of 34 candidates was investigated further in order to confirm the identification of these as 
electrons and to determine the number of $\delta$-electrons by estimating the background in the signal 
sample. This was done by performing a three-parameter, binned maximum-likelihood fit to the two-dimensional TR ratio vs.
$E/p$ distribution for the background sample and then fitting the resulting background shape to the ionization sample in the regions 
outside the signal acceptance. The results of this procedure are displayed in the lower plots of Figure~\ref{fig:electron2}. 
Note that the fit uses finer binning than is used for these projections. The plot on the left shows the distribution of the TR ratio 
for the 81 candidates passing the modified medium cuts (points with error bars) while the dashed histogram shows the 47 events 
in the background region and the solid curve shows the projection of the two-dimensional binned maximum likelihood fit, which 
provides a good description of the distribution from candidates in the background region (dashed histogram). The right-hand plot 
shows the distribution of $E/p$ for all 
candidates remaining after the additional application of the tight-selection cut on the TR ratio. The solid curve again shows the projection 
of the two-dimensional background fit leading to an estimate for the background contribution in the signal region (indicated by the dotted 
vertical lines) of ($8.3\pm3.0$) events. This is consistent with the hypothesis that the dominant background is muon bremsstrahlung, which 
should produce equal numbers of positive and negative candidates, and the 
observation of 4 positively charged candidates in the signal sample.

As a final check on the candidate events, several distributions related to shower profiles were compared to expectations
based on a Monte Carlo simulation of projective electrons (produced at the nominal IP) with transverse energy 
of 5~GeV, in the region $|\eta| < 0.8$ which is appropriate for comparison with the cosmic-ray electron sample obtained 
with this selection. These comparisons are shown in Figure~\ref{fig:electron4}. The upper left plot shows the lateral 
containment, in the $\phi$ direction, of energy in the cells of the second layer of the EM calorimeter, as defined by the 
ratio $E_{3\times{3}}/E_{3\times{7}}$ where $E_{\rm i\times{j}}$ represents the energy deposited in a collection of cells of 
size $\rm i\times j$ in $\eta\times\phi$. A large mean value is observed for both data and Monte Carlo, as expected since electrons
tend to have a small lateral shower width. The upper right plot shows the lateral extent of the shower in $\eta$, in the first 
layer of
the EM calorimeter, as measured by the sum of the cell-cluster $\eta$ separations, weighted by the cell energy. This also 
shows good agreement between data and Monte Carlo. The other quantities plotted are related to the longitudinal shower 
shape: the lower left plot shows the fraction of the total cluster energy deposited in the first layer of the EM calorimeter 
while the lower right plot shows the fraction of energy in the second layer.  In both cases the average value should be about 
40$\%$ for electrons, as these tend to start showering early in the calorimeter. There is reasonable agreement between data
and Monte Carlo, but both show some small discrepancies. 
These arise from the fact that several of the data events have much larger 
energies than were used for the Monte Carlo sample, which consists entirely of electrons with a transverse energy of 5~$\GeV$. 
The deviations are consistent with what would be expected from the bremsstrahlung background in the sample. Those events can 
be of higher energy than the electron events, affecting the energy distributions of the showers, particularly the longitudinal 
energy profiles. Distributions of the fractions of energy deposited in the presampler and in layer 3 of the EM calorimeter show a 
similar level of agreement with the distributions from the projective-electron Monte Carlo sample.

\subsection{Identification of low momentum electrons}
\label{sec:electrons2}

The majority of the electrons in the cosmic-ray data are expected to be of low energy, of the order of a few hundred \MeV. 
The probability of producing an electron with sufficiently high momentum to produce a standard e/$\gamma$ cluster in the EM 
calorimeter is rather small, as reflected in the relatively low statistics available using the selection described in the 
previous section. 

In addition to the sliding-window cluster used for the standard electron identification, ATLAS employs a topological clustering 
algorithm~\cite{Barillari:2009zza} which groups adjacent cells with energies above certain thresholds into clusters which are 
thus composed of varying number of cells, providing for better noise suppression. Each topological cluster is seeded by a cell 
having a signal-to-noise ratio ($|E|/\sigma_{\rm noise}$) above a threshold $t_{\rm seed}$, and is then expanded by  iteratively 
adding neighboring cells having $|E|/\sigma_{\rm noise}>t_{\rm neighbor}$. 
Following the iterative step, the cluster is completed 
by adding all direct neighbor cells along the perimeter having signal-to-noise above 
$|E|/\sigma_{\rm noise}>t_{\rm cell}$. Several types of topological clusters (differing in $t_{\rm seed}$, $t_{\rm neighbor}$ 
and $t_{\rm cell}$) are used by ATLAS, for the reconstruction of calorimeter energy deposits from hadrons, electrons and 
photons, over the full range of $\eta$. 

A selection based on the matching of an ID track to an EM topological cluster was applied to the cosmic-ray data.  
This analysis, run on data from both the 2008 and 2009 cosmic-ray data-taking periods, is similar to the one described in the 
previous section, also focusing on events in the barrel part of the detector. The topological signature of the electron 
events is the same as described in section~\ref{sec:electrons1}
and the data sample is separated into signal and background samples in a similar way, based on the number of tracks;
electrons are again searched for in events with at least 2 ID tracks, while events with only one reconstructed track are used 
as a background sample. Candidate tracks must match an EM cluster from the topological clustering algorithm with $t_{\rm seed}=4$, 
$t_{\rm neighbor}=3$ and $t_{\rm cell}=0$. This allows the reconstruction of electromagnetic clusters with energies down to about 500~MeV. 

\begin{figure*}[tbph]
  \begin{center}
    \resizebox{0.49\textwidth}{!}{\includegraphics{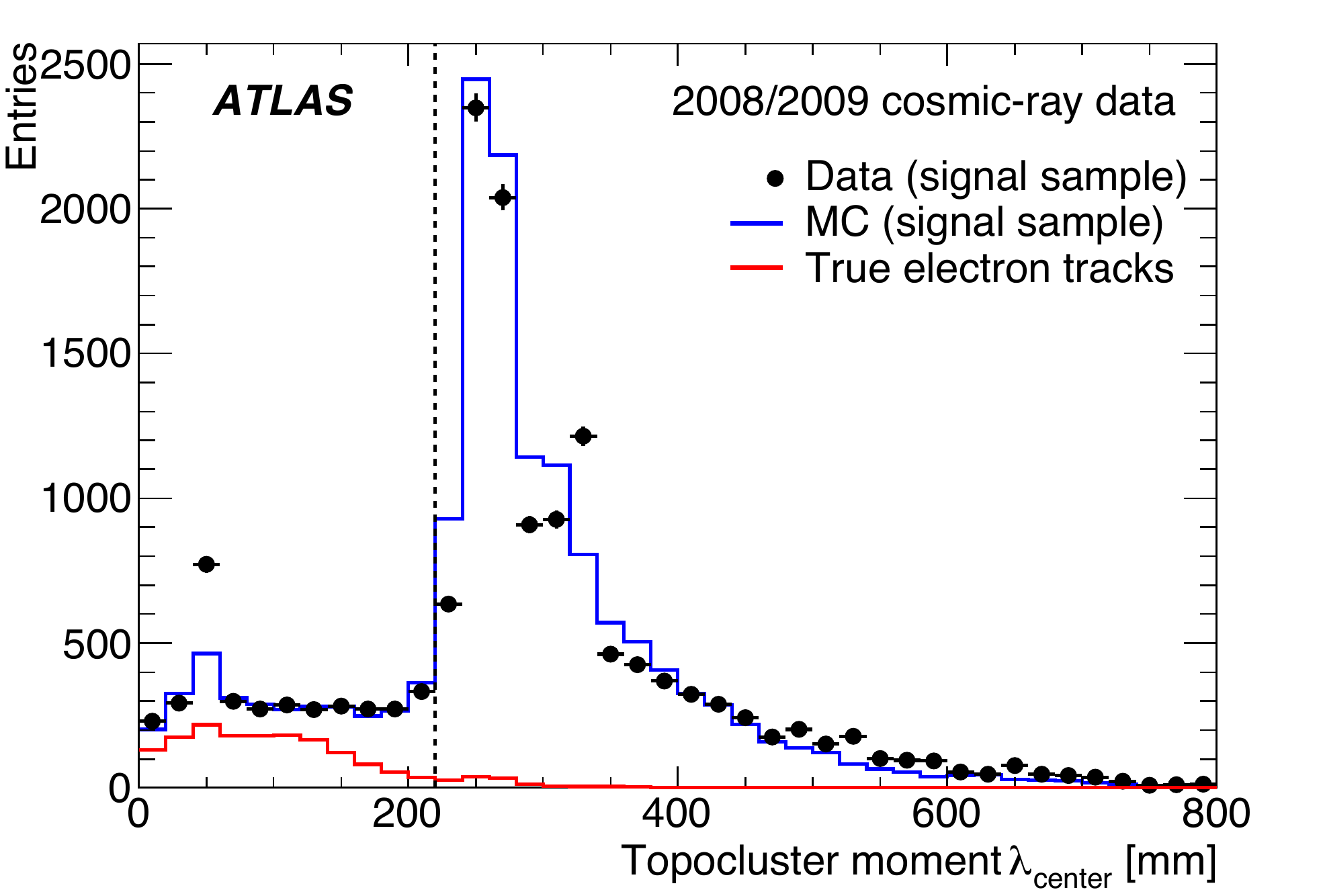}}
    \resizebox{0.49\textwidth}{!}{\includegraphics{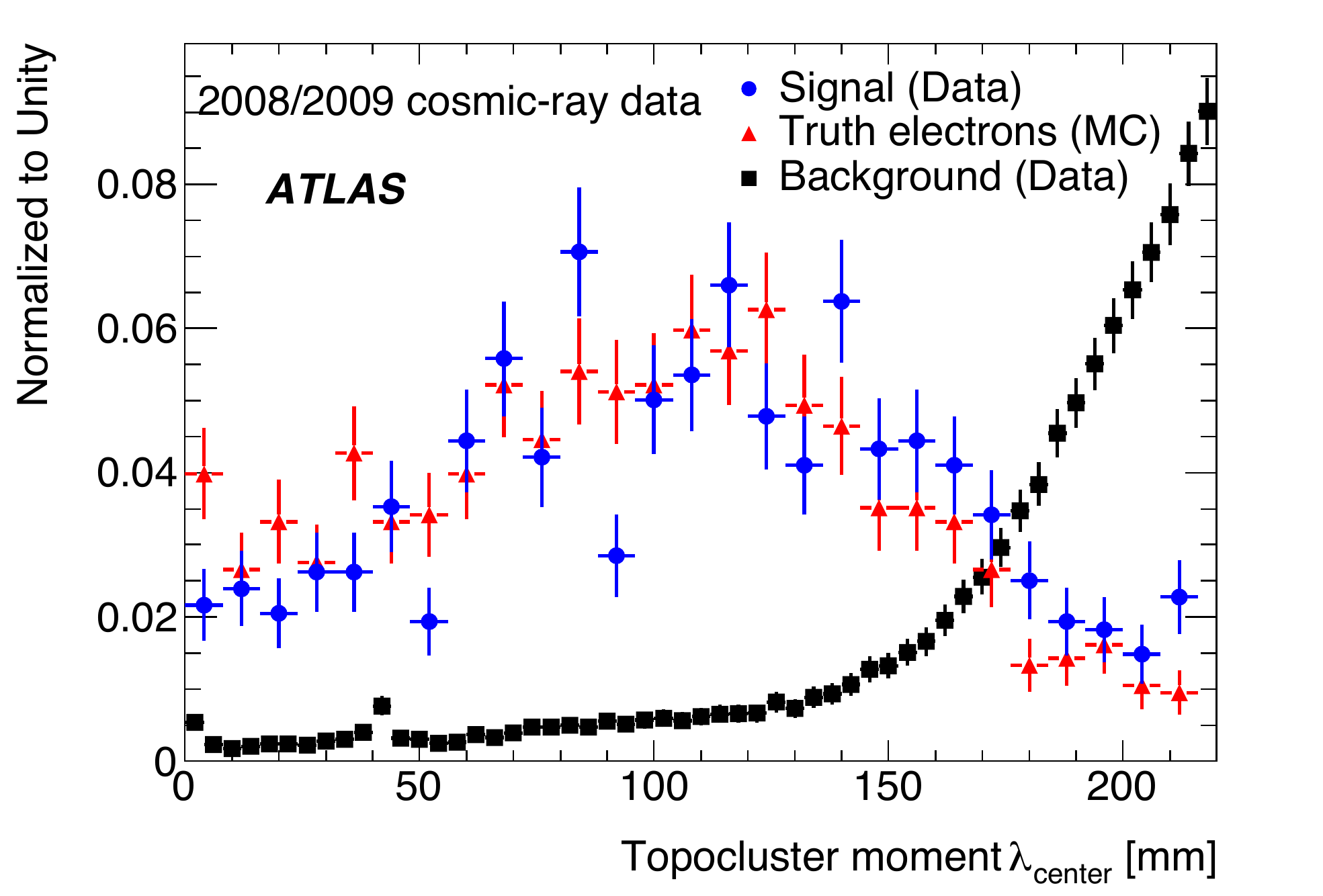}}
  \end{center}
  \caption{Data and Monte Carlo distributions for the topological cluster moment
$\lambda_{\rm center}$, for signal and background events from the low-$p_{\rm T}$ electron
selection. The left-hand plot shows the distribution obtained with the signal selection
applied to the cosmic-ray data and Monte Carlo, along with
the expected distribution for true electrons from the Monte Carlo. The distributions
are normalized to unity. The cut at $220\,{\rm mm}$ is indicated by the dotted vertical line.
For this plot, none of the cluster shape cuts have been applied. The right hand plot shows 
the selected region after all cuts, for the signal events, the events from the background
sample, and for the truth electron distribution from Monte Carlo.
    \label{fig:electron5}}
\end{figure*}

Electron candidate tracks are required to be in the barrel region of the TRT and to have at least 25 TRT hits to 
ensure good quality tracks. There is no requirement of silicon hits. The TR ratio is required to exceed 0.1. Further 
suppression of backgrounds is achieved using various moments of the calorimeter cluster designed to select the compact 
clusters typical of electromagnetic objects.
For example, Figure~\ref{fig:electron5} shows data and Monte Carlo distributions for the topological cluster 
moment $\lambda_{\rm center}$, defined as the distance from the calorimeter front face to the shower center, along
the shower axis.  The two plots show distributions for signal and background events accepted by the low-$p_{\rm T}$ 
electron selection, before (left) and after (right) application of the cluster-moment-based selection criteria.
The left-hand plot shows the distribution obtained with the signal selection applied to the cosmic-ray data and 
Monte Carlo along with the expected distribution for true electrons from the Monte Carlo. The MC
distribution has been normalized to the data. The cut of  $\lambda_{\rm center} < 220\,{\rm mm}$ is indicated by the dotted 
vertical line. Muons which traverse the calorimeter as MIPs leave their energy uniformly distributed in 
depth, producing a peak in the distribution at the point which corresponds to half the depth of the EM calorimeter.
The right-hand plot shows the selected region after all cuts, for the signal events, the events from the background 
sample, and for those events from the Monte Carlo which are matched  to real (``Monte Carlo truth'') electrons.
Good agreement is observed between data and Monte Carlo.

\begin{figure*}[tbph]
  \begin{center}
    \resizebox{0.49\textwidth}{!}{\includegraphics{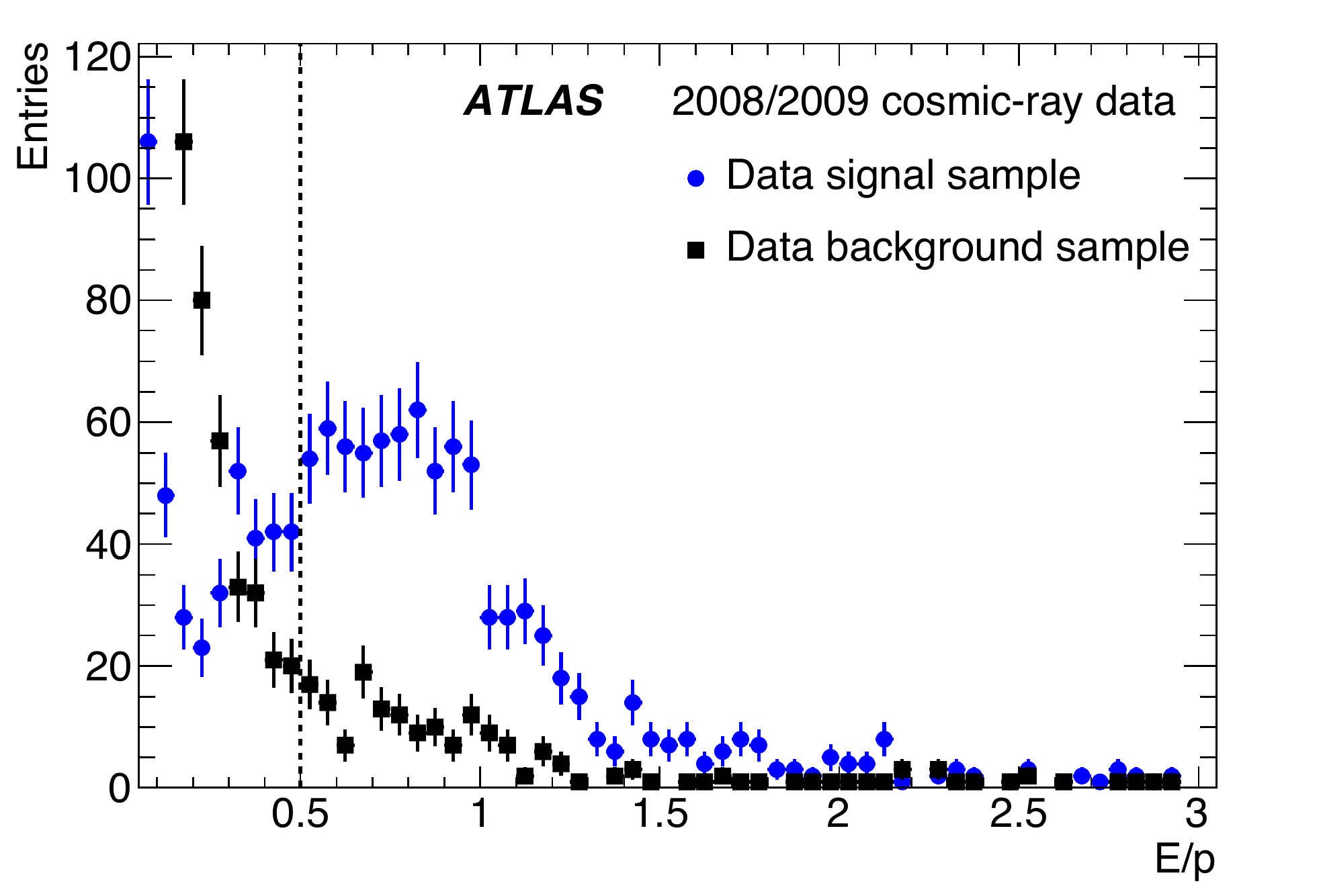}}
    \resizebox{0.49\textwidth}{!}{\includegraphics{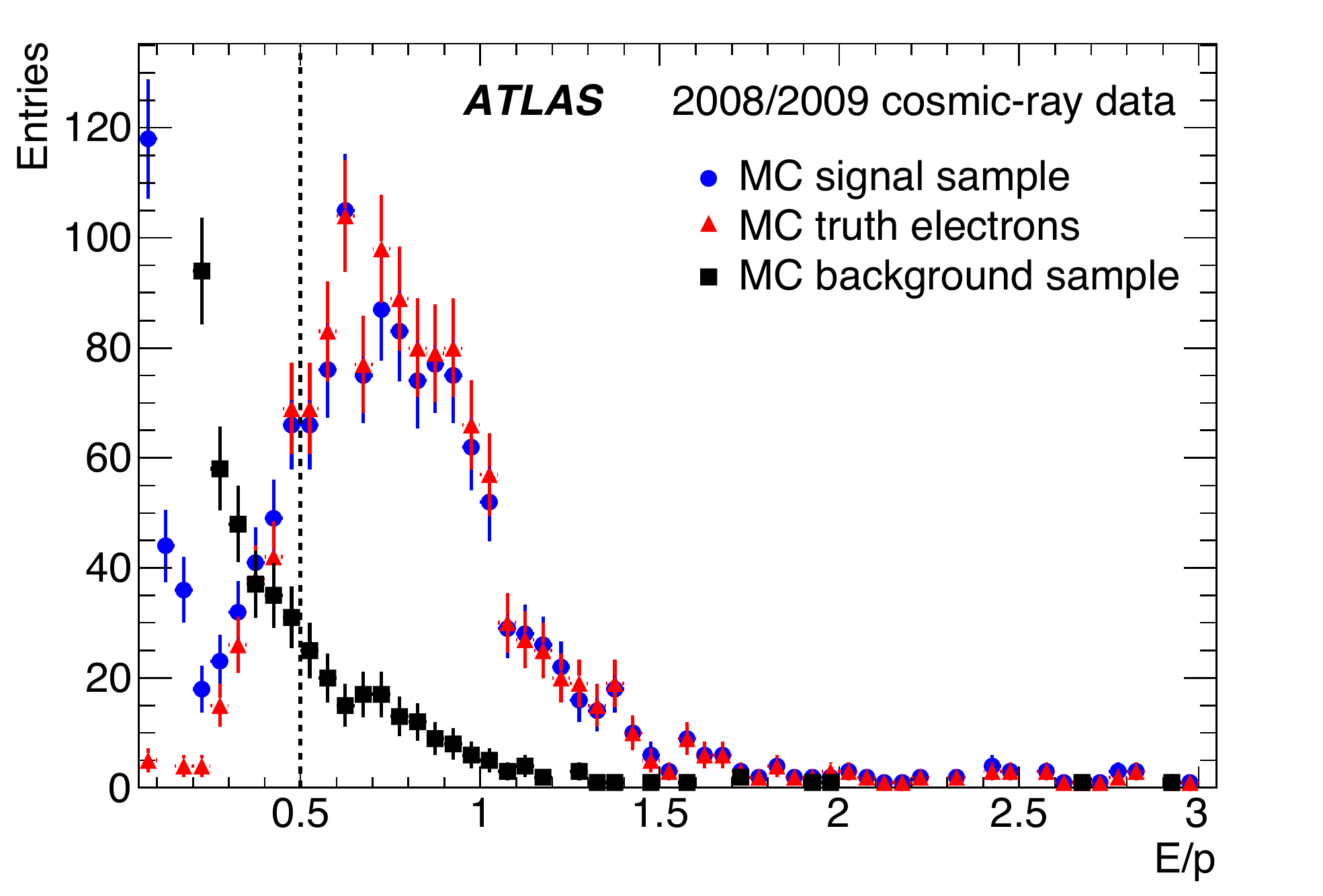}}
    \resizebox{0.49\textwidth}{!}{\includegraphics{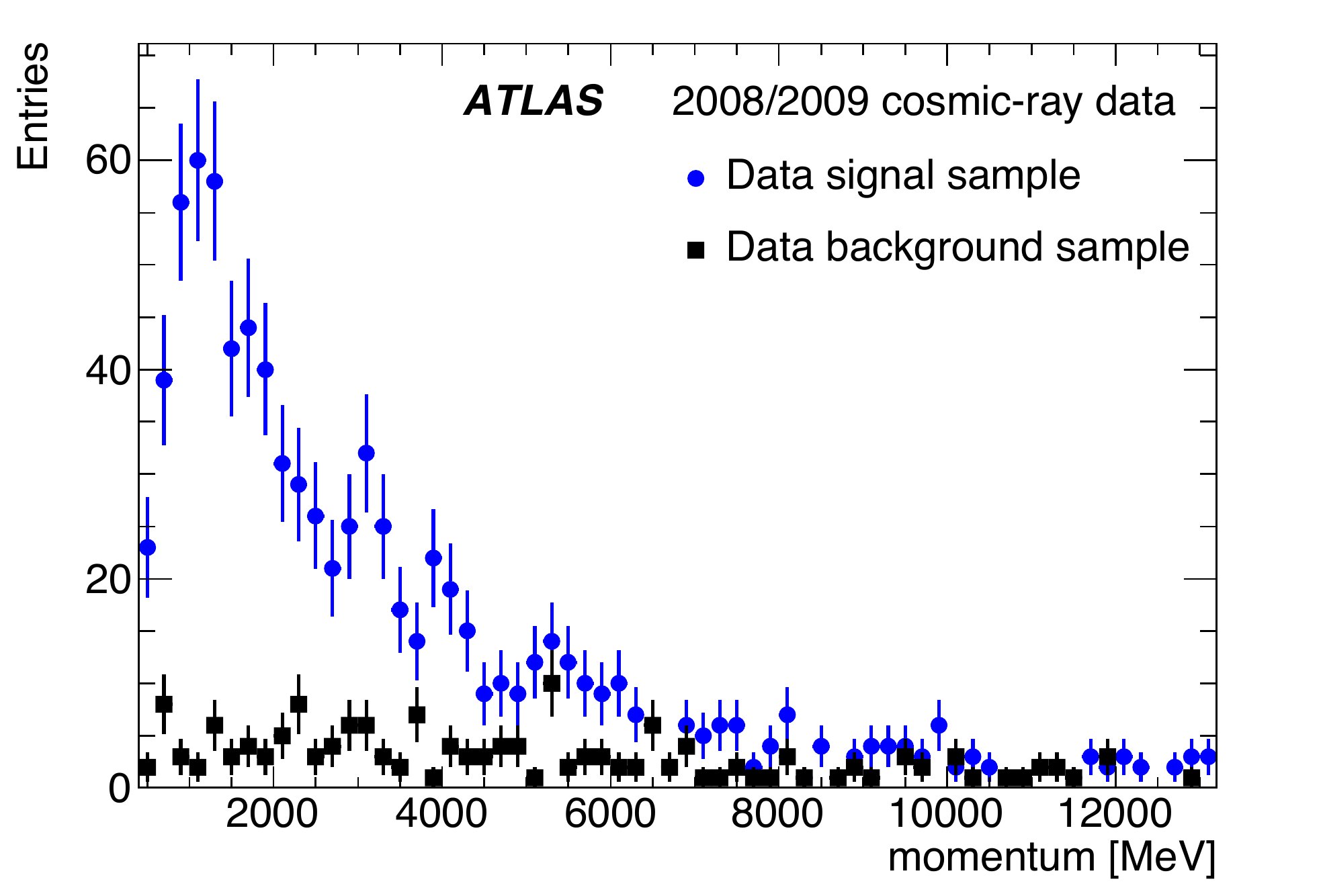}}
    \resizebox{0.49\textwidth}{!}{\includegraphics{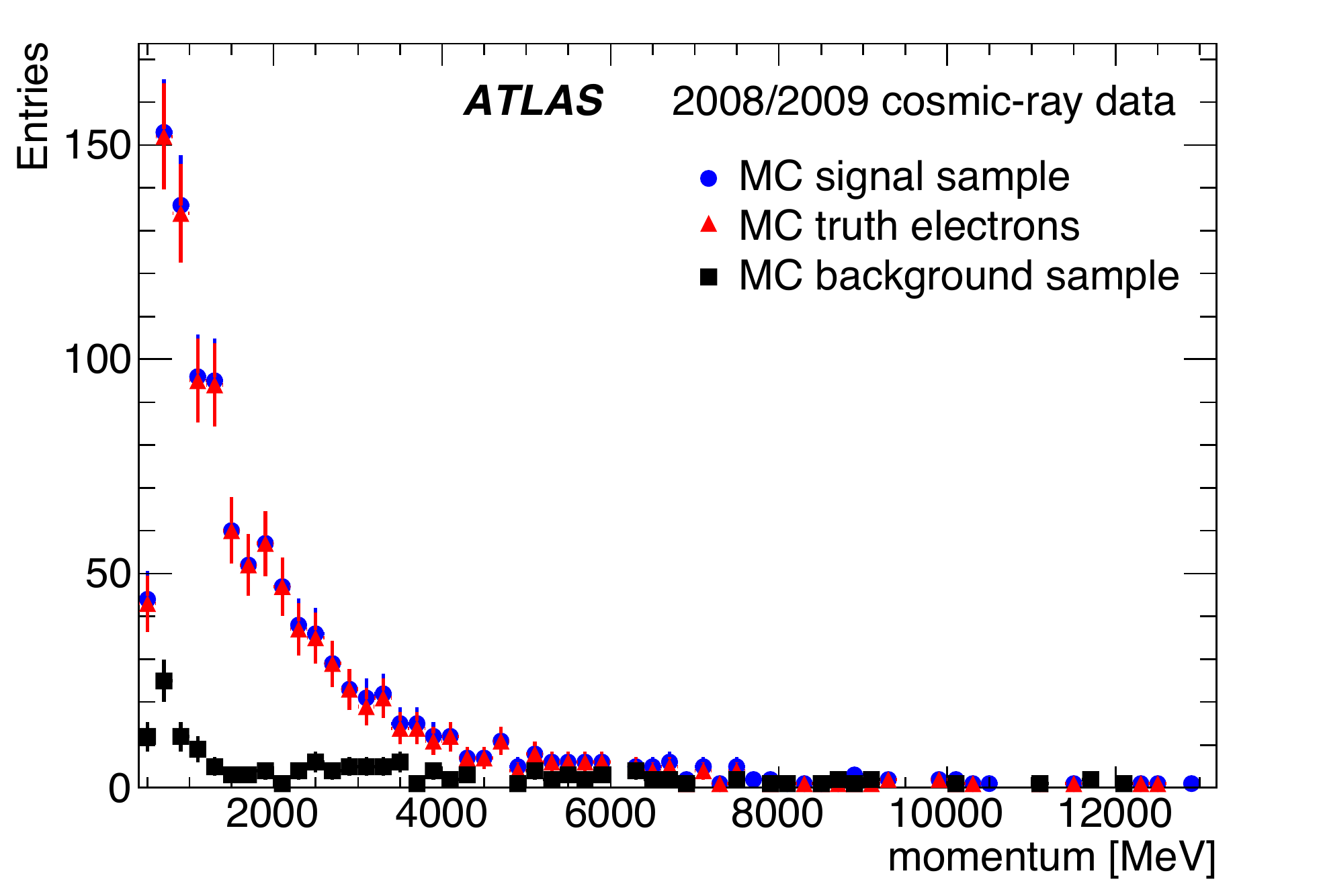}}
  \end{center}
  \caption{
Results of the selection of low $p_{\rm T}$ electrons from the cosmic-ray data samples. The upper plots show the 
$E/p$ distributions for selected events in data (left) and Monte Carlo (right), for both the signal (ionization) 
and background (muon-bremsstrahlung) samples, and for the signal candidates matched to true electrons in the case 
of the Monte Carlo. The lower plots show the corresponding momentum distributions, for events passing the $E/p$ cut,
illustrated by the dashed lines in the upper plots.
    \label{fig:electron6}}
\end{figure*}

As in the electron analysis described in the previous section, signal and background regions are defined in the plane 
of the TR ratio vs. $E/p$. A fit is performed to the data in the background region of the background sample and then 
used to estimate the background in the signal sample. Selected events from both samples are shown in Figure~\ref{fig:electron6} 
for data and Monte Carlo. The Monte Carlo plots also include the distributions of electron candidates that are matched to 
Monte Carlo truth electrons, corresponding to 97\% of the candidates selected from that sample. 
The upper plots show the $E/p$ distributions for the selected events. The final selection cut of $E/p > 0.5$ is illustrated
by the dashed line. This lower $E/p$ cut, relative to the analysis described in section~\ref{sec:electrons1}, is needed 
as the lower $p_{\rm T}$ electrons suffer relatively more energy loss in the detector material before reaching the calorimeter.
The lower plots show the momentum distributions of the electron candidates passing the full selection, and show acceptance 
down to $\sim 500$ MeV.

In general, ATLAS does not attempt to identify electrons down to such low energies. This commissioning analysis is intended
to illustrate the flexibility that exists for the identification of electrons. While the topological clustering technique 
discussed here is not part of the standard electron identification algorithm for most of the detector acceptance, it is the 
default technique 
in the forward region ($2.5<|\eta|<4.9$). This region is beyond the tracking acceptance, so in that case no matching is done to 
tracks. Instead, electrons are identified by topological clusters having properties (e.g. cluster moments) that are typical of 
electromagnetic energy deposits.

\subsection{Commissioning of the $\tau$ reconstruction and identification algorithms}

\begin{figure*}[htbp]
  \begin{center}
    \resizebox{0.49\textwidth}{!}{\includegraphics{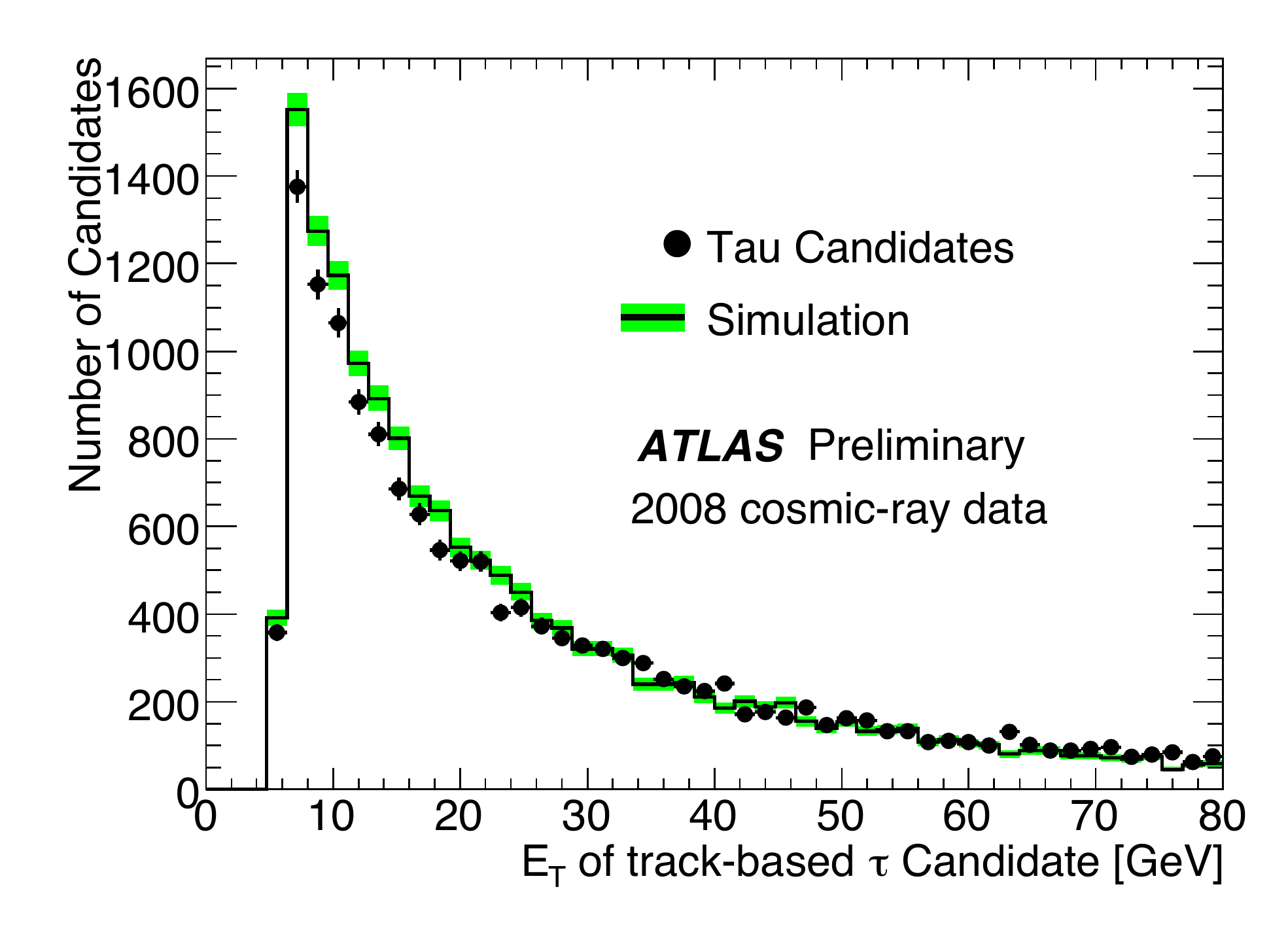}}
    \resizebox{0.49\textwidth}{!}{\includegraphics{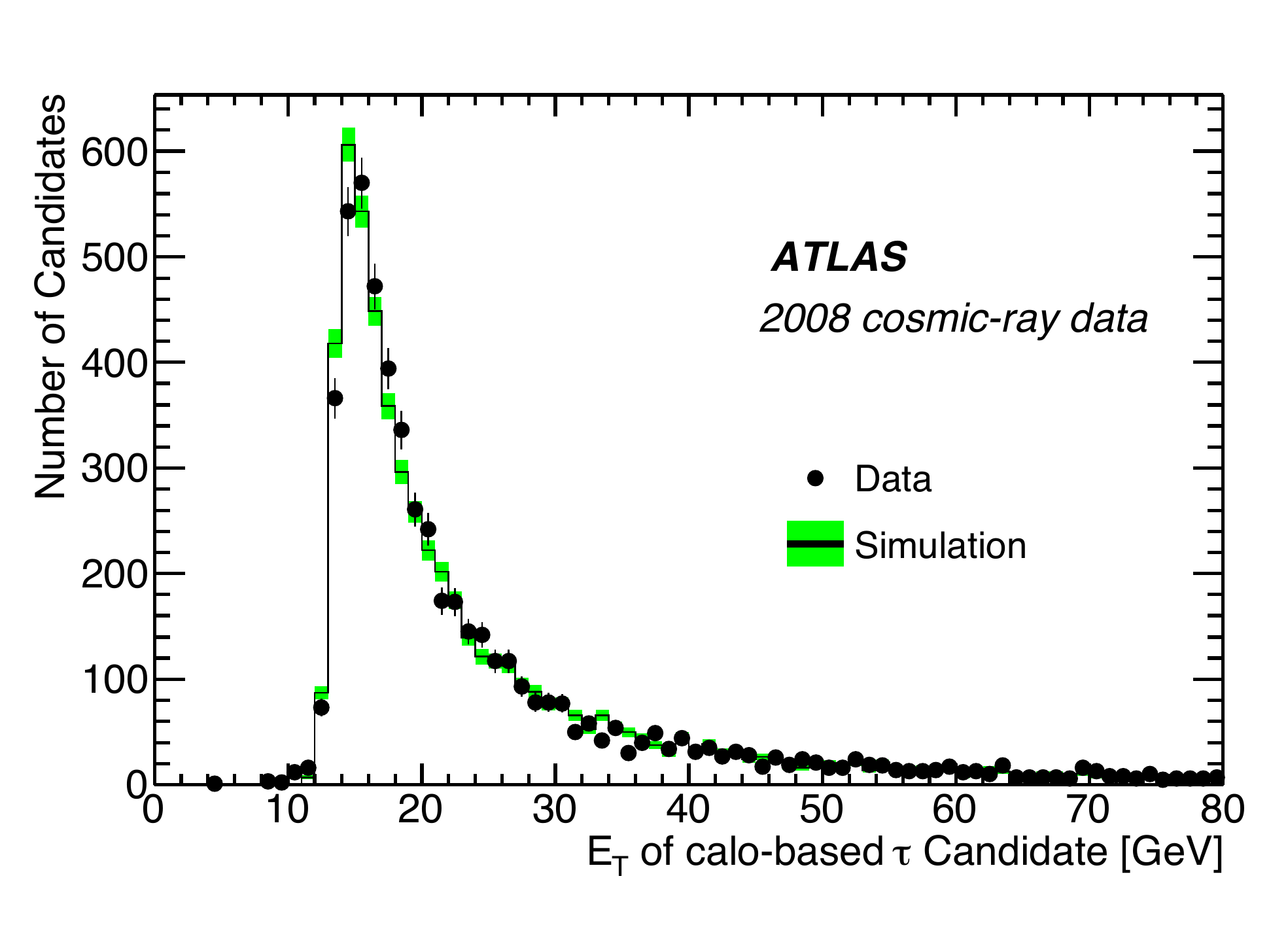}}
 \end{center}
  \caption{Data and Monte Carlo distributions for the transverse energy of $\tau$ candidates from the track-based (left)
and calorimeter-based (right) identification algorithms.
    \label{fig:tau1}}
\end{figure*}

\begin{figure*}[phtb]
  \begin{center}
    \resizebox{0.49\textwidth}{!}{\includegraphics{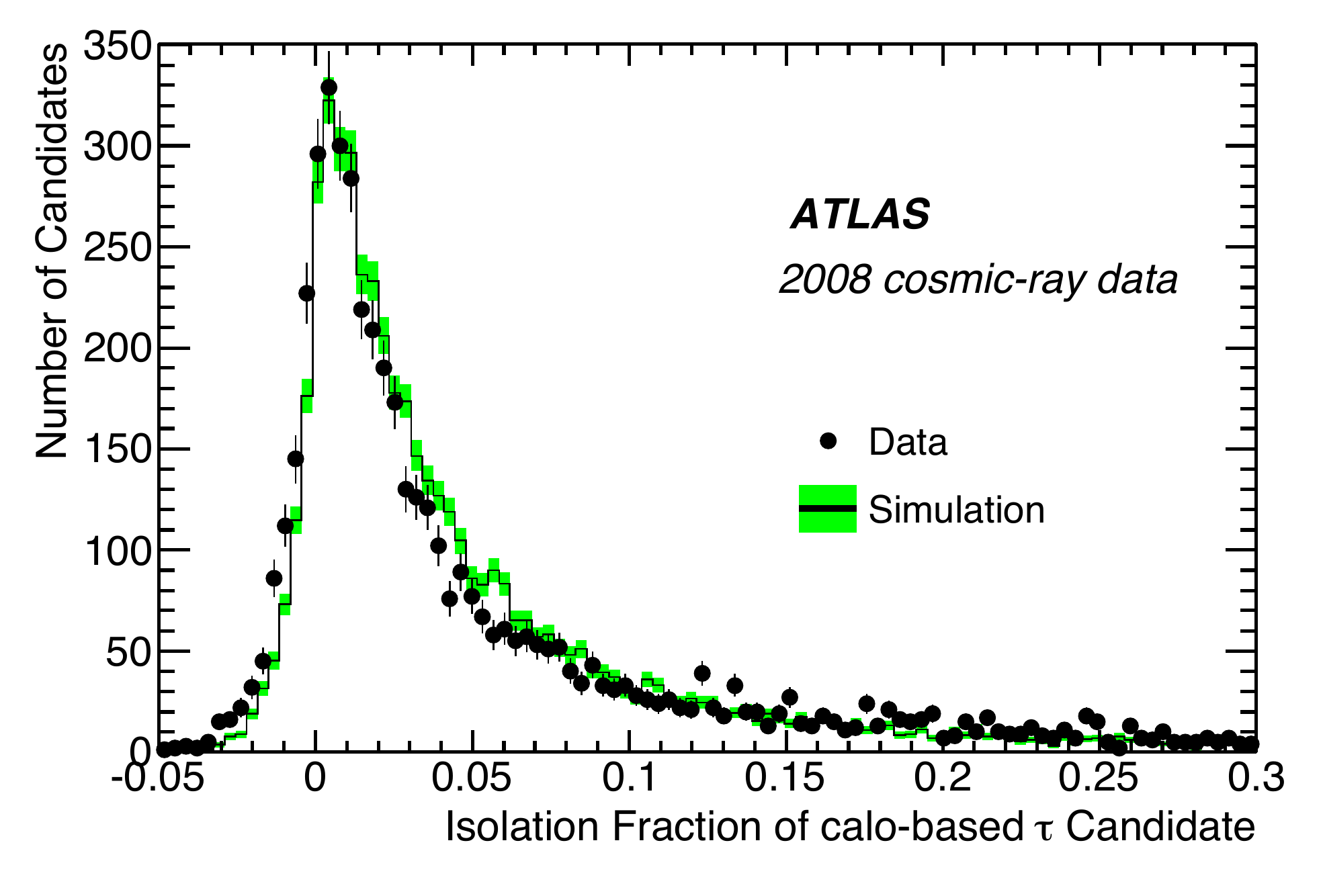}}
    \resizebox{0.49\textwidth}{!}{\includegraphics{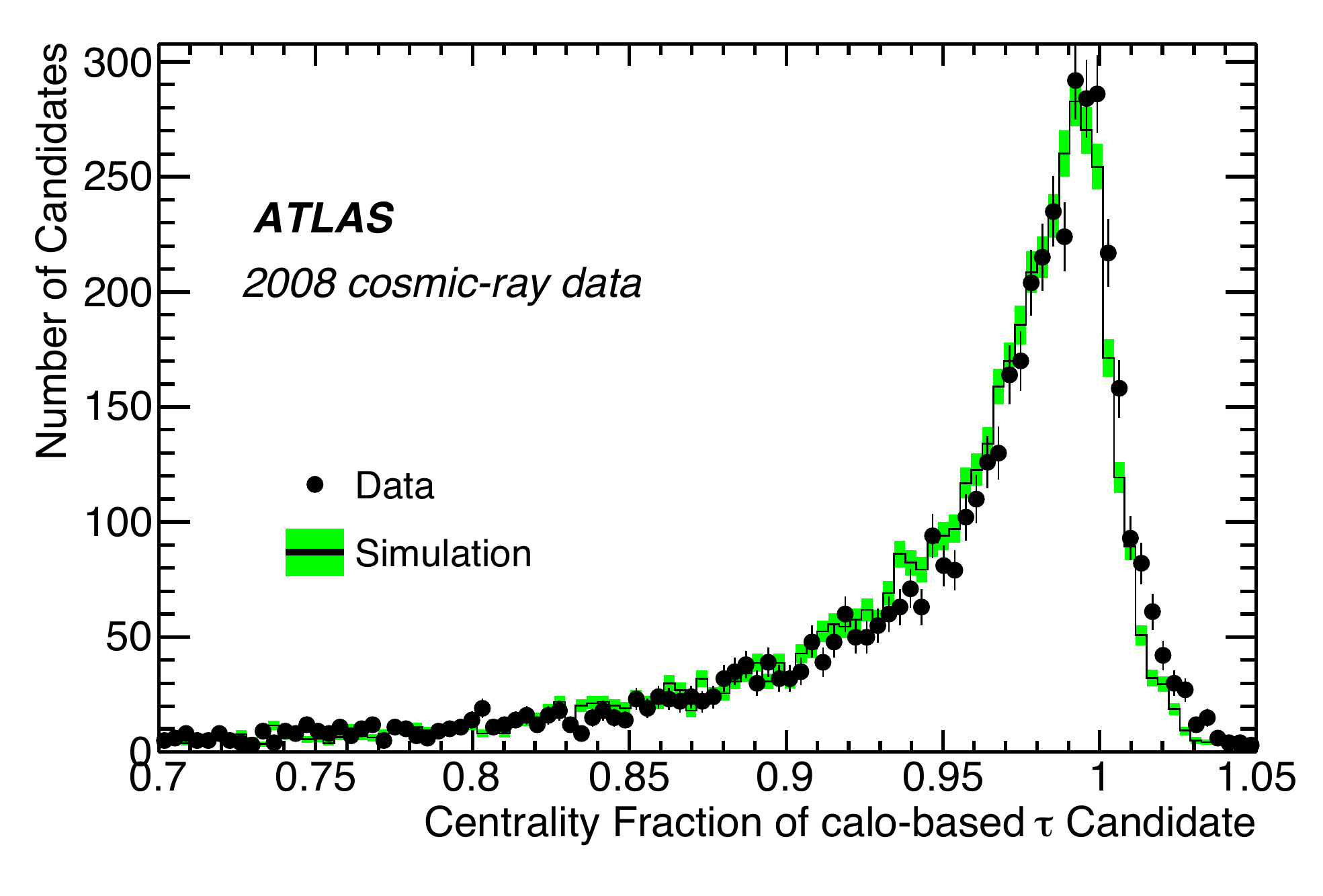}}
    \resizebox{0.49\textwidth}{!}{\includegraphics{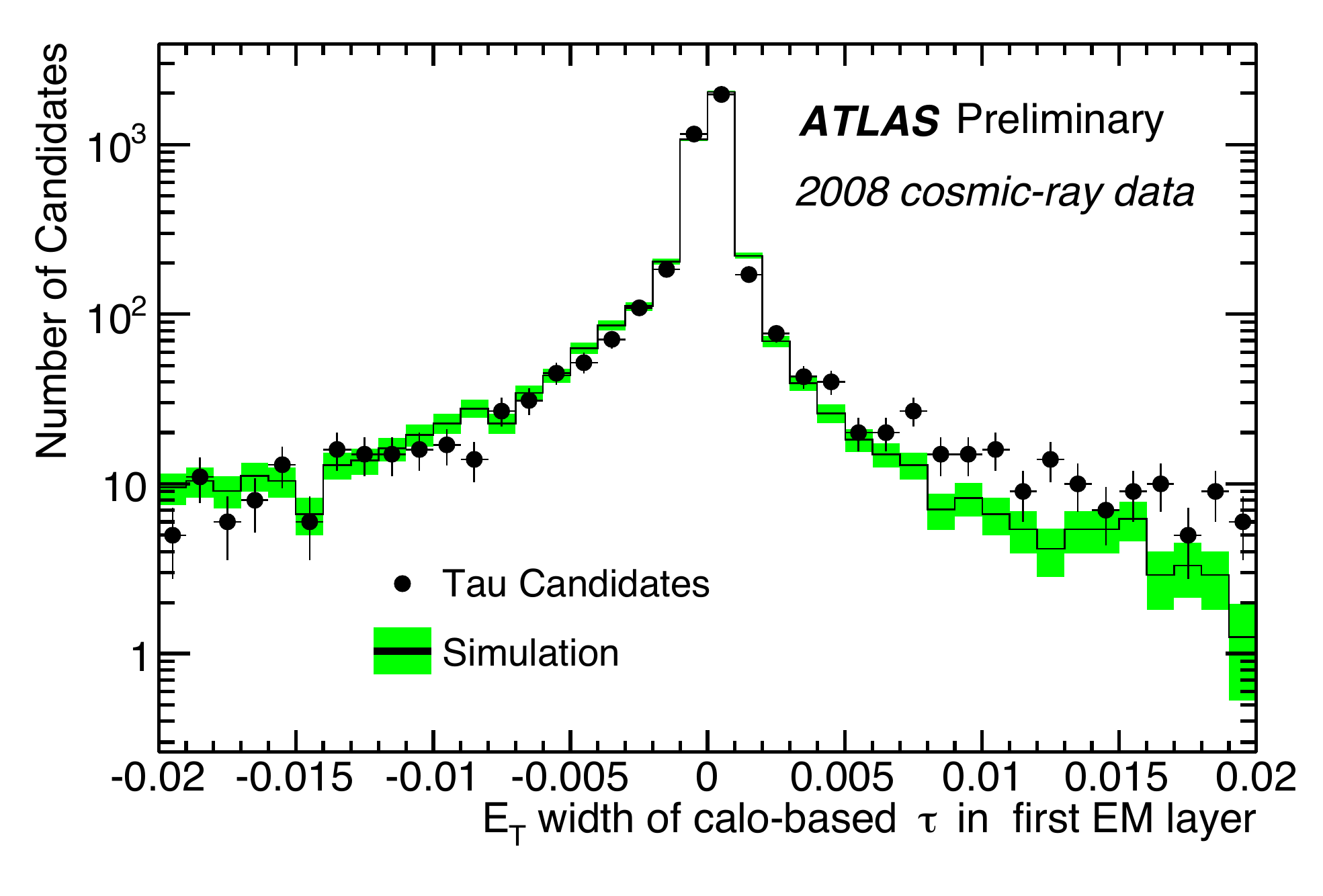}}
    \resizebox{0.49\textwidth}{!}{\includegraphics{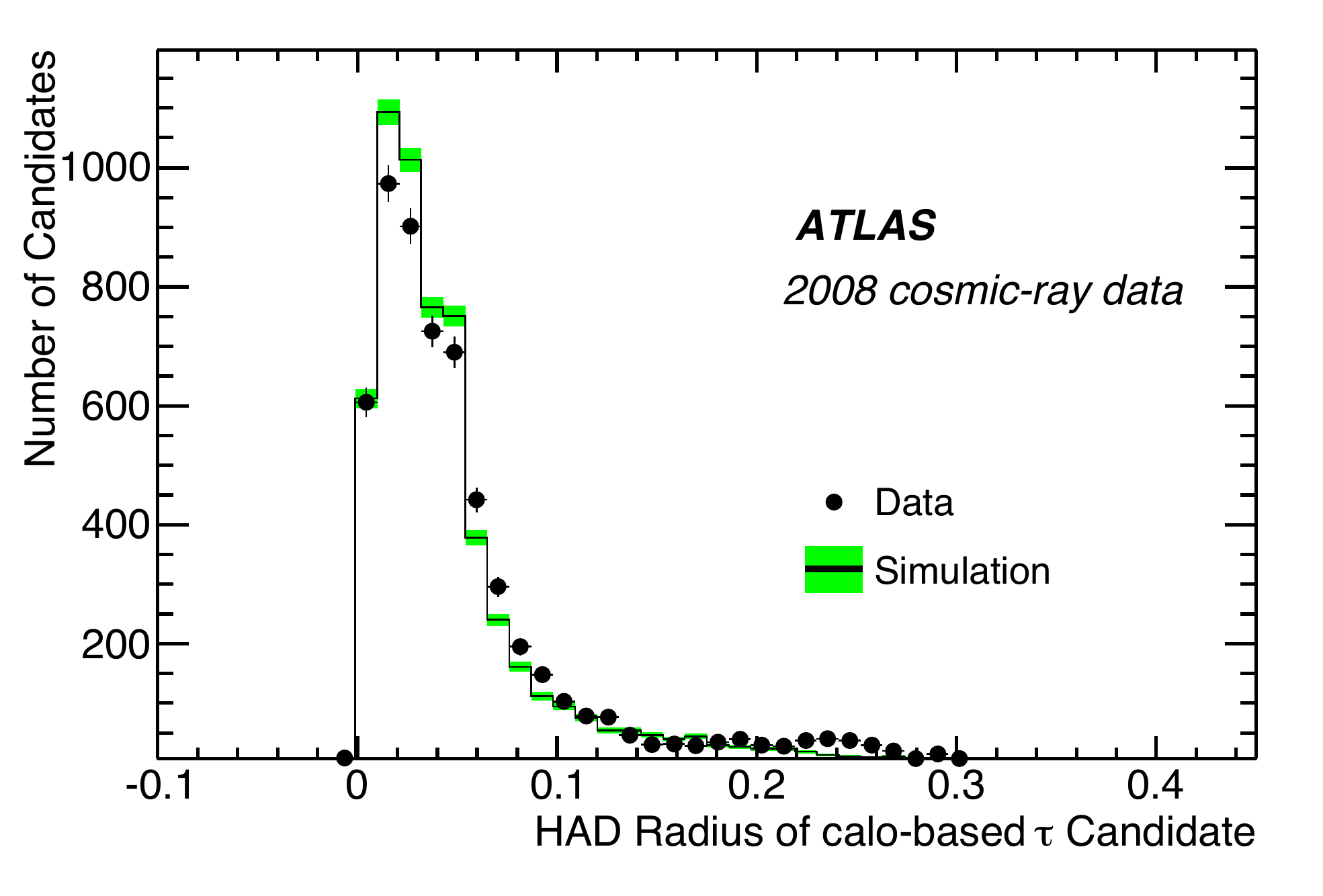}}
  \end{center}
  \caption{Data and Monte Carlo distributions of quantities used in the calorimeter-seeded $\tau$-identification 
algorithms. The upper left plot shows the isolation fraction, defined as the ratio in which the denominator is 
the energy deposited within a cone (around the $\tau$ direction) of $\Delta{R}<0.4$ and the numerator is the 
energy deposited in the region $0.1<\Delta{R}<0.2$. The upper right plot shows the centrality fraction, defined 
as the ratio of the energy within a cone of $\Delta{R}<0.1$ to that within a cone of $\Delta{R}<0.4$. The lower 
left plot shows the transverse energy-weighted width, in the $\eta$ direction, in the first layer of the EM 
calorimeter. The lower right plot shows the distribution of the hadronic radius, the energy-weighted width of 
the cluster, calculated from the energy and positions of the constituent calorimeter cells, relative to the cluster 
center.
    \label{fig:tau2}}
\end{figure*}

As discussed earlier, the cosmic-ray data have also been used to examine the tools used for the identification
of $\tau$ leptons.
A leptonically decaying $\tau$, where the visible final state is either an electron or muon, is difficult to distinguish 
from a primary electron or muon. The $\tau$ identification algorithm therefore focuses on hadronically 
decaying $\tau$ leptons, for which the dominant final states consists of either one or three charged hadrons and some number 
of neutrals. Reconstruction of these final states typically involves several subdetectors: one expects ID tracks associated 
with the charged hadrons and energy deposits in the calorimeter, from both charged and neutral hadrons. The neutrals are 
dominantly pions which decay to two photons and leave their energy in the EM calorimeter. Hadronically-decaying 
$\tau$ leptons are often referred to as $\tau$-jets. 

The identification of $\tau$ leptons is primarily concerned with distinguishing these from a large background due to QCD
jets. The identification algorithm relies upon features such as the track multiplicity, which should be low for
$\tau$ leptons, and the transverse profile of the energy deposits in the detector, which is typically narrower for $\tau$-jets 
than for those from QCD. A $\tau$ will almost always have a final state with either one or three tracks, though some allowance 
is made for imperfect track reconstruction in the ID. Finally, the $\tau$ final state will often result in a prominent deposit
in the electromagnetic calorimeter, associated with photons produced by the decays of neutral pions.

The identification of $\tau$ leptons is performed by an algorithm that can be seeded either by a
track from the ID or by an energetic jet in the calorimeter. 
Track-based $\tau$ candidates are seeded by one good quality track having $p_{\rm T}>6$~GeV and can incorporate
up to seven additional tracks with $p_{\rm T}>1$~GeV within $\Delta R < 0.2$ of the seed track. Once
the full set of tracks for a $\tau$ candidate is established, an associated calorimeter cluster is searched for within
$\Delta R < 0.2$ of the $p_{\rm T}$-weighted track barycenter. The existence of an associated cluster is not required.
Calorimeter-based candidates are seeded by jets reconstructed from calibrated 
topological clusters~\cite{Barillari:2009zza} 
with $\Delta R<0.4$ and $E_{\rm T}>10$~GeV. Once a seed jet is established the 
algorithm searches for associated ID tracks having $p_{\rm T}>1$~GeV, within a cone of radius $\Delta R < 0.3$. The existence 
of such accompanying tracks is not required.

Since no $\tau$ leptons are expected in the cosmic-ray data sample, the focus of the study described here was simply
to exercise the algorithms designed to identify them, and to investigate how 
well  the quantities used for the selection are modeled in the simulation. Since $\tau$ leptons produced in proton-proton
collisions originate from the interaction point, these algorithms normally impose tight requirements on 
the $d_0$ and $z_0$ parameters of the $\tau$ tracks.
However, since application of too tight a selection on these quantities (here with respect to the nominal IP) severely 
limits the available statistics, in this study acceptance cuts of $|d_{0}|\le 40\,{\rm mm}$ and $|z_{0}|\le 200\,{\rm mm}$ 
were used. These define a region which is well within the sensitive volume of the barrel part of the Pixel Detector, which 
extends to $r=123\,{\rm mm}$ and $z=\pm{400}\,{\rm mm}$.

The analysis described here uses the cosmic-ray data from the fall 2008 run. The PCM dataset was used as the starting point 
for each study. Additional requirements were placed on the presence of pixel
hits, differently for the two seeding methods. The track-based selection required that the seed track have at least one pixel hit.  
For studies of the calorimeter-seeded algorithm, while there was no explicit requirement on the association of a track to 
the seed jet, there was a requirement that there be at least one ID track in the event with at least one pixel hit. This track 
would normally be from the muon responsible for the calorimeter cluster around which the seed jet is formed. However, in 
cosmic-ray events these tracks are often not associated with the cluster. The pixel hit requirement is thus intended to 
ensure that the shower shapes (which are used by the identification algorithm) are approximately as expected for particles 
originating from the IP. 

The $\tau$-identification algorithm is designed to reconstruct $\tau$ leptons over a wide spectrum of energies.
However, the relative performance of the two seeding methods varies as a function
of energy with the track-seeding having better performance at lower energies while for higher energies, the calorimeter-seeding is superior.
Because of this, the type of cosmic-ray event producing fake $\tau$ candidates  differs for the two seed types. Most fake track-seeded 
candidates come from minimum-ionizing muons with low momentum, which produce an ID track that fakes a one-prong candidate. The dominant 
source of calorimeter-seeded fakes is cosmic-ray muons that undergo hard bremsstrahlung in the calorimeter. 
When considering real $\tau$ leptons reconstructed from collision data, ideally one would like to have candidates seeded simultaneously 
by the track and cluster-based algorithms. In cosmic-ray data, however, since the origin of fake $\tau$ leptons differs for each algorithm, 
very few candidates fulfil the criteria for both. For this reason, track-seeded and calorimeter-seeded $\tau$ candidates have been 
examined separately. 

Results are presented here to illustrate the agreement between data and cosmic-ray Monte Carlo for the properties of the  
two types of $\tau$ candidates, in particular for those quantities used in the identification algorithms. In what follows it should
be understood that ``tau candidate'' refers to a fake candidate that passes the selection described above, in which nominal
selection criteria have been loosened to ensure sufficient statistics to allow for a meaningful comparison of the data and the cosmic-ray
Monte Carlo simulation. Figure~\ref{fig:tau1} shows the $E_{\rm T}$ distributions of such candidates from the two types of seed. In the 
case of the track-seeded candidates this is reconstructed via an energy-flow algorithm~\cite{Aad:2009wy}. Good agreement is seen between 
the cosmic-ray data and the simulation.

Figure~\ref{fig:tau2} shows data versus Monte Carlo comparisons for some of the quantities used by the $\tau$-identification algorithm. 
The upper left plot shows the isolation fraction for calorimeter-seeded candidates, which is a measure of the collimation of the $\tau$-jet, 
defined as a ratio in which the denominator is the energy deposited within a cone (around the $\tau$ direction) of $\Delta{R}<0.4$ 
and the numerator is the energy deposited in the region $0.1<\Delta{R}<0.2$. For the same sample of calorimeter-seeded candidates, 
the upper right plot shows the centrality fraction, defined as the ratio of the energy within a cone of $\Delta{R}<0.1$ to 
that within a cone of $\Delta{R}<0.4$. The lower left plot shows the transverse energy-weighted width, in the $\eta$ direction, in 
the first (most finely segmented) layer of the EM calorimeter. The plot at the lower right shows the distribution of the hadronic
radius, which is the energy-weighted width of the cluster, calculated from the energy and positions of the constituent calorimeter 
cells, relative to the cluster center.
All distributions show good agreement between the data and the simulation. In the upper left plot of Figure~\ref{fig:tau2} there
are entries at negative values that are attributable to the noise. This is also the cause of the entries at values greater
than 1 in the plot of the centrality fraction. The agreement between data and simulation in these regions illustrates
that the modeling of the electronic noise in the simulation is reasonable.

\section{Jet and missing transverse energy studies using cosmic-ray events}
\label{sec:jets}

Numerous theories of physics beyond the Standard Model predict the existence of massive weakly interacting particles that 
escape detection and thus leave a large energy imbalance in the detector. For this reason, detailed understanding of the 
detector performance for missing transverse energy ($\met$) is extremely important. The most important input to the 
calculation of the $\met$  comes from the calorimeter, which provides coverage in the region of $|\eta|<4.9$. Cosmic-ray 
energy deposits in the calorimeter typically lead to an imbalance in the transverse energy in the event. This effect
can be large in the case of high-energy cosmic rays that lose a large amount of energy via bremsstrahlung. 
The energy deposits from cosmic-ray muons (or cosmic-induced air-shower events) 
can be reconstructed as jets, creating backgrounds to jet selections in many analyses. The properties of jets and $\met$ 
reconstructed from cosmic-ray data are presented below, along with a discussion of techniques that have been developed 
to suppress such contributions in the analysis of collision data. 

\subsection{Missing transverse energy in randomly-triggered events}

As is the case when running with proton-proton collisions, during cosmic-ray data-taking randomly triggered events are also recorded. 
The large sample of such events collected during the global cosmic-ray running allows 
investigations of the detector performance for the measurement of missing transverse energy. No energy imbalance is 
expected in these events. However, global quantities such as $\met$ and $\sum E_{\rm T}$ (defined below) result from the sum of 
energy deposits in $\sim$~200k calorimeter channels, each with its own electronic noise. A proper determination of these 
quantities relies on a good understanding of the cell-level noise in all calorimeter channels, and, in particular, a proper
treatment of a few very noisy cells and cells having non-nominal high-voltage. There are 
currently two standard methods for reconstructing  missing transverse energy in ATLAS. The first is a cell-level method 
that takes as input all calorimeter cells with $|E|>2\sigma_{\rm noise}$.
The second method takes as input calibrated 
topological clusters built with $t_{\rm seed}=4$, $t_{\rm neighbor}=2$ and $t_{\rm cell}=0$.  The reconstructed quantities are:

\begin{spaced_items}
\item $E_{\rm x}^{\rm miss} = -\sum{E}{\rm sin}\theta{\rm cos}\phi$ 
\item $E_{\rm y}^{\rm miss} = -\sum{E}{\rm sin}\theta{\rm sin}\phi$ 
\item $E_{\rm T}^{\rm miss} = ({(E_{\rm x}^{\rm miss})^2+(E_{\rm y}^{\rm miss})^2})^{1/2}$
\item $\sum E_{\rm T} = \sum{E}{\rm sin}\theta$ 
\end{spaced_items}
where in each case the sum is over all cells included in the cluster.

\begin{figure}[tbhp]
  \begin{center}
    \resizebox{0.49\textwidth}{!}{\includegraphics{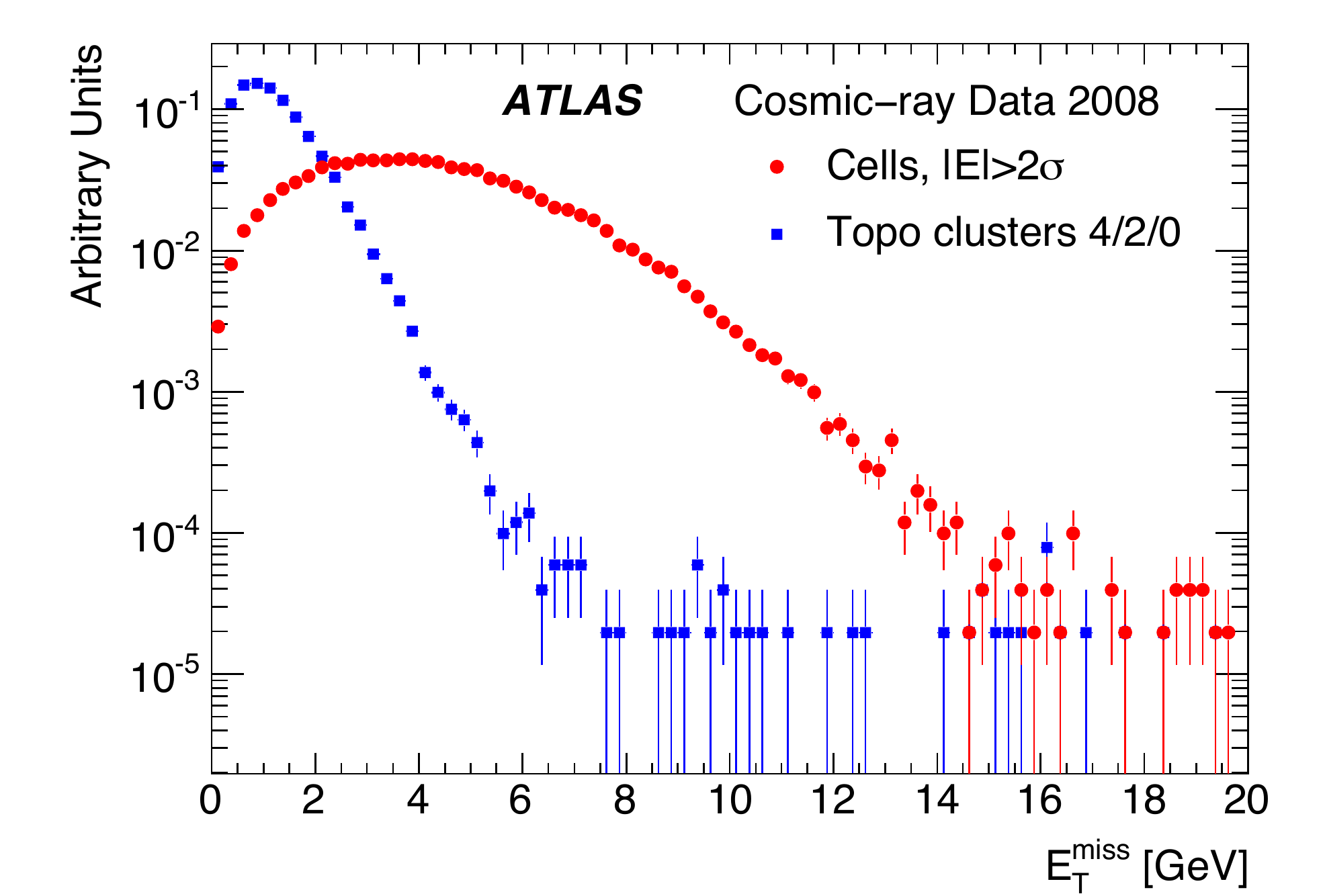}}
  \end{center}
  \caption{Distribution of $\met$ from analysis of random triggers recorded during the 
2008 global cosmic-ray running, for the two methods described in the text.
    \label{fig:met}}
\end{figure}

Figure~\ref{fig:met} shows the results of both calculations applied to the random triggers recorded during a 2008 
cosmic-ray run, illustrating the superior noise suppression of the method using the topological clustering. Tails in 
the distribution (beyond 8~GeV for topological-cluster-based, and 16~GeV for cell-based definition), contributing 
less than 0.1\% of events, are due to coherent noise in a specific region of the LAr presampler which has since been 
repaired.  The time stability of the \met ~calculation was also investigated and found to be very good. 
For the topological-cluster-based method, which provides the best resolution, the mean and width 
of the distributions of the $x$ and $y$ components of \met ~were stable to within about 100 MeV over the 
45 days of data-taking.

\subsection{Jets and missing transverse energy in cosmic-ray events}

\begin{figure*}[phtb]
  \begin{center}
    \resizebox{0.49\textwidth}{!}{\includegraphics{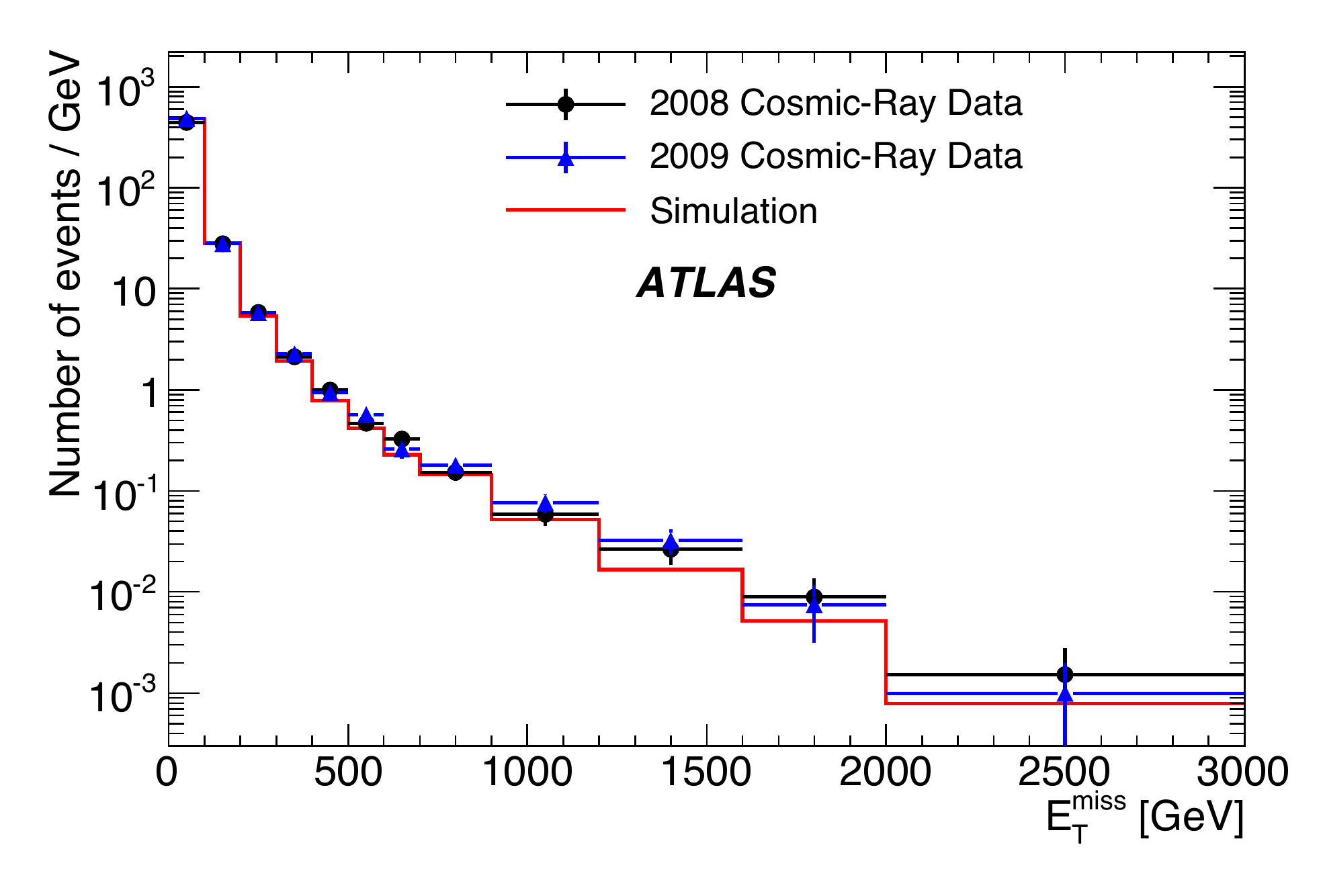}}
    \resizebox{0.49\textwidth}{!}{\includegraphics{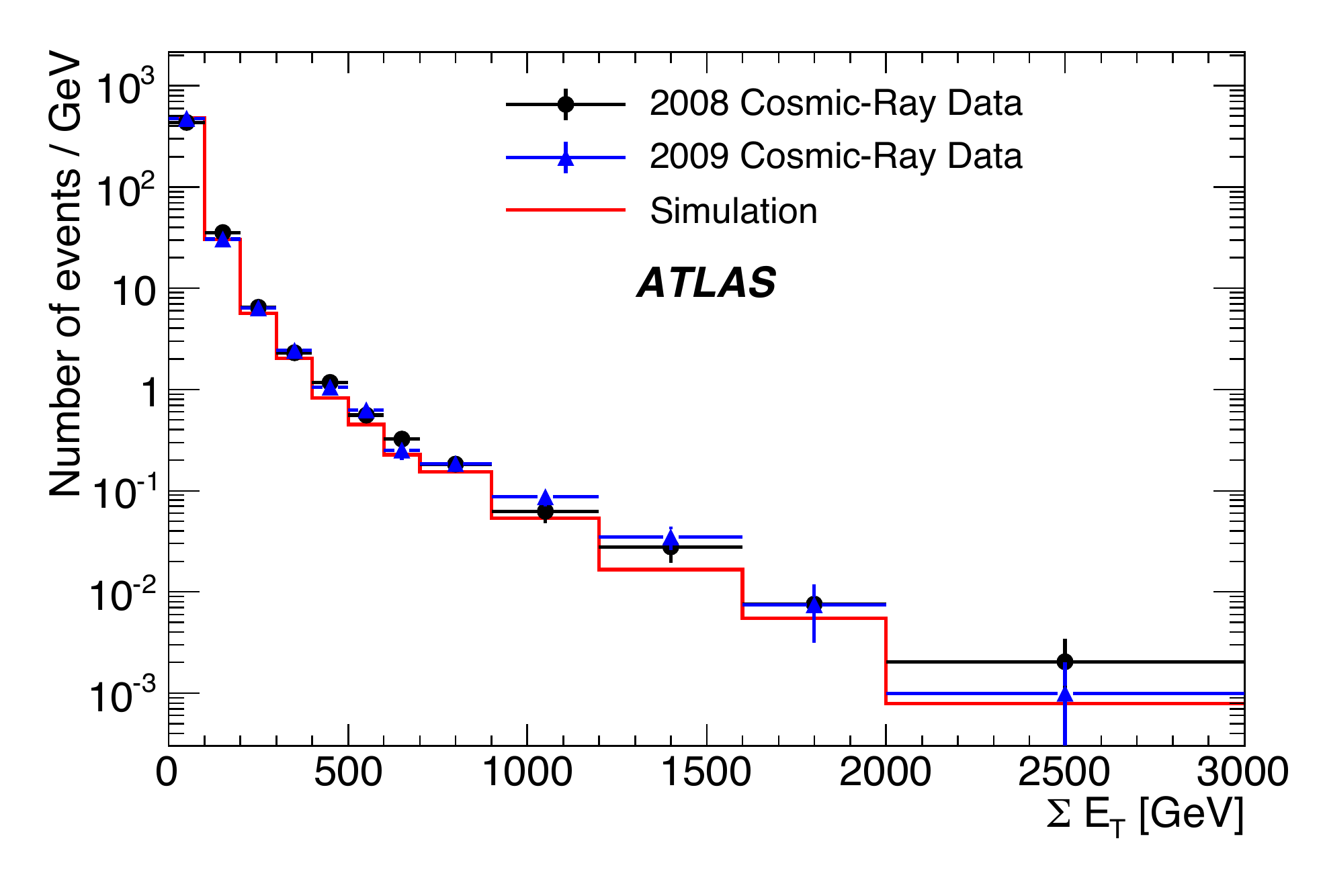}}
  \end{center}
  \caption{Distributions of $\met$ (left) and $\sum E_{\rm T}$ (right) from analysis of
the 2008 and 2009 L1Calo triggered cosmic-ray data and from the cosmic-ray 
Monte Carlo sample. The 2008 and Monte Carlo distributions are normalized to
the 2009 data distribution in the region $100 < \met < 300$~GeV.
    \label{fig:jet1}}
\end{figure*}

\begin{figure*}[phtb]
  \begin{center}
    \resizebox{0.49\textwidth}{!}{\includegraphics{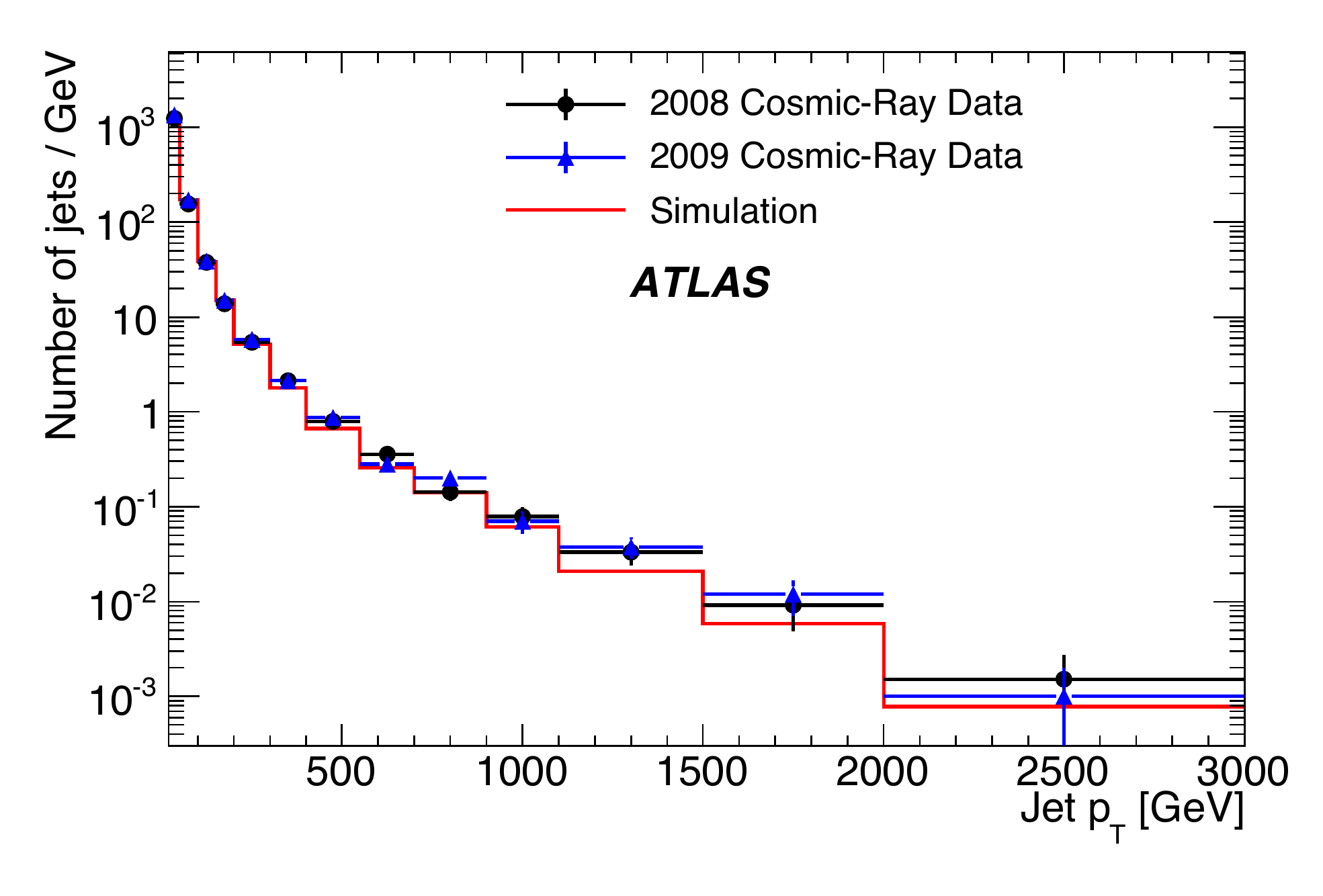}}
    \resizebox{0.49\textwidth}{!}{\includegraphics{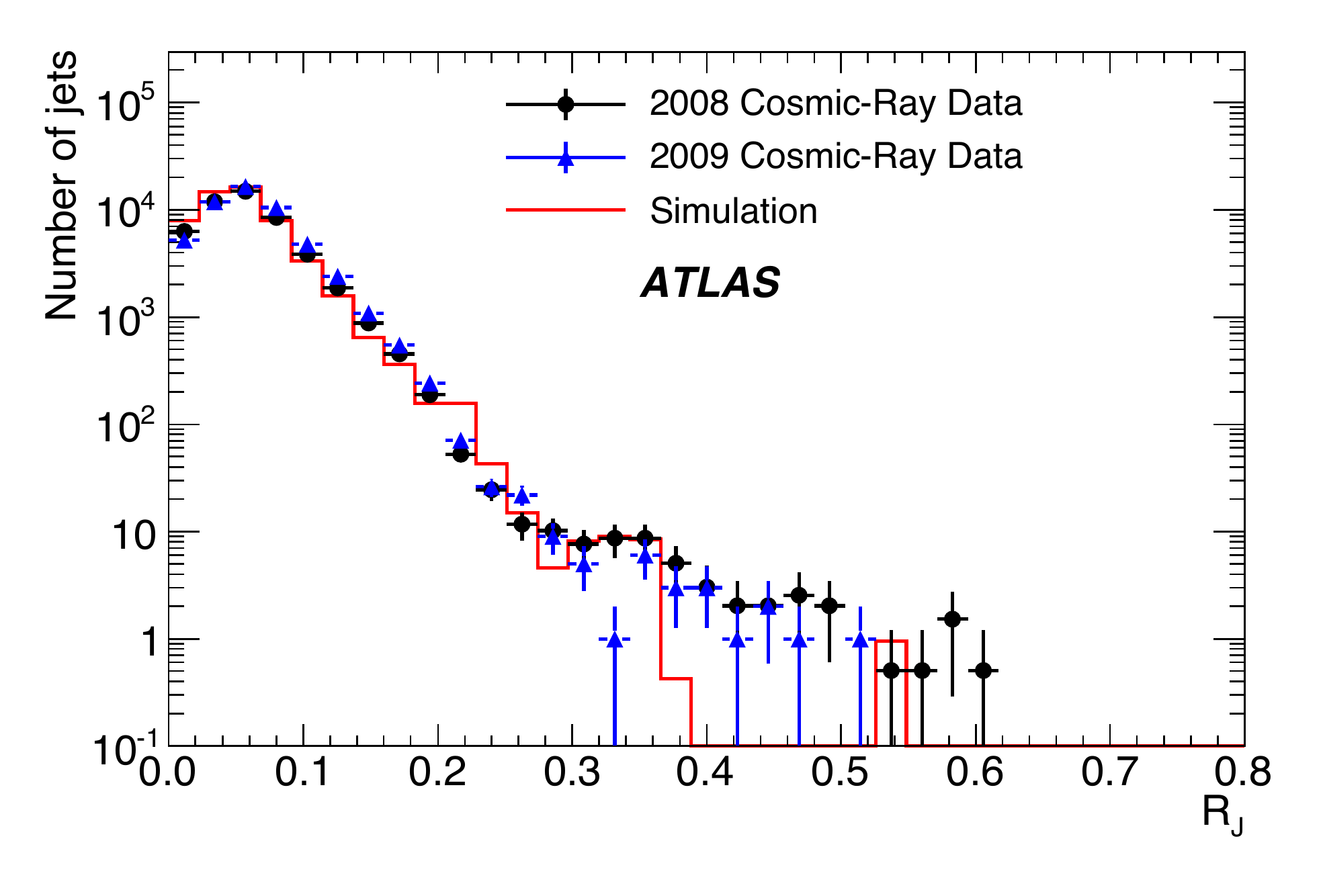}}
    \resizebox{0.49\textwidth}{!}{\includegraphics{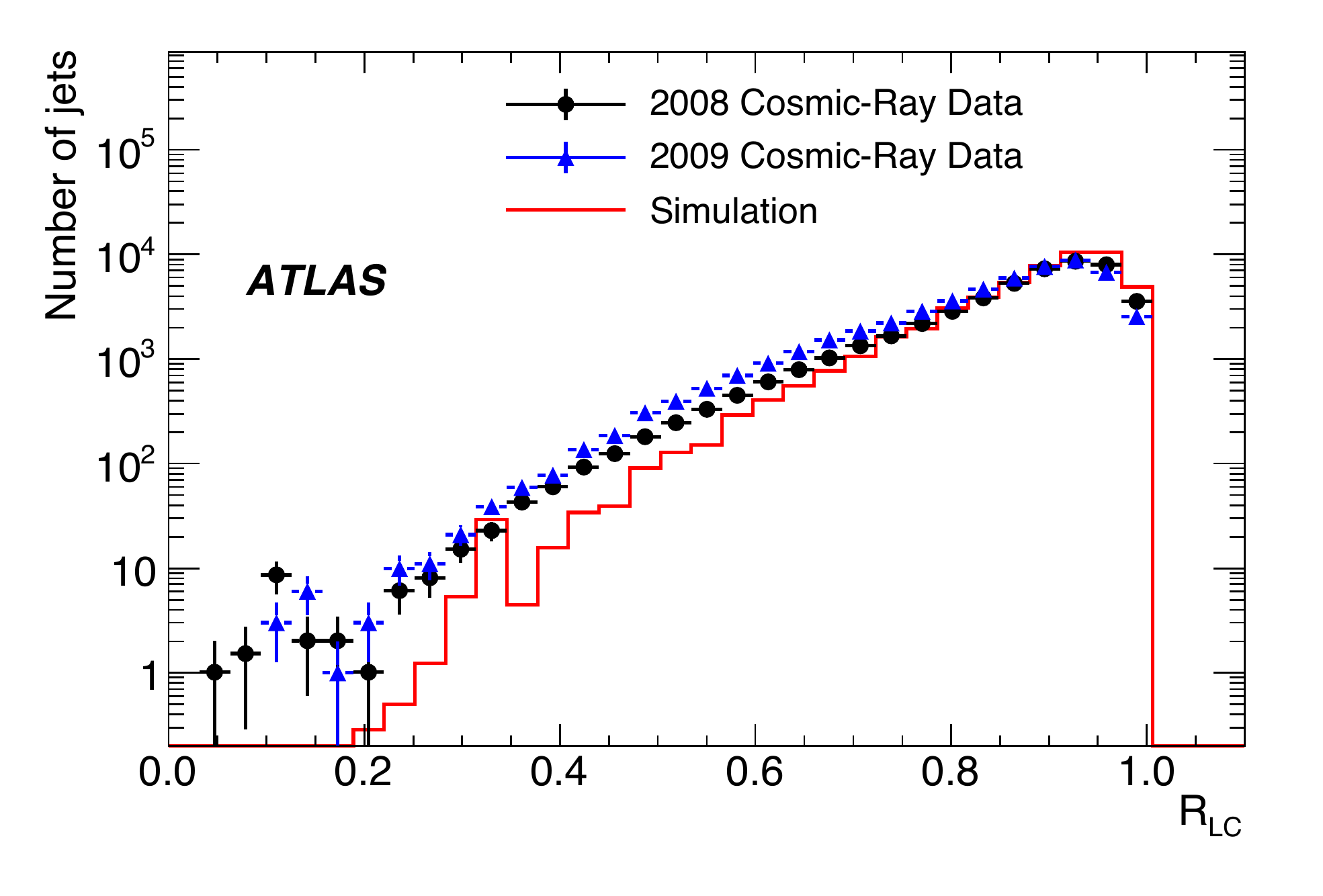}}
    \resizebox{0.49\textwidth}{!}{\includegraphics{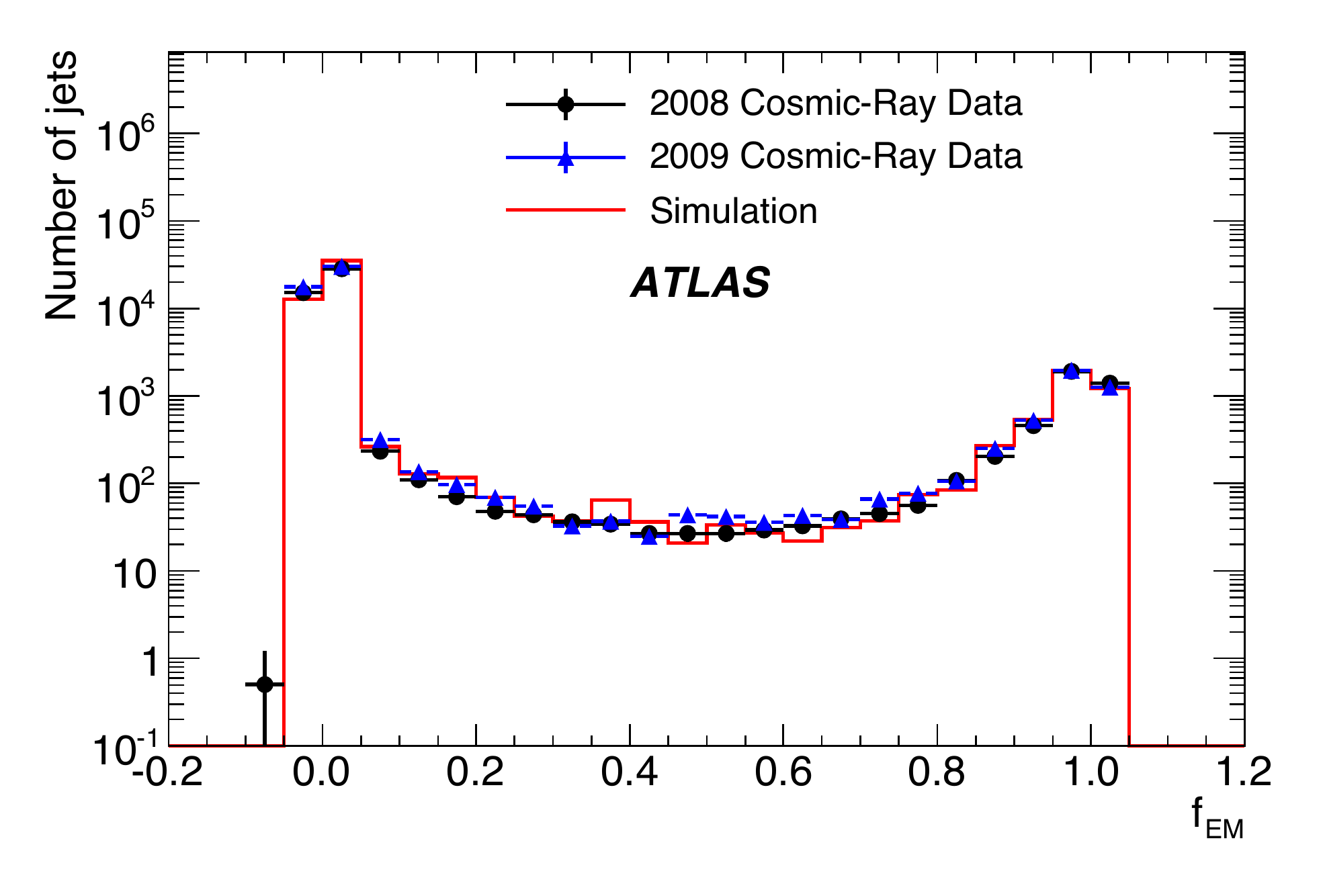}}
  \end{center}
  \caption{Properties of fake jets reconstructed from the 2008 and 2009 L1Calo triggered cosmic-ray
data samples: The upper left plot shows the jet $p_{\rm T}$ in the acceptance region above 20~GeV, 
while the other three plots show distributions in quantities used to suppress these contributions 
in collision data, as described in the text.  The normalizations are the same as used in Figure~\ref{fig:jet1}.
    \label{fig:jet2}}
\end{figure*}

The reconstruction of jets and \met ~in cosmic-ray events has been studied using the L1Calo-triggered data taken in the 
2008 and 2009 cosmic-ray runs. For jet reconstruction an anti-$k_{\rm t}$ algorithm\cite{Cacciari:2008gp} is employed, with 
calibrated topological clusters as input. Figure~\ref{fig:jet1} shows the distributions of missing transverse energy and summed 
transverse energy from cosmic-ray events having a reconstructed jet with $p_{\rm T} > 20$~GeV. The 2008 and 2009 data samples are 
shown separately to demonstrate the consistency of the two samples. The distributions from the 2008 data and the cosmic-ray Monte 
Carlo are normalized to that of the 2009 data in the region of $100 < \met < 300$~GeV. This is in order to avoid any threshold 
effects, since the trigger was not simulated in the cosmic-ray Monte Carlo sample. In each case there is agreement with the 
shape expected from the Monte Carlo, which requires an understanding of the electronic noise in each calorimeter channel. The 
upper left plot in Figure~\ref{fig:jet2} shows the corresponding $p_{\rm T}$ distribution of the jets reconstructed in this sample.

Suppression of these fake jet candidates can be performed using a selection based on three quantities:
\begin{spaced_items}
\item $R_{\rm J} =\sum \limits_{i=1}^{N}\sqrt{(\eta_{i}-\eta_{\rm jet})^2+(\phi_{i}-\phi_{\rm jet})^2}\cdot E_{i}
        /\sum \limits_{i=1}^{N}E_{i}$ 
\item $R_{\rm LC} = (\sum\limits_{i=1}^{2}E_{i}^{\rm Had}+\sum\limits_{i=1}^{32}E_{i}^{\rm EM})/\sum\limits_{i=1}^{N}E_{i}$
\item $f_{\rm EM}= E_{\rm EM}/(E_{\rm EM}+E_{\rm Had})$
\end{spaced_items}
Here $R_{\rm J}$ represents the energy-weighted lateral extent of the jet, in $\eta\times\phi$ space. $R_{\rm LC}$ represents the 
fraction of the jet energy contained in the ``leading cells'', defined as the two most energetic cells in the hadronic 
calorimeter and the 32 most energetic cells in the EM calorimeter, where the sum in the denominator is over all $N$ 
calorimeter cells associated with the jet candidate. Finally, $f_{\rm EM}$ represents the electromagnetic fraction of the jet, 
defined as the fraction of the jet energy that is deposited in the EM calorimeter. The distributions of these three quantities
for the selected jets are also shown in Figure~\ref{fig:jet2}. Again there is good agreement between the 
2008 and 2009 cosmic-ray data as well as reasonable agreement with the cosmic-ray Monte Carlo. The normalization of the 
distributions in Figure~\ref{fig:jet2} is the same as used in Figure~\ref{fig:jet1}.

\begin{figure*}[phtb]
  \begin{center}
    \resizebox{0.49\textwidth}{!}{\includegraphics{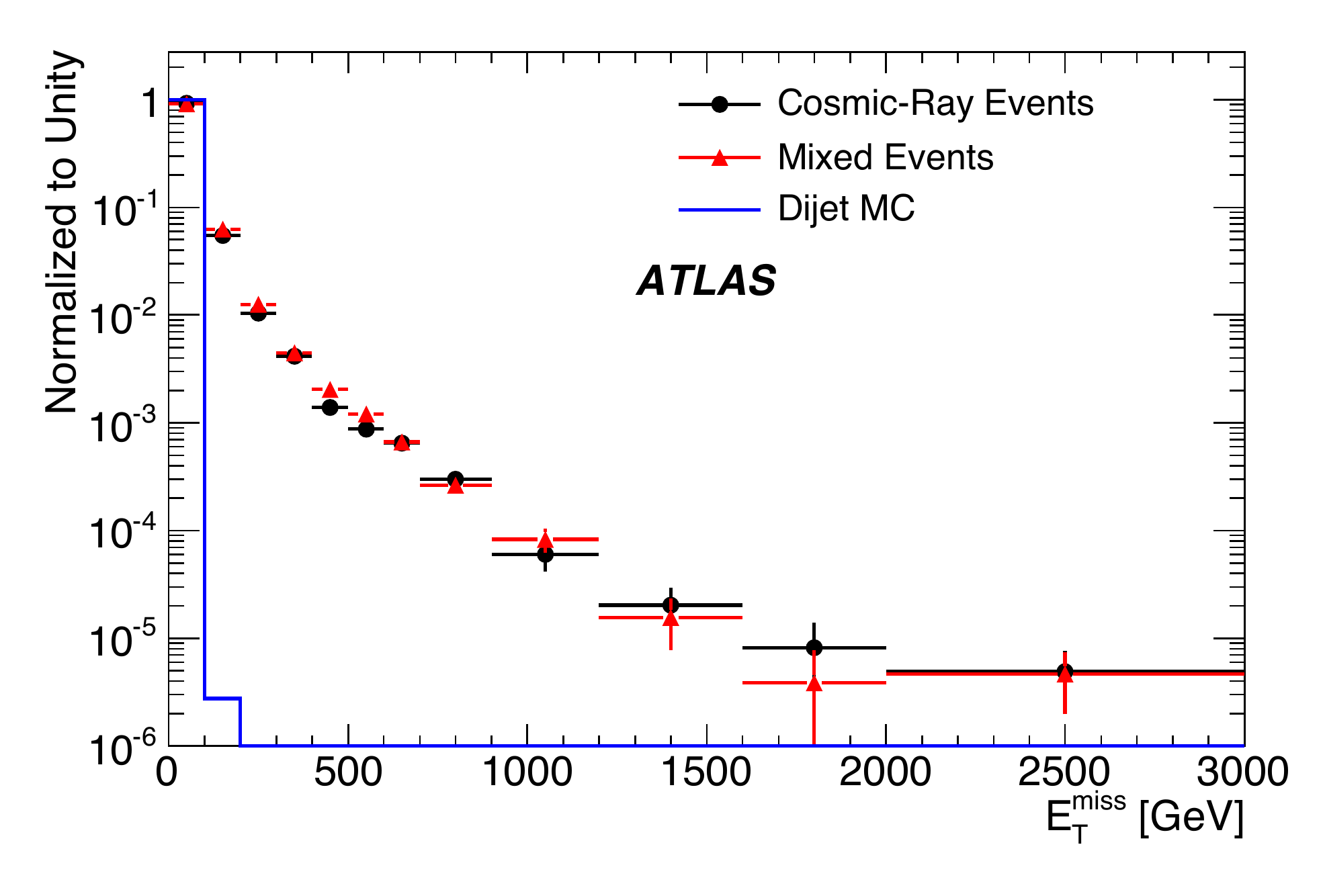}}
    \resizebox{0.49\textwidth}{!}{\includegraphics{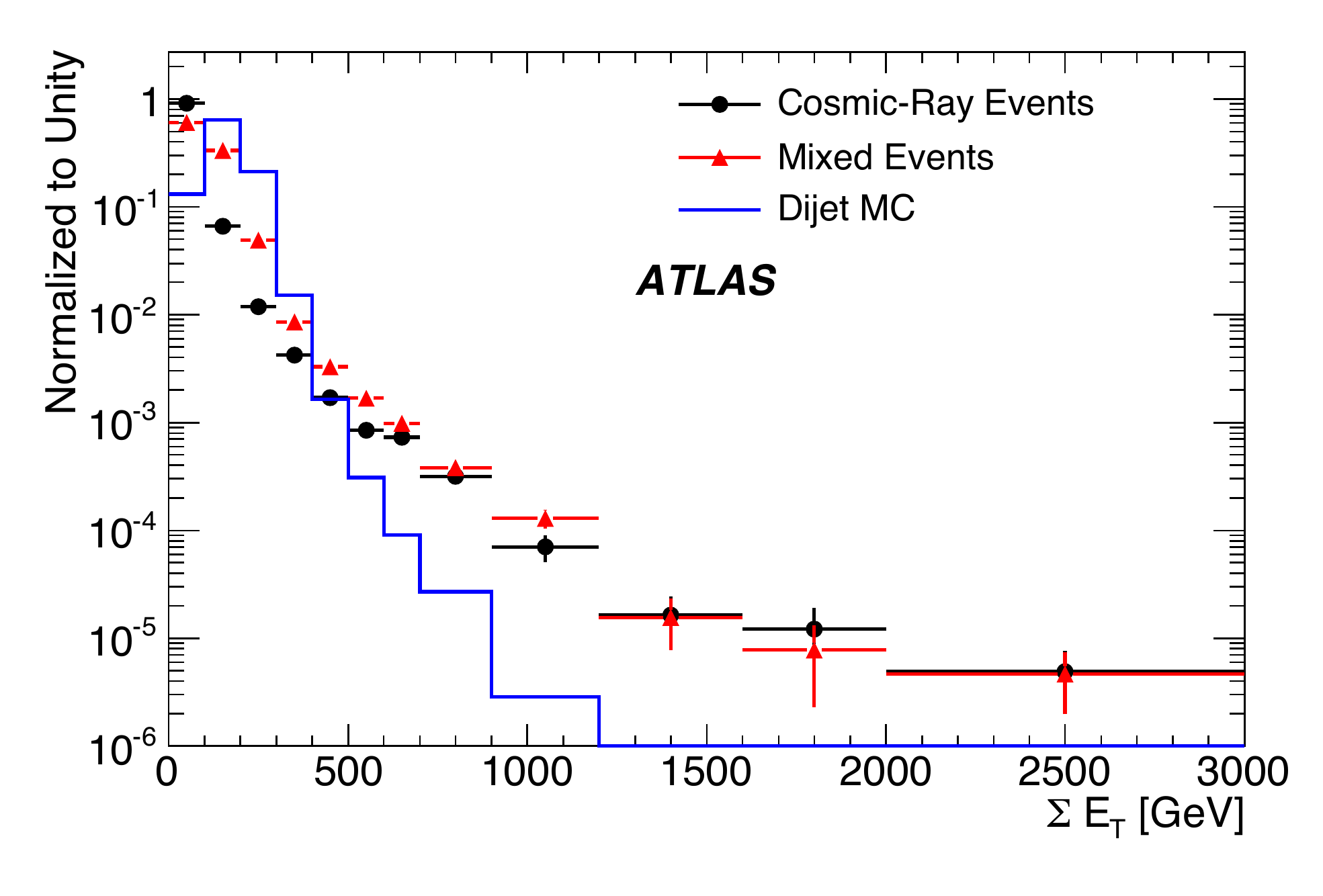}}
  \end{center}
  \caption{The same distributions as presented in Figure~\ref{fig:jet1}, obtained from
the cosmic-ray data and from the mixed data sample described in the text. The plots are normalized to allow comparison
of the shapes of the two distributions. Also shown are the corresponding distributions from dijet
Monte Carlo events.
    \label{fig:jet3}}
\end{figure*}

\begin{figure*}[phtb]
  \begin{center}
    \resizebox{0.49\textwidth}{!}{\includegraphics{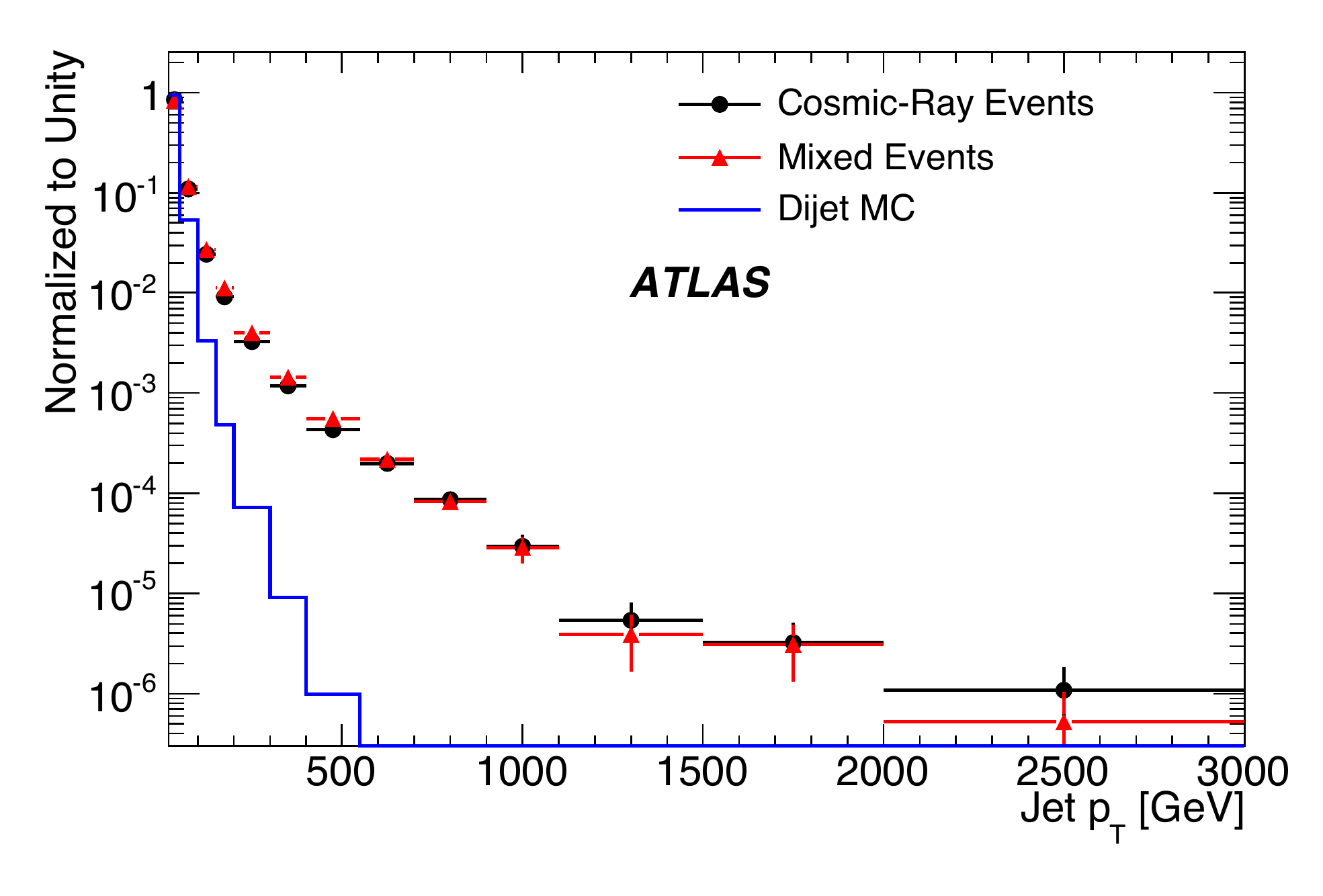}}
    \resizebox{0.49\textwidth}{!}{\includegraphics{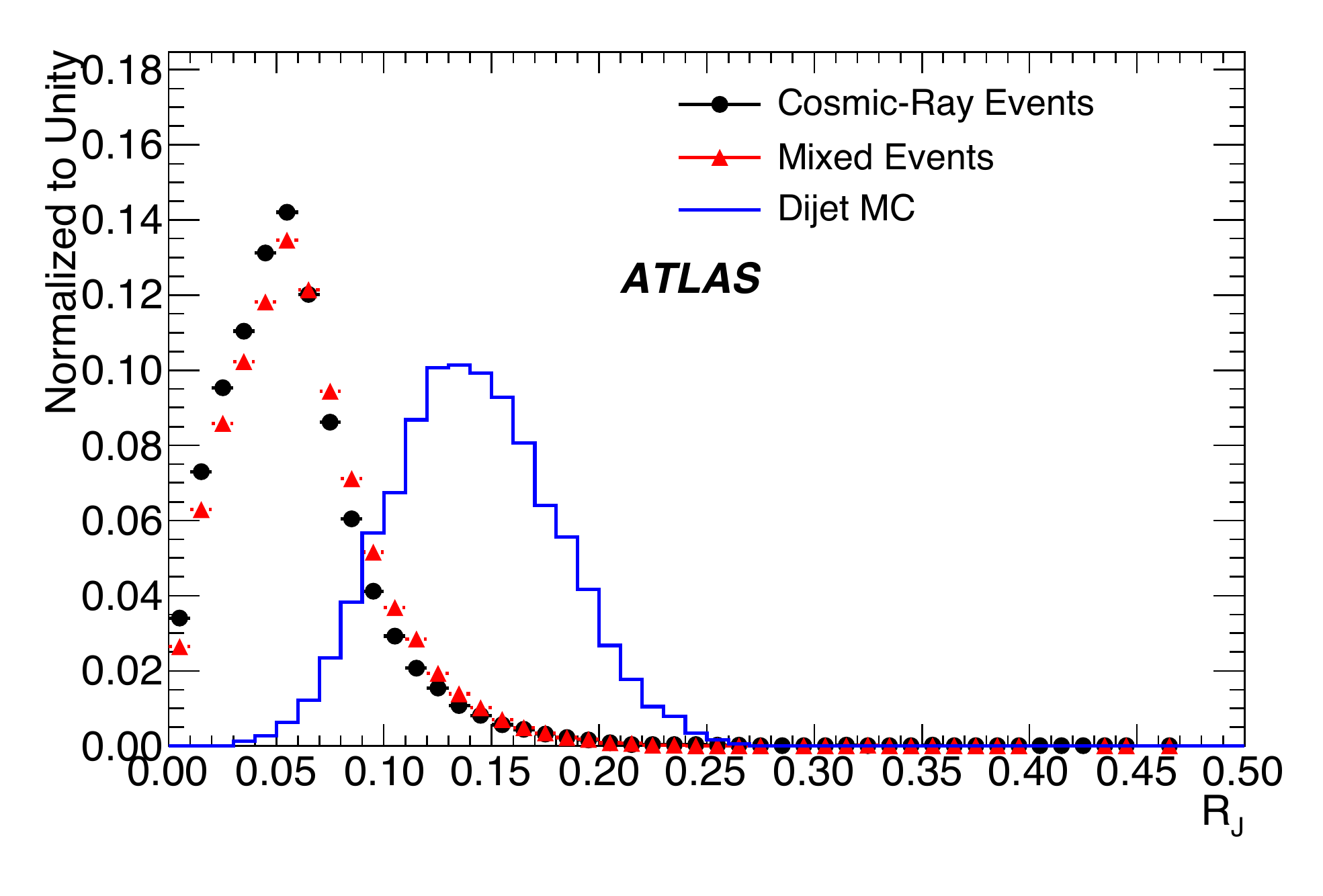}}
    \resizebox{0.49\textwidth}{!}{\includegraphics{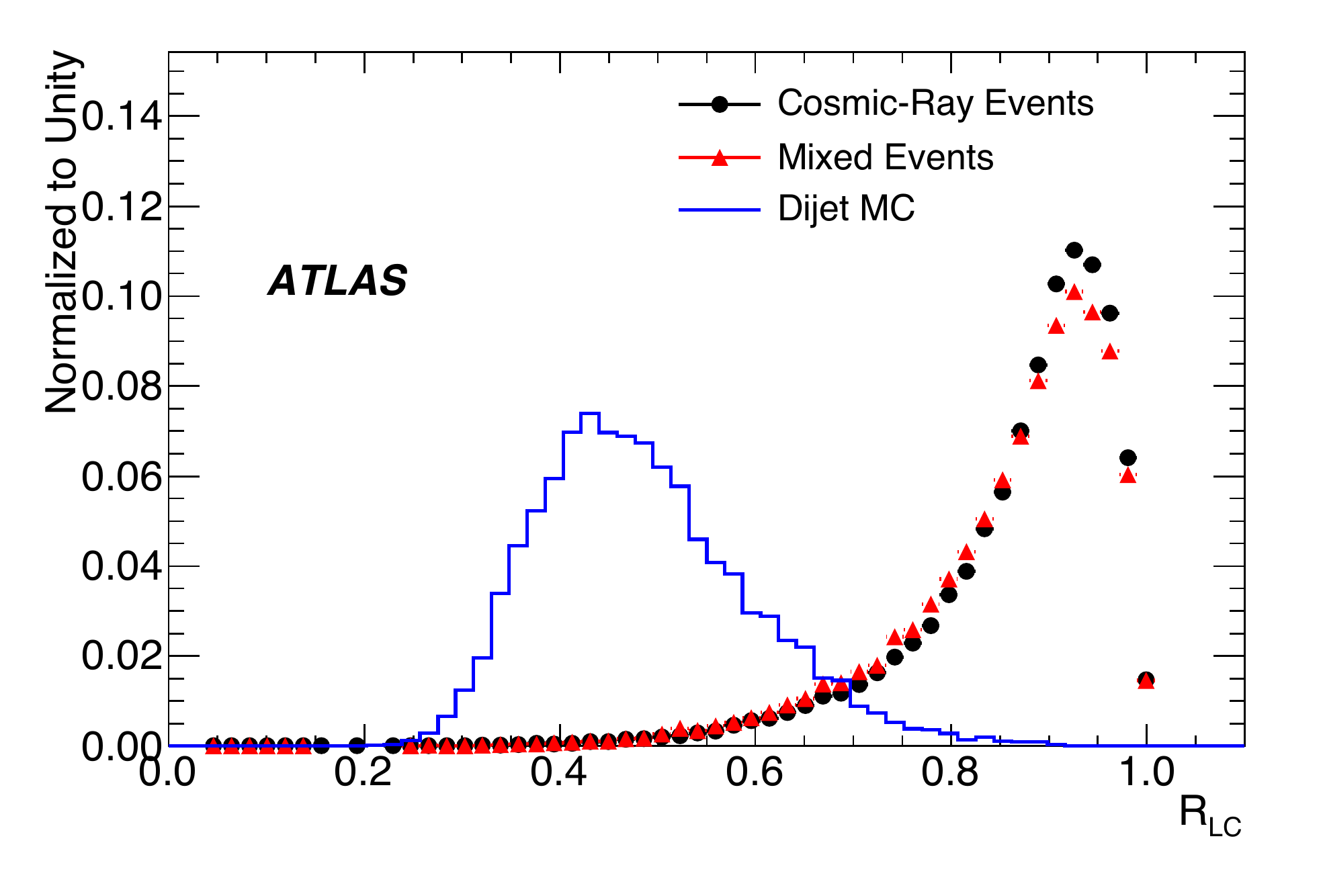}}
    \resizebox{0.49\textwidth}{!}{\includegraphics{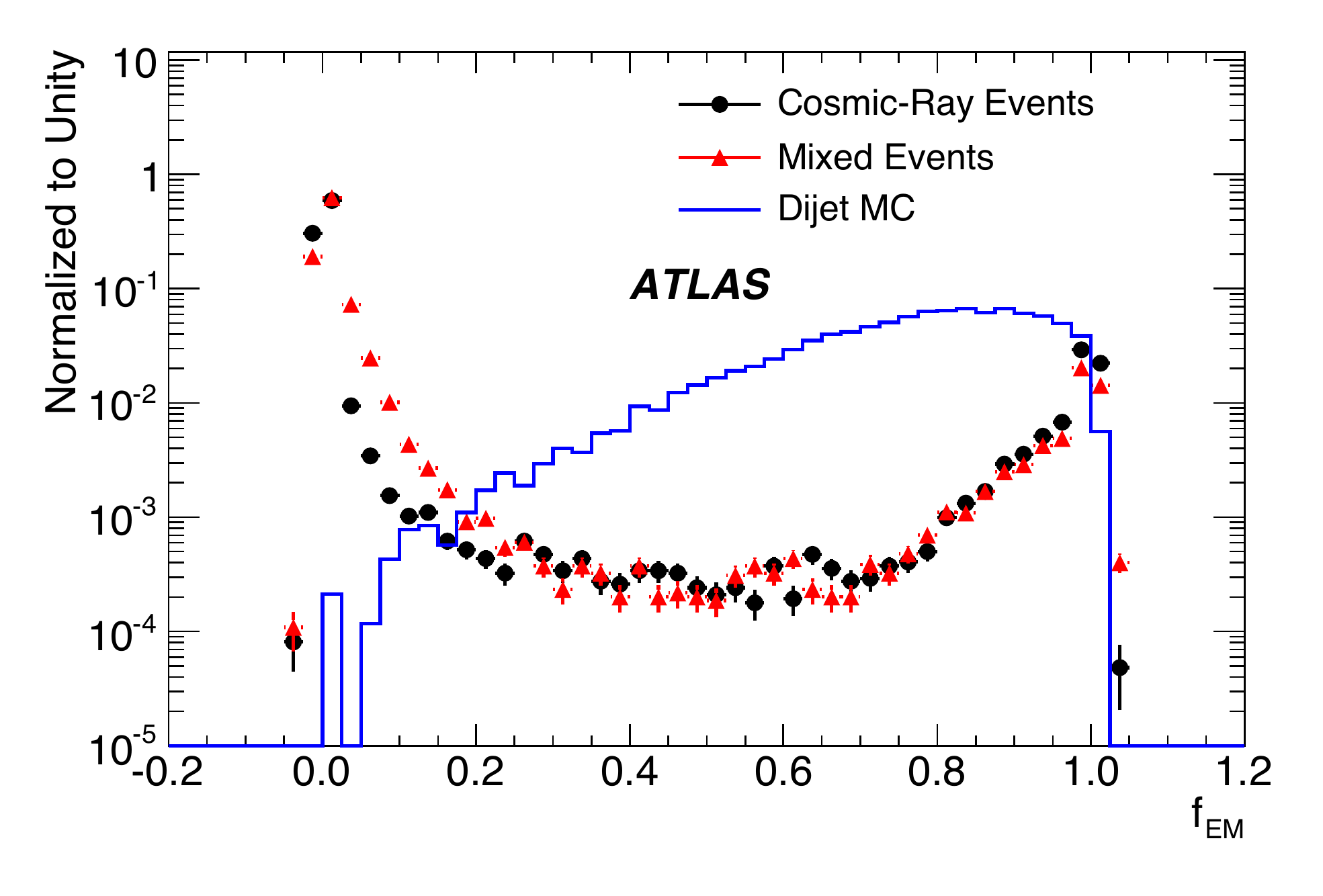}}
  \end{center}
  \caption{The same distributions as presented in Figure~\ref{fig:jet2}, obtained from the 
cosmic-ray data and from the mixed data sample. The plots are normalized to allow comparison
of the shapes. Also shown are the corresponding distributions from dijet
Monte Carlo events.
    \label{fig:jet4}}
\end{figure*}

When operating ATLAS for proton-proton collisions, contributions from cosmic-ray events can either trigger readout of the 
detector, or overlap with a triggered collision event. Since cosmic-ray energy deposits in the latter category may be more 
difficult to identify, this scenario  has been studied using a special data sample in which cosmic-ray events 
from the 2008 data were overlaid with Monte Carlo minimum-bias events. The overlay is done only with single 
minimum-bias events, so cannot account for events with pileup. However, in terms of faking a missing $E_{\rm T}$ signal, one might 
expect that the relative contribution, of a single cosmic-ray event, to a collision event would be highest in the case of
overlap with a single collision. The effect of this additional energy on the $\met$ 
and $\sum E_{\rm T}$ distributions is illustrated in Figure~\ref{fig:jet3} which compares the distributions obtained from the 
mixed sample to those obtained from cosmic-ray data alone. The corresponding distributions obtained from 
a dijet Monte Carlo sample are also shown. In each case the distributions are obtained from all events having a jet with 
$p_{\rm T} > 20$~GeV and $|\eta|<2.5$. They are shown normalized to unity to allow better comparison of the shapes.
The effect of the additional energy from the minimum-bias event is apparent in the 
$\sum E_{\rm T}$ distribution, at low values.

The mixed data sample was used to investigate the robustness of the jet-discrimination variables in the case where a 
cosmic-ray event is overlaid with a minimum-bias event. The distributions shown in Figure~\ref{fig:jet4} are for the same
quantities shown in Figure~\ref{fig:jet2}, now normalized to unity. Each plot shows 
the distribution obtained from cosmic-ray data, from the mixed sample and from a sample of Monte Carlo dijet events. 
For the three variables introduced earlier, comparison of the distributions obtained from the two samples shows these variables 
to be robust against the presence of the additional energy due to the minimum-bias event. This was not the case for other 
discriminating variables ({\it e.g.}, the number of clusters or tracks included in jets) that were also investigated. Rejection of 
fake jets from cosmic-ray events can be performed using a log-likelihood ratio (LLR) based on input probability distribution
functions (pdfs) from the mixed sample 
and the Monte Carlo dijet sample. As investigations of the three discriminating variables showed a high 
degree of correlation between $R_{\rm J}$ and $R_{\rm LC}$, a 2-dimensional pdf for these two variables was employed along with a 
one-dimensional pdf for $f_{\rm EM}$.

\begin{figure}[phtb]
  \begin{center}
    \resizebox{0.49\textwidth}{!}{\includegraphics{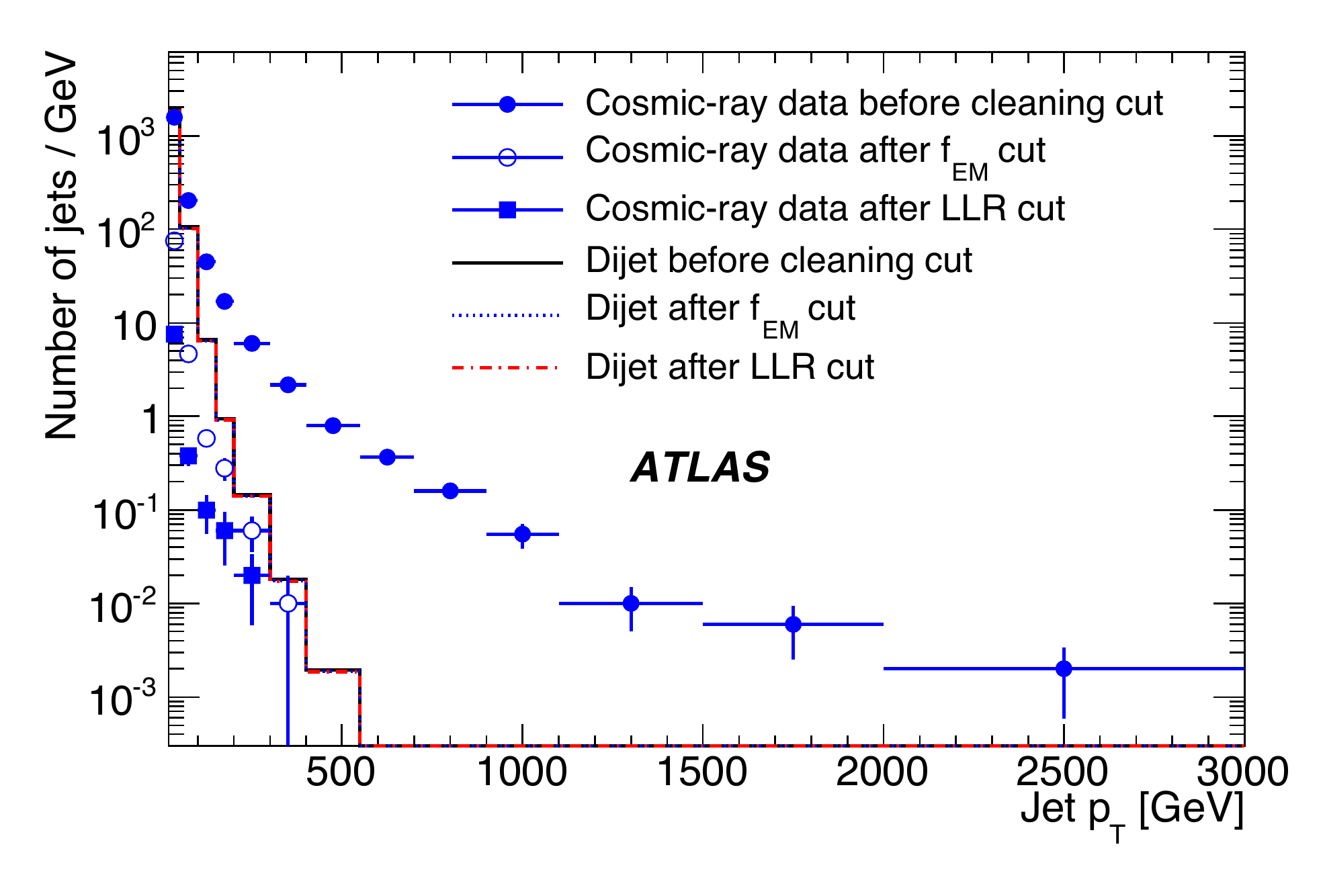}}
    \resizebox{0.49\textwidth}{!}{\includegraphics{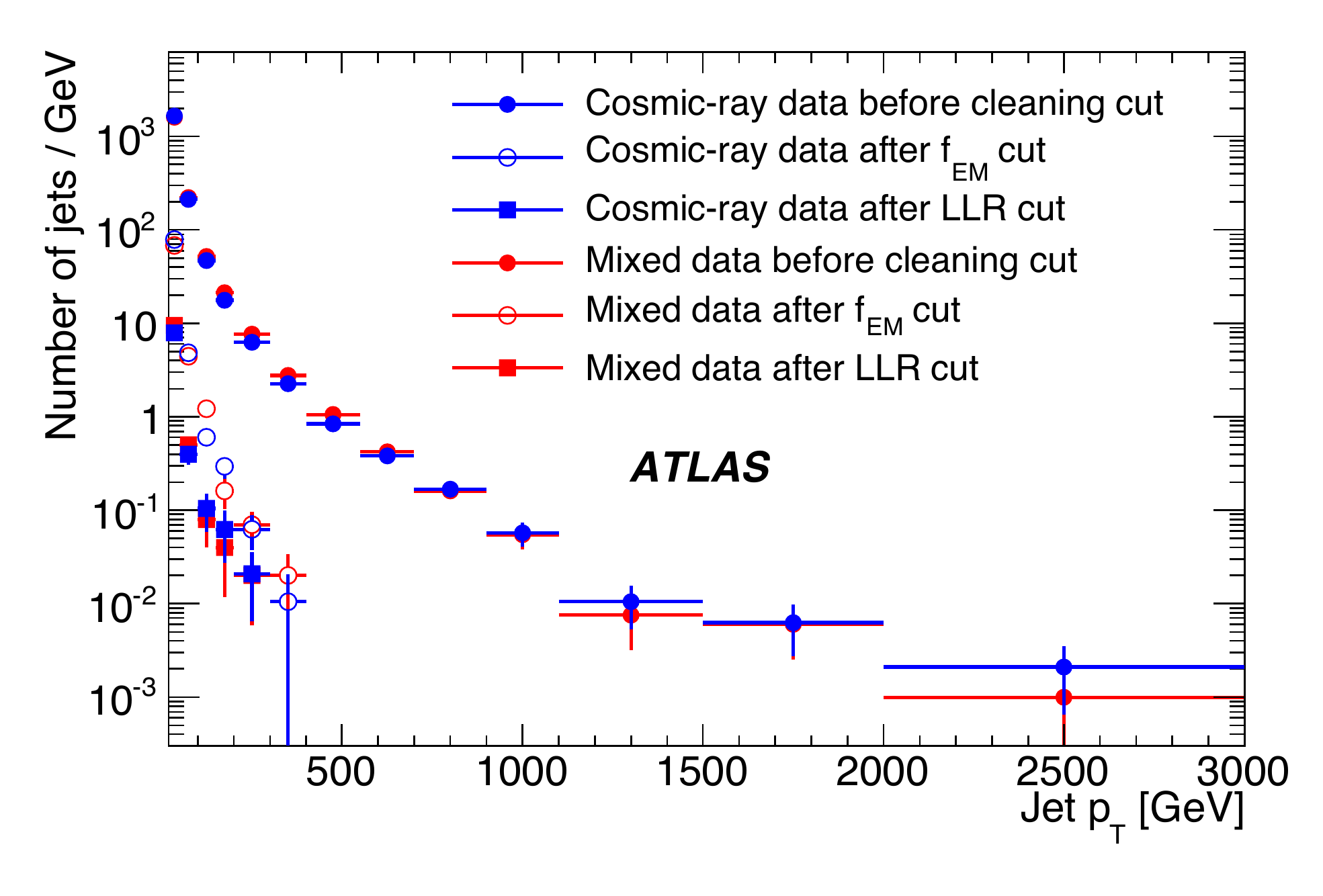}}
    \resizebox{0.49\textwidth}{!}{\includegraphics{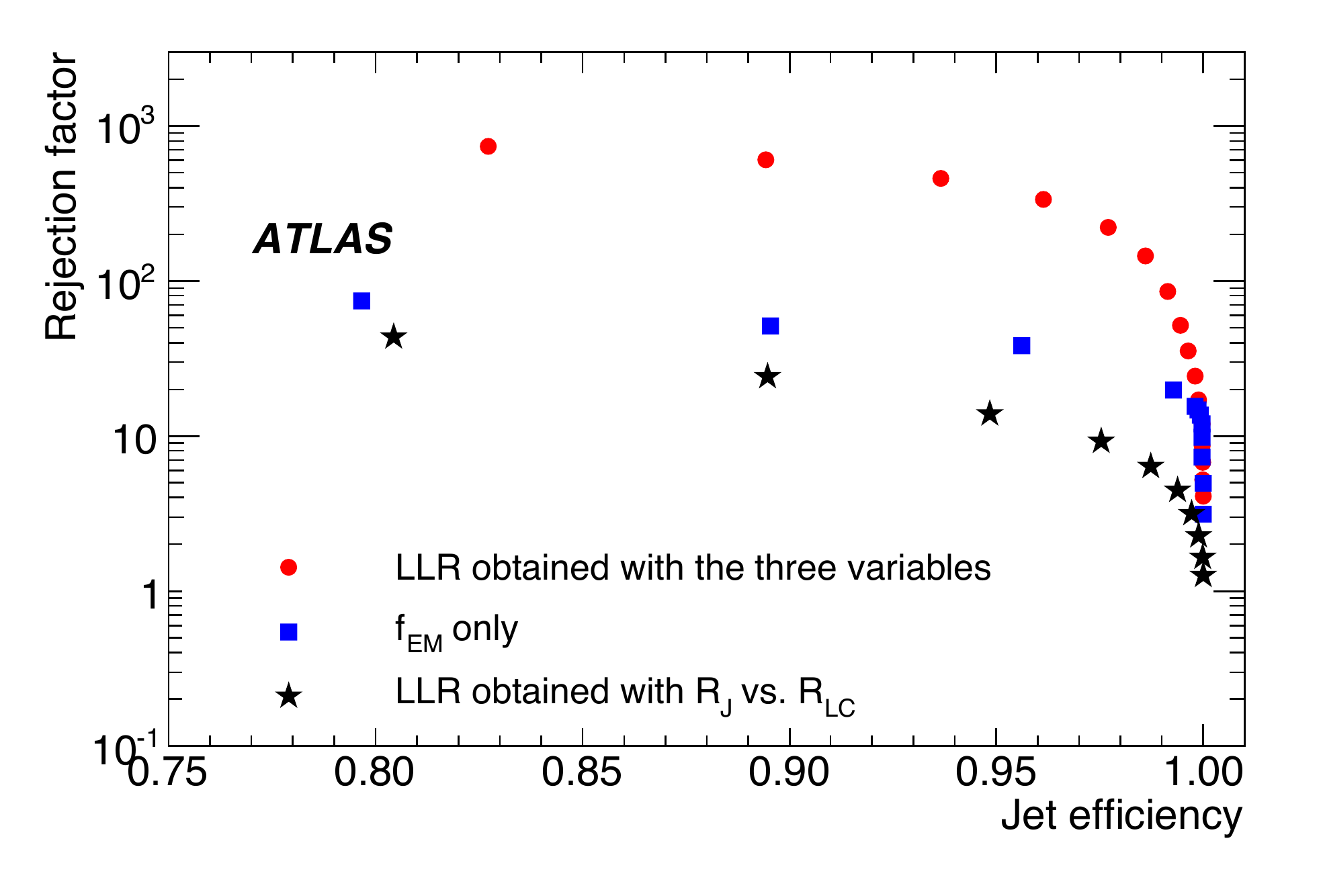}}
  \end{center}
  \caption{Performance of the cleaning cuts for the suppression of fake jets from cosmic-ray events. The upper plot shows the effect 
of the cleaning cuts on the $p_{\rm T}$ distribution of accepted jets, for different cleaning cuts, for the cosmic-ray and
dijet Monte Carlo samples. The middle plot shows similar distributions, this time comparing the mixed events to those obtained using 
only cosmic-ray data. The lower plot shows the achievable cosmic-ray rejection vs. the efficiency for dijet events.
    \label{fig:jet5}}
\end{figure}

Figure~\ref{fig:jet5} illustrates the effects of different applications of ``cleaning cuts'' based on these pdfs.  
The upper plot shows the cumulative effect of successive applications of the two LLR cuts 
on the $p_{\rm T}$ distribution from the dijet sample and on the fake jet $p_{\rm T}$ distribution from 
cosmic-ray events. For the chosen cuts, the effect of each cut on the dijet sample is at the 2\% level in each of the $p_{\rm T}$ bins.
The middle plot compares the effect of the same cuts on the mixed and cosmic-ray data samples. 
The lower plot shows the rejection factor for events with jets produced by cosmic-ray interactions plotted against 
the efficiency for the selection of Monte Carlo dijet events, in the acceptance region previously defined, for three different 
scenarios:
\begin{itemize}
\item application of an LLR cut based on $f_{\rm EM}$ only
\item application of a LLR cut based only on $R_{\rm J}$ and $R_{\rm LC}$
\item application of the full, three-variable LLR.
\end{itemize}
The rejection factor is obtained from an analysis of the mixed sample while the efficiency is derived from application of 
the selection to the dijet Monte Carlo sample. An overall rejection factor of about 400 can be obtained with 95\% efficiency for 
jets from the dijet Monte Carlo sample. For cosmic-ray events without overlaid minimum-bias energy, the situation is somewhat 
better, with a rejection factor (again for 95\% efficiency for jets in dijet events) of around 550.

\section{Summary}

Cosmic-ray interactions provide a source of physics signals in the ATLAS detector that have 
allowed for investigations of the detector alignment, calibration and performance prior to 
the arrival of first LHC beams. Such events have been used to exercise the detector readout 
and associated data-handling infrastructure, and the accumulated datasets have been exploited 
for both standalone and combined performance studies of the detector subsystems. 
Cosmic-ray data will continue to be relevant to the commissioning of the muon spectrometer until
a sufficient number of high-$p_{\rm T}$ muons have been accumulated from proton-proton collisions.
In this paper, results relevant to lepton identification and reconstruction as well as the measurement of missing transverse 
energy were presented, along with studies related to the rejection of background from cosmic-ray events in
collision data. These results, along with those presented in the publications describing results from 
subsystem-specific cosmic-ray commissioning, demonstrate that ATLAS was prepared for the first collisions
from the LHC. Measured distributions obtained from analysis of the cosmic-ray data agree well with the predictions 
of the detector simulation and a dedicated cosmic-muon event generator, demonstrating that the modeling of the 
detector response was also in good shape prior to first collisions.
%
%
\section{Acknowledgements}

\input{Acknowledgement-08Oct10}

%
%
\bibliographystyle{atlasstylem}
\bibliography{cpcosmics}

\newpage
\input{cpcosmics_authlist_final}

\end{document}

%% file: Acknowledgement-08Oct10.tex
We are greatly indebted to all CERN's departments and to the LHC project for their immense efforts not only in building the LHC, but also for their direct contributions to the construction and installation of the ATLAS detector and its infrastructure. We acknowledge equally warmly all our technical colleagues in the collaborating Institutions without whom the ATLAS detector could not have been built. Furthermore we are grateful to all the funding agencies which supported generously the construction and the commissioning of the ATLAS detector and also provided the computing infrastructure.

The ATLAS detector design and construction has taken about fifteen years, and our thoughts are with all our colleagues who sadly could not see its final realisation.

We acknowledge the support of ANPCyT, Argentina; YerPhI, Armenia; ARC, Australia; BMWF, Austria; ANAS, Azerbaijan; SSTC, Belarus; CNPq and FAPESP, Brazil; NSERC, NRC and CFI, Canada; CERN; CONICYT, Chile; CAS, MOST and NSFC, China; COLCIENCIAS, Colombia; MEYS (MSMT), MPO and CCRC, Czech Republic; DNRF, DNSRC and Lundbeck Foundation, Denmark; ARTEMIS, European Union; IN2P3-CNRS, CEA-DSM/IRFU, France; GNAS, Georgia; BMBF, DFG, HGF, MPG and AvH Foundation, Germany; GSRT, Greece; ISF, MINERVA, GIF, DIP and Benoziyo Center, Israel; INFN, Italy; MEXT and JSPS, Japan; CNRST, Morocco; FOM and NWO, Netherlands; RCN, Norway;  MNiSW, Poland; GRICES and FCT, Portugal;  MERYS (MECTS), Romania;  MES of Russia and ROSATOM, Russian Federation; JINR; MSTD, Serbia; MSSR, Slovakia; ARRS and MVZT, Slovenia; DST/NRF, South Africa; MICINN, Spain; SRC and Wallenberg Foundation, Sweden; SER,  SNSF and Cantons of Bern and Geneva, Switzerland;  NSC, Taiwan; TAEK, Turkey; STFC, the Royal Society and Leverhulme Trust, United Kingdom; DOE and NSF, United States of America.

%% file: cpcosmics_authlist_final.tex
\begin{flushleft}
{\Large The ATLAS Collaboration}
\bigskip

G.~Aad$^{\rm 48}$,
B.~Abbott$^{\rm 111}$,
J.~Abdallah$^{\rm 11}$,
A.A.~Abdelalim$^{\rm 49}$,
A.~Abdesselam$^{\rm 118}$,
O.~Abdinov$^{\rm 10}$,
B.~Abi$^{\rm 112}$,
M.~Abolins$^{\rm 88}$,
H.~Abramowicz$^{\rm 153}$,
H.~Abreu$^{\rm 115}$
% ...
\bigskip

% ...
$^{10}$ Institute of Physics, Azerbaijan Academy of Sciences, H. Javid Avenue 33, AZ 143 Baku, Azerbaijan\\
$^{11}$ Institut de F\'isica d'Altes Energies, IFAE, Edifici Cn, Universitat Aut\`onoma  de Barcelona,  ES - 08193 Bellaterra (Barcelona), Spain\\
% ...
$^{48}$ Albert-Ludwigs-Universit\"{a}t, Fakult\"{a}t f\"{u}r Mathematik und Physik, Hermann-Herder Str. 3, D - 79104 Freiburg i.Br., Germany\\
$^{49}$ Universit\'e de Gen\`eve, Section de Physique, 24 rue Ernest Ansermet, CH - 1211 Geneve 4, Switzerland\\
% ...
$^{88}$ Michigan State University, Department of Physics and Astronomy, High Energy Physics Group, East Lansing, MI 48824-2320, United States of America\\
% ...
$^{111}$ University of Oklahoma, Homer L. Dodge Department of Physics and Astronomy, 440 West Brooks, Room 100, Norman, OK 73019-0225, United States of America\\
$^{112}$ Oklahoma State University, Department of Physics, 145 Physical Sciences Building, Stillwater, OK 74078-3072, United States of America\\
% ...
$^{115}$ LAL, Univ. Paris-Sud, IN2P3/CNRS, Orsay, France\\
% ...
$^{118}$ Oxford University, Department of Physics, Denys Wilkinson Building, Keble Road, Oxford OX1 3RH, United Kingdom\\
% ...
$^{153}$ Tel Aviv University, Raymond and Beverly Sackler School of Physics and Astronomy, Ramat Aviv, IL - Tel Aviv 69978, Israel\\
% ...
\end{flushleft}
%